\documentclass{article}  

\usepackage[left=1.25in,top=1.25in,right=1.25in,bottom=1.25in,head=1.25in]{geometry}

\RequirePackage{amsthm,amsmath,amsfonts,amssymb}
\RequirePackage[authoryear]{natbib}
\RequirePackage[colorlinks,citecolor=blue,urlcolor=blue]{hyperref}
\RequirePackage{graphicx}

\theoremstyle{remark}

\def\bP{\mathbb{P}}

\def\one{\mathds{1}}

\def\P{\mathbb{P}}

\def\R{\mathbb{R}}

\def\gh
\def\cI{\mathcal{I}}
\def\cL{\mathcal{L}}

\def\cN{\mathcal{N}}

\def\one{\mathds{1}}

\newcommand{\icol}[1]{
  \left(\begin{smallmatrix}#1\end{smallmatrix}\right)%
}

\def\y{\mathbf{\color{black}y}}
\def\X{\mathbf{\color{black}X}}
\def\C{\mathbf{\color{black}C}}
\def\bmu{\boldsymbol{\color{black}\mu}}
\def\bSigma{\boldsymbol{\color{black}\Sigma}}
\def\balpha{\boldsymbol{\color{black}\alpha}}
\def\bbeta{\boldsymbol{\color{black}\beta}}

\def\bI{\boldsymbol{\color{black}I}}

\def\bW{\boldsymbol{\color{black}W}}
\def\bG{\boldsymbol{\color{black}G}}
\def\bF{\boldsymbol{\color{black}F}}
\def\br{\boldsymbol{\color{black}r}}

\def\bU{\boldsymbol{\color{black}U}}

\def\bb{\boldsymbol{\color{black}b}}

\def\bD{\boldsymbol{\color{black}D}}
\def\bE{\boldsymbol{\color{black}E}}
\def\bZ{\boldsymbol{\color{black}Z}}
\def\bA{\boldsymbol{\color{black}A}}
\def\bB{\boldsymbol{\color{black}B}}
\def\bs{\boldsymbol{\color{black}s}}

\def\bzero{\boldsymbol{\color{black}0}}
\def\bGamma{\boldsymbol{\color{black}\Gamma}}
\def\bP{\boldsymbol{\color{black}P}}
\def\bQ{\boldsymbol{\color{black}Q}}
\def\bepsilon{\boldsymbol{\color{black}\epsilon}}
\def\be{\boldsymbol{\color{black}e}}
\def\ba{\boldsymbol{\color{black}a}}
\def\bb{\boldsymbol{\color{black}b}}

\def\bU{\boldsymbol{\color{black}U}}
\def\bC{\boldsymbol{\color{black}C}}
\def\bD{\boldsymbol{\color{black}D}}
\def\bT{\boldsymbol{\color{black}T}}

\usepackage{accents}
\newcommand{\dbtilde}[1]{\accentset{\approx}{#1}}

\usepackage{pbox}

 \def\one{\mathds{1}}
\usepackage{dsfont}
\usepackage[long]{optidef}            \newcommand{\argmin}{\mathop{\mathrm{argmin}}}
\newcommand{\argmax}{\mathop{\mathrm{argmax}}}

\usepackage{siunitx}
\usepackage{booktabs}
\usepackage{subfig}
\usepackage{rotating} \usepackage{algorithm}
\usepackage[noend]{algpseudocode}

\usepackage{booktabs}
\usepackage{array}

\usepackage{natbib}

\usepackage{authblk}
  \title{Modeling Cell Populations Measured By Flow Cytometry With
    Covariates Using Sparse Mixture of Regressions}
\author[1]{Sangwon Hyun\thanks{sangwonh@ucsc.edu}}
\author[2]{Mattias Rolf Cape\thanks{mcape@uw.edu}}
\author[2]{Francois Ribalet\thanks{ribalet@uw.edu}}
\author[1]{Jacob Bien\thanks{jbien@usc.edu}}
\affil[1]{Department of Data Sciences and Operations, University of Southern
  California}
\affil[2]{School of Oceanography, University of Washington}
\begin{document}
\maketitle

\begin{abstract}
The ocean is filled with microscopic microalgae called phytoplankton, which
together are responsible for as much photosynthesis as all plants on land
combined. Our ability to predict their response to the warming ocean relies on
understanding how the dynamics of phytoplankton populations is influenced by
changes in environmental conditions. One powerful technique to study the
dynamics of phytoplankton is flow cytometry, which measures the optical
properties of thousands of individual cells per second. Today, oceanographers
are able to collect flow cytometry data in real-time onboard a moving ship,
providing them with fine-scale resolution of the distribution of phytoplankton
across thousands of kilometers. One of the current challenges is to understand
how these small and large scale variations relate to environmental conditions,
such as nutrient availability, temperature, light and ocean currents.  In this
paper, we propose a novel sparse mixture of multivariate regressions model to
estimate the time-varying phytoplankton                 subpopulations while simultaneously
identifying the specific environmental covariates that are predictive of the
observed changes to these subpopulations.  We demonstrate the usefulness and
interpretability of the approach using both synthetic data and real
observations collected on an oceanographic cruise conducted in the north-east
Pacific in the spring of 2017.

\end{abstract}

\section{Introduction}

Marine phytoplankton are responsible for as much photosynthesis as all plants on
land combined, making them a crucial part of the earth's biogeochemical cycle
and climate \citep{Field237}. A better understanding of the ecology of marine
phytoplankton species and their relationship with the ocean environment is
therefore important both to basic biology and to shedding light on their role in
carbon dioxide uptake.
In order to study these single cell organisms in the ocean, flow cytometry has
been instrumental for the past three decades \citep{Sosik2010}.

Flow cytometry measures light scatter and fluorescence emission of individual
cells at rates of up to thousands of cells per second. Light scattering is
proportional to cell size, and fluorescence is unique to the emission spectra
of pigments; these parameters can be used to identify populations of
phytoplankton with similar optical properties. Over the two decades, automated
environmental flow cytometers such as CytoBuoy \citep{Dubelaar1999}, FlowCytoBot
\citep{Olson2003}, and SeaFlow \citep{seaflow} have provided an unprecedented
view of dynamics of phytoplankton across large temporal and spatial scales.

\begin{figure}[t!]
\centering
\centering
\includegraphics[width=\linewidth]{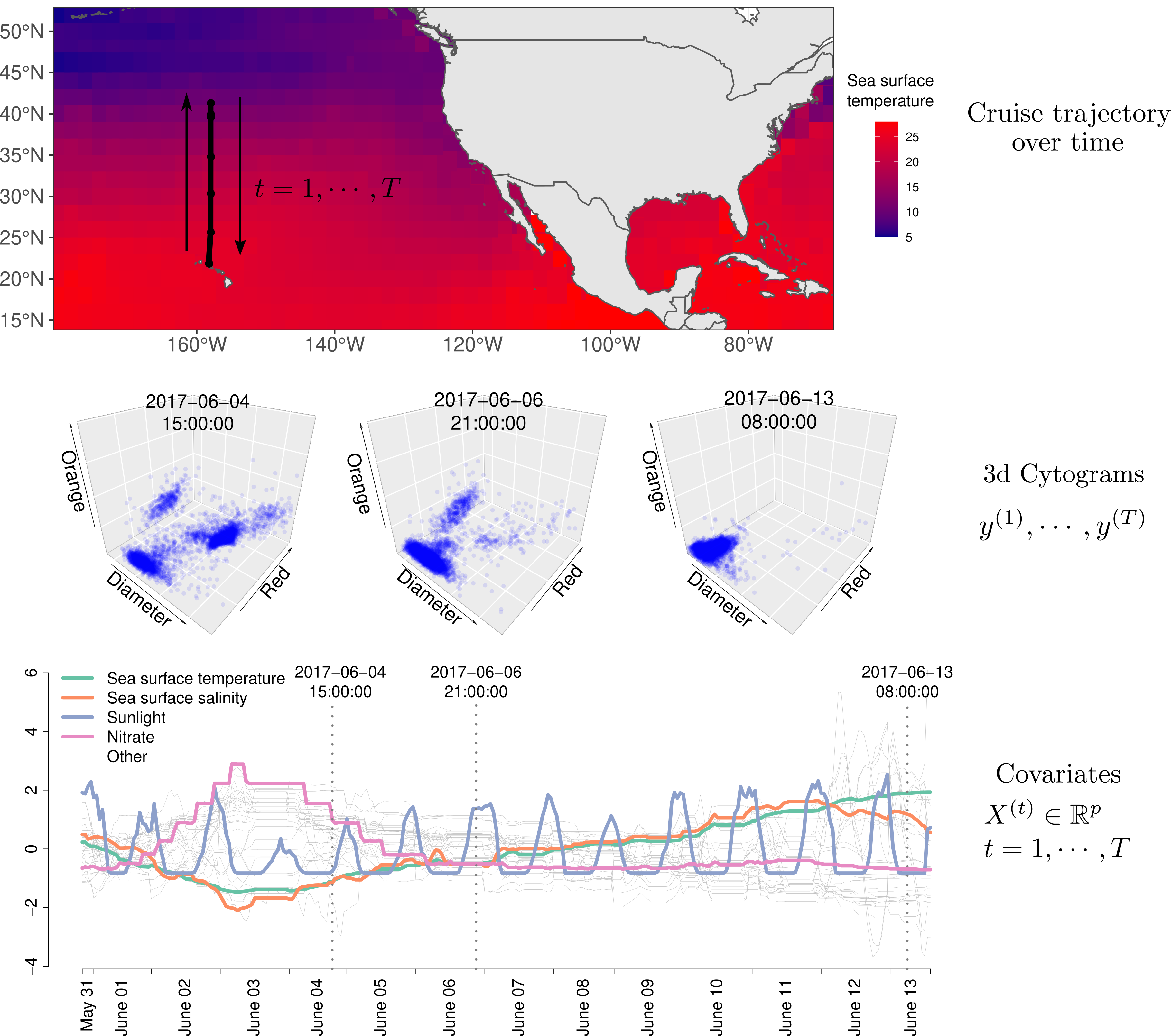} 
\caption{\it A schematic showing the data setup. (Top) This figure shows the
  trajectory of the Gradients 2 cruise, which moves North and then South along a
  trajectory starting at Hawaii. (Middle) The individual $3$-dimensional
  particles are measured rapidly and continuously. From this, we form $T$
  cytograms $\y^{(t)}, t=1,\cdots, T$ at an hourly time resolution. The three
  data dimensions have simplified labels Red, Orange, and Diameter; the first
  two represent fluorescence emission, and the last measures cell
  diameter. (Bottom panel) At each time $t=1,\cdots, T$, environmental
  covariates $\X^{(t)} \in \mathbb{R}^p$ are also available through remote
  sensing and on-board measurements.  Only a few of the 30+ normalized
  covariates are highlighted here. Our proposed model identifies subpopulations
  by modeling them as Gaussian clusters whose means and probabilities are driven
  by environmental covariates.  }
\label{fig:intro}
\end{figure}

Automated in-situ flow cytometry data can be represented as a scatterplot-valued
time series, $\y^{(1)},\ldots, \y^{(T)}$, where an $n_t$ by $d$ matrix $\y^{(t)}$
whose rows are vectors $\{ \y^{(t)}_i\in\R^d : i=1, \cdots, n_t\}$ is called a
{\em cytogram} and can be thought of as a $d$-dimensional scatterplot
representing $n_t$ particles observed during time interval $t$.  The $d$
dimensions of the scatterplot represent $d$ optical properties that are useful
in distinguishing different cell types from each other. Figure~\ref{fig:intro}
shows an example of three cytograms collected by SeaFlow in June 2017 during a
two-week cruise conducted in the Northeast Pacific.  With SeaFlow, cytograms are
of dimension $d = 3$.

As apparent in the figure, the points within the cytograms display clear
clustering structure.  These different clusters correspond to cell populations
of different types of phytoplankton.  As the environmental conditions change,
the populations change over time. In particular, in optical space, two
noticeable phenomena over time are:
\begin{enumerate}
\item The number of cells in a given population can increase or decrease, with
  populations sometimes even appearing or disappearing entirely.
\item The centers of the cell populations are not fixed, but rather
  move over time.
\end{enumerate}
Using expert knowledge and close manual inspection, oceanographers have been
able to explain how some of these phenomena can be attributed to specific
changes in environmental factors (e.g., oscillations in cell size due to
sunlight and cell division) \citep{Vaulot1999, Sosik2003, Ribalet2015}.

Our goal is to develop a statistical approach for identifying how environmental
factors can be predictive of changes to the cytograms.  The promise of such a
tool would be to discover new relationships between cell populations and
environmental factors beyond those that may be known, or visible to the human
eye.

Based on these observations and with this goal in mind, our statistical model
for time-varying cytograms postulates a finite mixture model in which both the
cluster probabilities and centers are allowed to vary over time.  Changes to the
cluster probabilities over time can capture the growing/shrinking and
appearing/disappearing described above, while changes to the centers over time
can capture the drifting/oscillating.

To be clear, our method does not explicitly incorporate the time (or space)
aspect of the data. Instead, in our model, these cluster probabilities and
centers are controlled by $p$ time-varying covariates $\X^{(t)}\in\R^p$.  While
our model can accommodate features that are purely functions of time (e.g.,
$\sin t$, $t^2$, spline basis functions, etc.), our focus here is on
environmental covariates. Our analysis uses biological, physical, and chemical
variables, shown in the bottom panel of Figure \ref{fig:intro}, that were
retrieved from the Simons Collaborative Marine Atlas Project (CMAP) database
(\url{https://simonscmap.com}), which is a public database compiling various
oceanographic data over space and time.

\begin{figure}[ht!]
  \includegraphics[width=\linewidth]{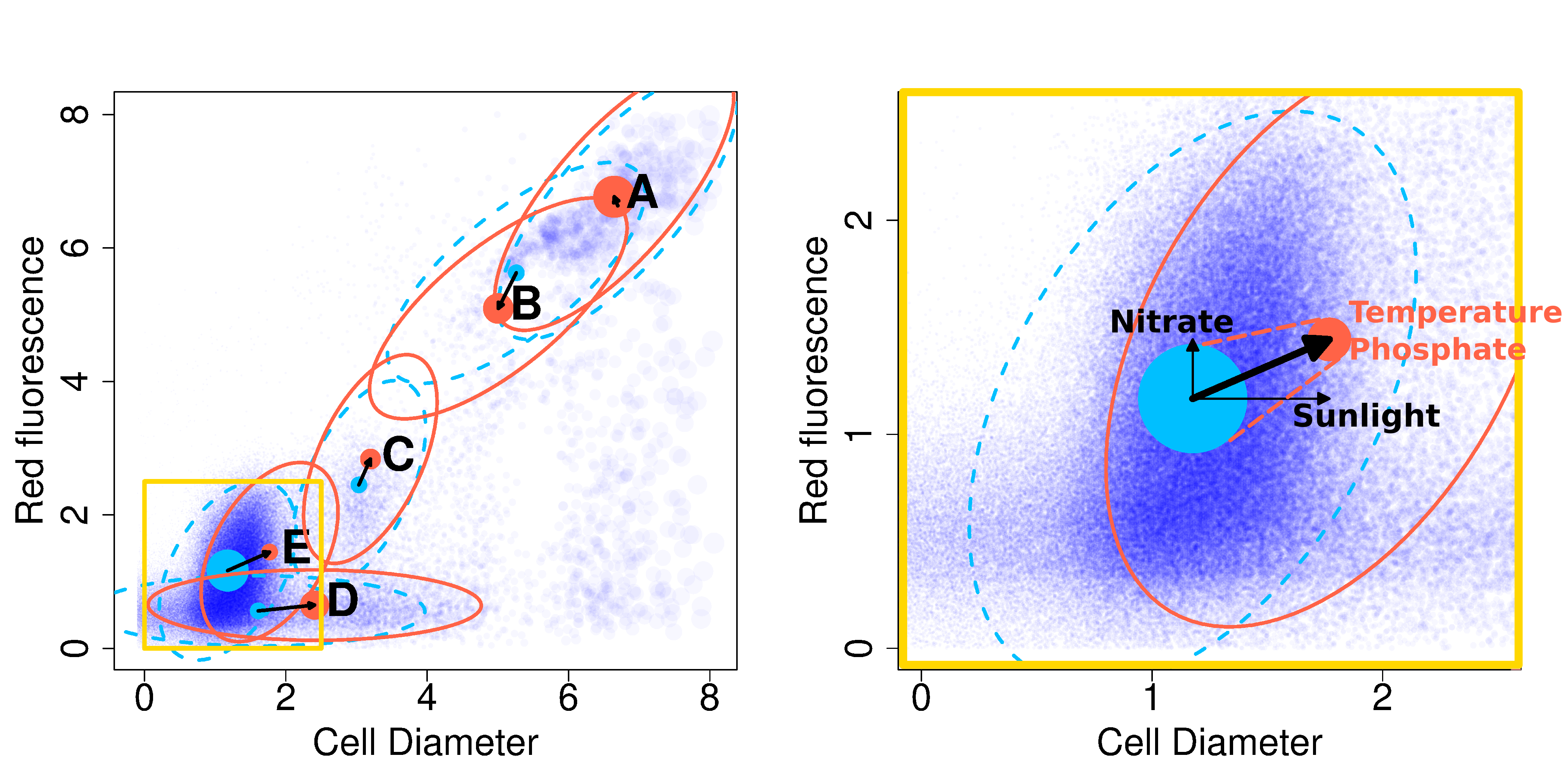}
  \caption{\it Our method produces estimates of cluster centers (shown as disks)
    and cluster probabilities (represented by the size of the disk) for every
    time point.  The covariance of each mixture component (represented by an
    ellipse) is assumed to be constant over time. Blue and red show parameter
    estimates at two different time points. In the background, particles from
    only one time point are shown in (partially transparent) dark blue with the
    size of a point proportional to the particle's biomass.  The right figure
    takes a closer look at a subregion of the cytogram shown in the lower left
    corner of the left figure, focusing on cluster E which is a
    \textit{Prochlorococcus} population.  The change in the probability of
    cluster E is well predicted by sea surface temperature and phosphate, and
    the horizontal and vertical movement of cluster E's center are each
    predicted by time-lagged sunlight and nitrate. Note, we are showing only
    five of the ten clusters used for estimation.}
\label{fig:intro-model}
\end{figure}

One key strength of our method is the \textit{variable selection} property,
allowing the analyst to identify the subset of covariates that are the strongest
predictor of each cluster's mean and probability movement over time. For
instance, in Figure~\ref{fig:intro-model}, the estimated coefficients reveal
that higher sea surface temperature and lower phosphate can predict a decrease
in probability of cluster E located in the lower-left corner, and time-lagged
sunlight and nitrate well predict the horizontal and vertical movement of
cluster E's center.

Our framework represents a substantial improvement in the detail and richness of
how this data can be modeled and analyzed.  Flow cytometry data are traditionally
analyzed by a technique called \textit{gating}, which counts the number of cells
falling into certain fixed, expert-drawn polygonal regions of $\R^d$
corresponding to each cell population \citep{gating}, reducing each scatterplot
into several counts (giving the number of cells in each gated region)
\citep{seaflow-clustering}. Subjectivity in manual gating has been shown to be
an obstacle to reproducibility \citep{flowcore}. Furthermore, the presence of
overlapping cell communities suggests that hard assignments to fixed disjoint
regions may not be advisable.  These and other shortcomings have led multiple
authors to develop mixture model based approaches, as discussed in
\citet{flow-cytometry-techniques-review}.  While such models are an improvement
over traditional gating, they do not naturally extend to oceanography in which
we have a time series of cytograms.  Naively, one might think one could get away
with fitting a separate mixture model to each individual cytogram.  However,
doing so leaves one with the problem of matching clusters from distinct
clusterings, a task made particularly challenging since these clusters can move,
change in size, and appear/disappear.  Our approach fits a single mixture model
{\em jointly} across the entire time series while integrating information from
the covariates.  By using all data sources in a single mixture model, our method
is able to estimate the distinct components, even in cases where two
populations' centers may be nearby or a cluster may sometimes vanish.

In the statistics literature, the term \textit{finite mixture of regressions} is
used to refer to mixture models in which (univariate) means are modeled as
functions of covariates (see, e.g., \citealt{finite-mixture-model-book}). Early
works such as \citet{poisson-mixture-regression} use information criteria and
exhaustive search while more modern approaches have used penalized sparse models
\citep{finite-mixture-regression-variable-selection,
  finite-mixture-regression-high-dim}.  Our methodology differs from these
methods in three respects: first, our means are multivariate ($d$-dimensional);
second, the mixture weights are also modeled as functions of the covariates;
third, the model coefficients are penalized. Of these, the first two aspects are
shared by \citet{flexmix-v2}, but without penalization. The idea of allowing the
mixture weights to be functions of the features is more common in the machine
learning literature, where such models are called {\em mixtures of experts}
\citep{mixture-of-experts-original}.

To the best of our knowledge, this is the first attempt to extend mixture
modeling of flow cytometry data by directly linking mixture model parameters
with environmental covariates via sparse multivariate regression models. In
Section \ref{sec:methodology}, we describe our proposed model in detail.  In
section \ref{sec:numerical}, we use our proposed model to draw rich new insights
from a marine data source. We also conduct two realistic numerical simulations
based on some pseudo-real ocean flow cytometry data. We provide an R package
called \texttt{flowmix} that can be run both on a single machine, and also on
remote high performance servers that use a parallel computing environment. While
our focus is on time-varying flow cytometry in the ocean, our method can be
applied more broadly to any collection of cytograms with associated
covariates. For example, in biomedical applications each cytogram could
correspond to a blood sample from a different person, and person-specific
covariates could model the variability in cytograms.

 \section{Methodology}
\label{sec:methodology}

\subsection{Likelihood function of cytogram}\label{sec:likelihood}

We model the $n_t$ particles $\{\y^{(t)}_i\}_{i=1}^{n_t}$ measured at time $t$ as
i.i.d. draws from a probabilistic mixture of $K$ different $d$-variate Gaussian
distributions, conditional on the covariate vector $\X^{(t)} \in \R^p$. The
latent variable $Z_i^{(t)}$ determines the cluster membership,
\begin{equation} \label{eq:latent}
  P(Z_i^{(t)}  = k|\X^{(t)}) = \pi_{kt},\; k=1,\cdots, K,
\end{equation} 
and the data is drawn from the $k$'th Gaussian distribution,
\begin{equation*} 
  (\y^{(t)}_i | \X^{(t)}, Z_i^{(t)}=k) \sim \cN_d \left(\bmu_{kt}, \bSigma_k\right),
\end{equation*}
where the cluster center $\bmu_{kt} \in \R^d$ and cluster
probability $\pi_{kt}$ at time $t$ are modeled as functions of $\X^{(t)}$:
\begin{align*}
  \bmu_{kt}(\beta) &= \bbeta_{0k} + \bbeta_k^T \X^{(t)}\\
  \pi_{kt}(\alpha) &= \frac{\exp(\alpha_{0k} + {\X^{(t)}}^T \balpha_k)}{\sum_{l=1}^K \exp(\alpha_{0l} + {\X^{(t)}}^T \balpha_l)}
\end{align*}
for regression coefficients $\bbeta_{0k} \in \R^d$,
$\bbeta_{k} \in \R^{p \times d}$, $\balpha_k \in \R^p$, and $\alpha_{0k} \in \R$;
throughout, we use $\alpha$, $\beta$, and $\Sigma$ to denote the collection of
coefficients $\{\alpha_{0k}, \balpha_k\}_{k=1}^K$,
$\{\bbeta_{0k}, \bbeta_k\}_{k=1}^K$, and $\{\bSigma_k\}_{k=1}^K$ for
brevity. Since all random variables
are conditional on the covariates $\X^{(t)}$, we will omit it hereon for
brevity.  Denoting the density of the $k$'th Gaussian component of data at time
$t$ as $\phi(\cdot; \bmu_{kt}, \bSigma^{(k)})$, the log-likelihood function is
\begin{equation}
  \label{eq:loglik}
  \log \cL(\alpha, \beta, \Sigma;\{\y_i^{(t)}\}_{i,t}) = \sum_{t=1}^T\sum_{i=1}^{n_t} \log \left( \sum_{k=1}^K \pi_{kt}(\alpha) \cdot \phi\left(\y_i^{(t)}; \bmu_{kt}(\beta), \bSigma_{k}\right)\right).
\end{equation}
By modeling the Gaussian means $\{\bmu_{kt}\}_{k,t}$ and mixture probabilities
$\{\pi_{kt}\}_{k,t}$ as regression functions of $\X^{(t)}$ at each time point
$t=1,\cdots, T$, our model directly allows environmental covariates to predict
the two main kinds of cell population changes over time -- movement in optical
space, and change in relative population abundance. Furthermore, the signs and
magnitudes of the entries of $\balpha$ and $\bbeta$ directly quantify the contribution of
environment covariates to each population's abundance and direction of movement
in cytogram space.

\subsection{Penalties and constraints}\label{sec:penalty}

In practice, there are a large number of environmental covariates that may in
principle be predictive of a cytogram. Also, the number of regression parameters
is $(p+1)(d+1)K$, which can be large relative to the number of cytograms
$T$. Furthermore, we would prefer models in which only a small number of
parameters is nonzero. Therefore, we penalize the log-likelihood with lasso
penalties \citep{orig-lasso-paper} on $\alpha$ and $\beta$.

In our application, each cell population has a limited range in optical
properties, due to biological constraints. We incorporate this domain knowledge
into the model by constraining the range of $\bmu_{k1}, \cdots, \bmu_{kT}$ over
time. Since $\bbeta_k^T\X^{(t)} = \bmu_{kt} - \bbeta_{0k}$, limiting the size of
$\bbeta_k^T\X^{(t)}$ is equivalent to limiting the \textit{deviation} of the
$k$'th cluster mean at all times $t=1,\cdots,T$ away from the overall center
$\bbeta_{0k}$. Motivated by this, we add a \textit{hard} constraint so that
$\|\bbeta_k^T \X^{(t)}\|_2 \le r$ for some fixed radius value $r>0$.

The choice of $r$ should be specific to the data application. For 1-dimensional
cytograms of cell diameter measurements used in the analysis in
Section~\ref{sec:1d-real-data}, the size of $r$ holds the intuitive meaning of
not allowing the average optical properties of a particular cell population to
deviate more than a multiplicative upper and lower bound over time compared to
an overall average.

The constraint also plays an important role for model interpretability. We wish
for the $k$'th mixture component to correspond to the same cell population over
all time. When a cell population vanishes we would like $\pi_{kt}$ to go to zero
rather than for $\bmu_{kt}$ to move to an entirely different place in cytogram
space.

Our estimator is thus a solution to the following optimization problem:
\begin{mini}|l| {
    \alpha, \beta, \Sigma}{-\frac{1}{N} \log \cL(\alpha, \beta, \Sigma;
\{\y_i^{(t)}\}_{i,t}
    ) + \lambda_{\alpha}  \sum_{k=1}^K\| \balpha_{k} \|_1 + \lambda_{\beta} \sum_{k=1}^K\|\bbeta_{k}\|_1.}{}{}
  \addConstraint{
    \| \bbeta_k^T \X^{(t)} \|_2 \le r \;\; \forall t=1,\cdots, T \;\;\forall k =
  1, \cdots, K}.
\label{eq:objective}
\end{mini}
We divide the log-likelihood term by $N := \sum_{t=1}^T n_t$ to make the
scale consistent with that of a single particle.

\subsection{Multiplicity generalization}\label{sec:multiplicity}

Cytogram datasets can be extremely large, and cell populations can have highly
imbalanced probabilities.  To overcome the computational and methodological
difficulties posed by these issues, we generalize the model to assign to
particle $\y_i^{(t)}$ a multiplicity factor $C_i^{(t)}$ (which defaults to
$1$).The log-likelihood in \eqref{eq:loglik} becomes,
\begin{multline}
  \label{eq:loglik-mult}
  \ell_{(n_1,\cdots, n_T)}(\alpha, \beta, \Sigma; (\y^{(1)}, \cdots, \y^{(T)}), (\C^{(1)}, \cdots, \C^{(T)})) = \\\sum_{t=1}^T\sum_{i=1}^{n_t} C_i^{(t)}\log \left( \sum_{k=1}^K \pi_{kt}(\alpha) \cdot \phi\left(\y_i^{(t)}; \bmu_{kt}(\beta), \bSigma_{k}\right)\right).
\end{multline}
where $\y^{(t)} \in \R^{n_t \times d}$ and $\C^{(t)} \in \R^{n_t}$.  Furthermore,
the scaling by $N$ in the optimization objective \eqref{eq:objective} is
generalized to $N := \sum_{t=1}^T \sum_{i=1}^{n_t} C_i^{(t)}$, the overall
sum of the multiplicities.

The multiplicity generalization is useful for an approximate data representation
by placing particles in bins and dealing with bin counts. We discretize cytogram
space along a lattice of $B=D^d$ $d$-dimensional cubes $\{E_b\}_{b=1}^B$ whose
centers $\tilde \y_b\in\R^d$ can be arranged as the rows of a matrix
$\tilde \y \in \R^{B \times d}$.  This coarsened data representation involves
counts $\{C_b^{(t)}\}_{b,t}$ of the number of particles in each fixed bin
$E_b$:
\begin{equation*}
  C_b^{(t)} = \sum_{i=1}^{n_t} \one \{ i: \y_i^{(t)} \in E_b\},
\end{equation*}
whose collection is $\C^{(t)} \in \R^{B}$. Using $C_b^{(t)}$ and
$\tilde \y_b^{(t)} := \tilde \y_b$ to replace $C_i^{(t)}$ and $\y_i^{(t)}$ in
\eqref{eq:loglik-mult}, we obtain the log-likelihood of the binned data,
\begin{multline}
  \label{eq:loglik-bin}
  \ell_{(B,\cdots, B)}\left(\alpha,\beta, \Sigma; (\tilde  \y, \cdots, \tilde \y) , (\C^{(1)}, \cdots, \C^{(T)})\right)
\\  = \sum_{t=1}^T\sum_{b=1}^{B} C_b^{(t)}\log \left( \sum_{k=1}^K \pi_{kt}(\alpha) \cdot \phi\left(\tilde \y_b; \bmu_{kt}(\beta), \bSigma_{k}\right)\right).
\end{multline}

Whereas before each cytogram required its own set of $n_t$ particle locations,
in the binned data representation, the same set of locations are shared across
all $t$, which is indicated by the notation $(\tilde y, \dots, \tilde y)$.

This binned likelihood is identical to the original log-likelihood
\eqref{eq:loglik} after replacing each particle by its bin center. The
computational savings are apparent from noticing that
$\sum_{b=1}^B \one\{C_b^{(t)} \neq 0\} \ll n_t$ since typically only a small
subset of the bins $\{E_b\}$ contain any particles. Additionally, the number of
Gaussian density calculations are reduced by a factor of $T$, since the
particles $\tilde \y_b$ do not depend on $t$.

There is no finite value of $B$ for which the binned log-likelihood in
\eqref{eq:loglik-bin} is equal to the log-likelihood calculated on the original
data, due to the nonzero distance between bin centers $\tilde \y_b$ and data
$\y_i^{(t)}$ even for very large $B$. However, the following
proposition~\ref{prop:bin} establishes that parameter estimation from the binned
data is asymptotically equivalent to parameter estimation from the original
data, as the number of bins $B$ grows to $\infty$. The proof is provided in
Supplement A. As for what occurs for finite values of $B$, a simulation study
in Supplement~F suggests that even using a relatively small number of bins can
achieve similar predictive performance as using the original data.

\newtheorem{prop}{Proposition} \begin{prop}\label{prop:bin}
  Let
\begin{equation}
  \label{eq:obj1}
 \tilde \Theta_B := \argmin_{(\alpha, \beta, \Sigma) \in \Theta} \;\; -\frac{1}{N}
  \ell_{(B,\cdots, B)}\left(\alpha,\beta, \Sigma; (\tilde  \y, \cdots, \tilde \y), (\C^{(1)}, \cdots, \C^{(T)})\right)
                    + g(\alpha,\beta)
 \end{equation}
 be the set of minimizers of the penalized negative log-likelihood of the binned
 data, and let
 \begin{align}
  \label{eq:obj2}
  \hat \Theta := \argmin_{(\alpha, \beta, \Sigma) \in \Theta} \;\; -\frac{1}{N}\log \cL (\alpha, \beta, \Sigma; \{\y_i^{(t)}\}_{i,t}) + g(\alpha,\beta),
\end{align}
be that of the original data.  The term $g(\alpha, \beta)$ encapsulates the
penalties on $\alpha$ and $\beta$ and the constraint on $\beta$ in
\eqref{eq:objective}. Assume the following:
\begin{enumerate}
\item The parameter space $\Theta$ of $(\alpha, \beta, \Sigma)$ is compact, and
  $\{\lambda_{\text{min}}(\bSigma_k) < c \} \cap \Theta = \emptyset$ for all $k=1,\cdots, K$, for
  some constant $c>0$.
\item The data belongs to a compact set $\mathcal{Y}$ with $\max_{\y,\y' \in \mathcal{Y}} \| \y - \y'
    \|_\infty \le R$ for some positive constant $R<\infty$.
  \item 
    The log likelihood $\log \cL (\alpha, \beta, \Sigma; \{\y_i^{(t)}\}_{i,t}) < \infty$ for all
    $(\alpha, \beta, \Sigma) \in \Theta$.
  \end{enumerate}
  Then, given any sequence $\tilde \theta_B \in \tilde \Theta_B (B=1,2,\cdots)$
  of minimizers of the penalized negative log-likelihood of the binned data, a
  sequence $s_B$ exists such that the subsequence $\tilde \theta_{s_B}$
  converges to an element in $\hat \Theta$:
  \begin{equation}\label{eq:theory}
    \lim_{B\to\infty} \tilde \theta_{s_B} \in \hat \Theta.
  \end{equation}
\end{prop}

\begin{figure}[ht!]
  \makebox[\textwidth]{
    \includegraphics[ width=.35\linewidth]{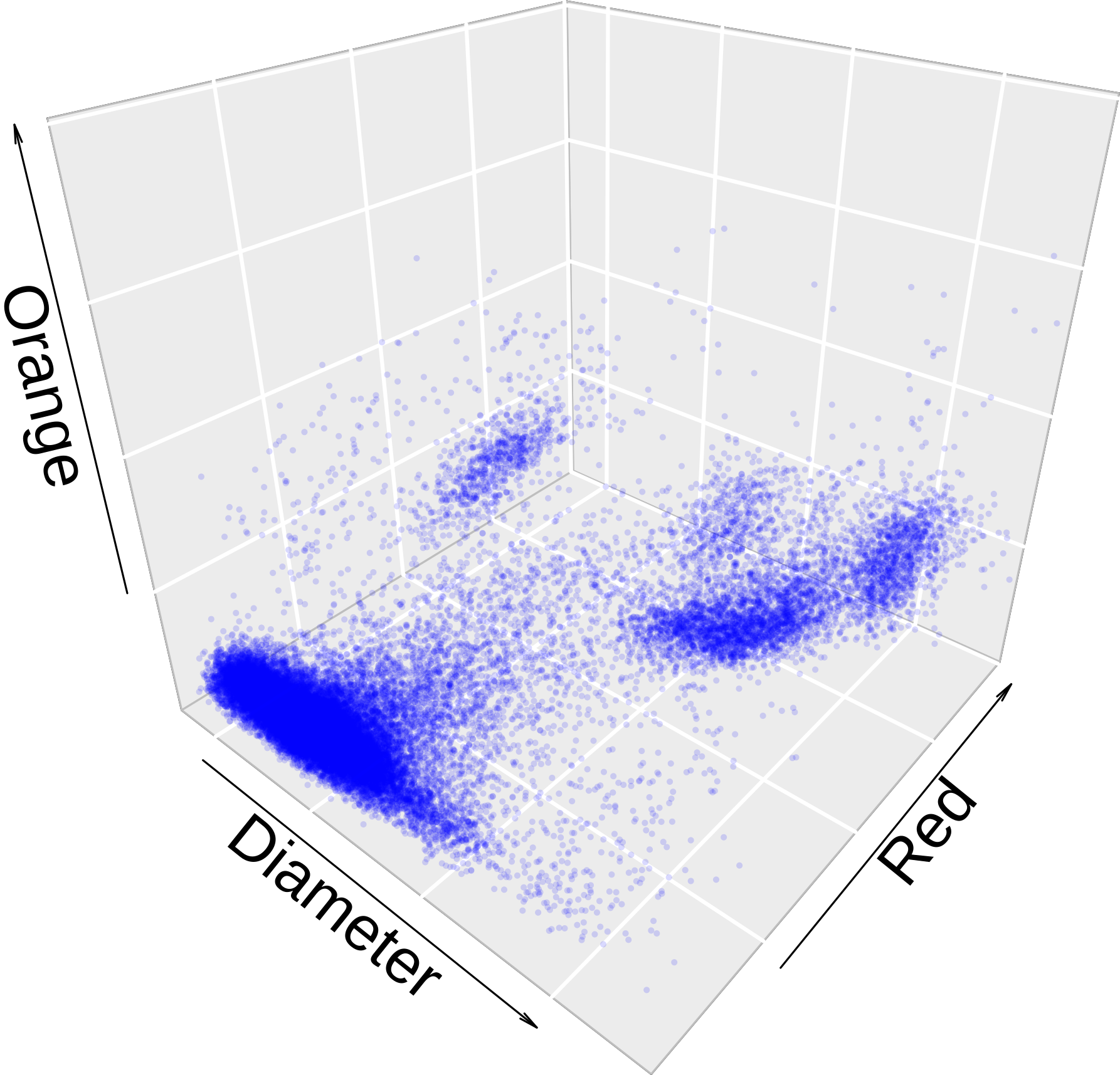}
    \includegraphics[ width=.35\linewidth]{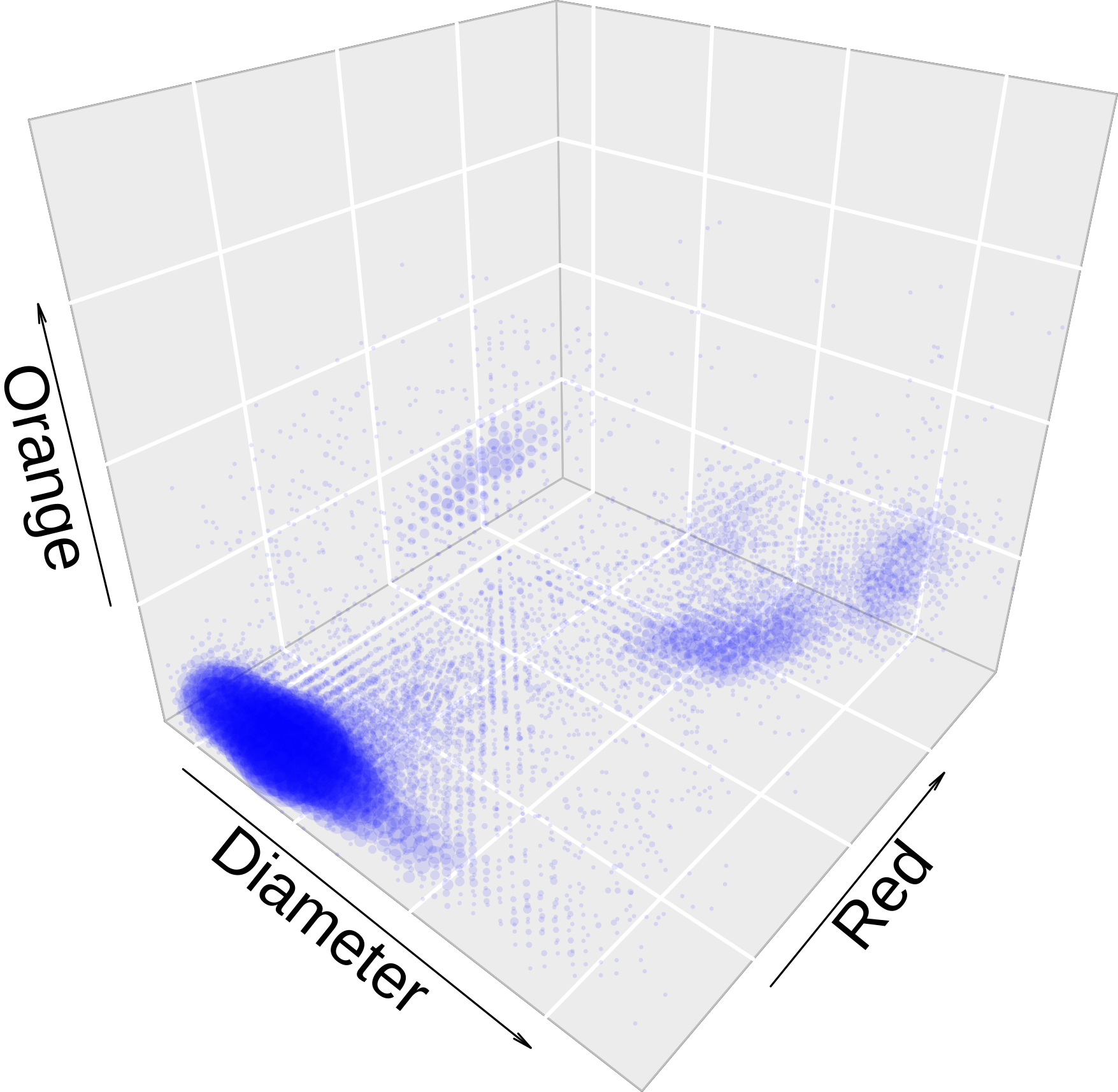}
    \includegraphics[ width=.35\linewidth]{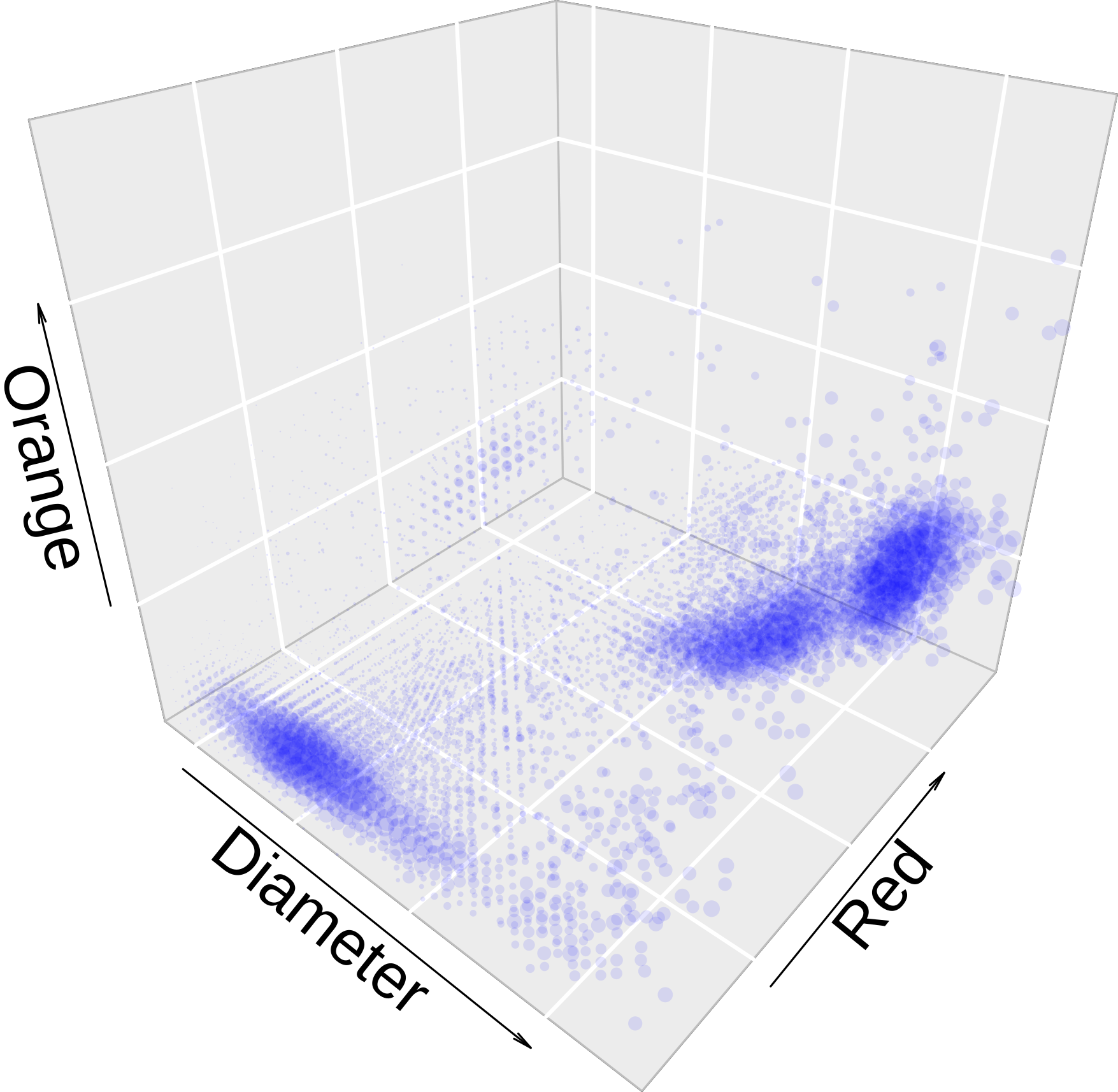}
    }
    \caption{\it Original particles (left) and binned counts with $D=40$
      (middle), and binned biomass (right). In the middle and right plots, the
      size of the points are proportional to the multiplicity. The
      left-hand-side original cytogram contain one hour's worth of particles,
      for a total of $n_t = 36,757$ points, occupying a total of $0.86$ Mb of
      memory.  The binned cytogram in the middle occupies about $1/8$'th the
      memory. The right hand side shows binned biomass data, which has lesser
      imbalance in cluster distribution than the binned count data in the
      middle.}
\label{fig:raw-vs-binned}
\end{figure}
This generalization to a binned data representation can be thought of as trading
off some data resolution for significant computational savings in practice. To
illustrate, the entire set of 3d particles collected during the Gradients 2
cruise, divide into about $T=300$ hourly cytograms containing
$n_t \simeq 100,000$ particles each. This occupies $d \cdot \sum_{t=1}^T n_t$
doubles, or $800$ Mb in memory for $d=3$. Equally burdensome is the size of the
responsibilities $\{\gamma_{itk}\}_{i,t,k}$ (to be defined shortly in
Section~\ref{sec:em}) and densities of each particle with respect to all $K$
clusters, which are each $\sum_t (n_t \cdot K \cdot d)$ doubles, or $2.5$ Gb in
memory for $K=10$. By contrast, when binned with $D=40$, this becomes $40$ Mb in
memory.

The \textit{biomass} representation of data uses carbon quotas -- the amount of
carbon in each particle, in pgC per cell --
$C_i^{(t)}:=\text{Biomass}(\y_i^{(t)}),$ instead of repeated particle counts as
multiplicities, and the \textit{binned} biomass representation of data
aggregates the total carbon biomass in each bin, as
$ C_b^{(t)} = \sum_{i: i\in A_b^{(t)}} \text{Biomass}(\y_i^{(t)})$.  The data
analysis in our paper uses the binned biomass representation.

From a modeling viewpoint, the carbon biomass representation is an attractive
alternative to the particle count representation because our cytograms have
highly imbalanced particle clusterings, a setting in which mixture models
generally perform poorly \citep{em-convergence-properties}. From a
biogeochemical standpoint, biomass distributions are meaningful since cell count
is usually inversely proportional to particle size: small cells tend to dominate
numerically the ocean due to their smaller size and lesser expenditure of
biochemical resources \citep{Maranon2015}.

However, representing data with biomass is not without complication. Using
biomass as multiplicities requires an additional assumption that carbon atoms
can be treated in the same way we have treated particles. However, we know that
carbon atoms arrive in bundles (according to particle sizes) and therefore
treating them as independent is an unrealistic assumption.  That said, in
practice, we see that this simplifying assumption still produces useful and
interpretable estimated models.

\subsection{Penalized Expectation-Maximization Algorithm}\label{sec:em}

Directly maximizing the penalized log-likelihood \eqref{eq:objective},
generalized with multiplicities, is difficult due to its nonconvexity.  We
outline a \textit{penalized} EM algorithm \citep{penalized-em}
for indirectly maximizing the
objective.

Recall from \eqref{eq:latent} that latent variable $Z_i^{(t)}$ encodes the
particle's cluster membership:
\begin{equation*}
Z_i^{(t)} \in \{1,\cdots, K\}.
\end{equation*}
Also define the joint log-likelihood of the data and the latent variables to be:
\begin{multline}
  \log \cL_c(\alpha, \beta, \Sigma; \{\y_i^{(t)}\}_{i,t}, \{Z_i^{(t)}\}_{i,t}, \{C_i^{(t)}\}_{i,t})
\\
  = \sum_{t=1}^T \sum_{i=1}^{n_t} C_i^{(t)}\sum_{k=1}^K  \one\{Z_i^{(t)} = k\} \cdot \log \left( \pi_{kt}(\alpha)\cdot \phi(\y_i^{(t)}; \bmu_{kt}(\beta), \bSigma_k) \right).
\end{multline}
Now, denote the conditional probability of membership as:
$$\gamma_{itk}(\alpha, \beta, \Sigma) = \P_{\alpha, \beta, \Sigma}(Z^{(t)}_i = k | \y^{(t)}, \X^{(t)}),$$
sometimes called \textit{responsibilities} in the literature.

Given some latest estimates of the parameters
$(\hat \alpha, \hat \beta, \hat \Sigma)$, we make use of the surrogate objective
$Q(\alpha, \beta, \Sigma| \hat \alpha, \hat \beta, \hat \Sigma)$ defined as the
penalized conditional expectation (in terms of the conditional distribution of
$Z^{(t)}|\y^{(t)}, \X^{(t)}$) of the joint penalized log-likelihood,
\begin{multline}\label{eq:qfun}
  Q(\alpha, \beta, \Sigma| \hat \alpha, \hat \beta, \hat \Sigma) = \frac{1}{N}\sum_{t=1}^T \sum_{i=1}^{n_t}  C_i^{(t)}\sum_{k=1}^K  \gamma_{itk}(\hat \alpha, \hat \beta, \hat \Sigma)\log \left( \pi_{kt}(\alpha)\ \cdot \phi(\y_i^{(t)} ; \bmu_{kt} (\beta), \bSigma_k)\right)\\
  - \lambda_{\alpha} \sum_{k=1}^K\| \balpha_{k}\|_1 - \lambda_{\beta} \sum_{k=1}^K\|\bbeta_{k}\|_1 - \sum_{k=1}^K \sum_{t=1}^T \one_\infty\{\|\bbeta_k^T \X^{(t)}\|_2 \le r \}.
\end{multline}
The algorithm alternates between estimating the conditional membership
probabilities $\gamma_{itk}$, and updating the latest parameter estimates
$(\hat \alpha, \hat \beta, \hat \Sigma)$ by the maximizer of the penalized Q
function in \eqref{eq:qfun}.
\begin{enumerate}
\item \textbf{E-step} Given $(\hat \alpha, \hat \beta, \hat \Sigma)$, estimate
  the conditional membership probabilities as
  \begin{equation}
    \label{eq:responsibility}
    \gamma_{itk}(\hat \alpha, \hat \beta, \hat \Sigma) = \frac{\phi\left(\y_i^{(t)}; \bmu_{kt}(\hat \beta),
        \hat \bSigma_k\right) \cdot \pi_{kt}(\hat \alpha) }{\sum_{l=1}^L \phi\left(\y_i^{(t)} ;
        \bmu_{lt}(\hat \beta), \hat \bSigma_l\right) \cdot
      \pi_{lt}(\hat \alpha) },
  \end{equation}
  for $k=1,\dots, K$; $t=1,\dots, T$; $i=1,\dots, n_t$.  For the first
  iteration, choose some initial values for means
  $\bmu_{kt} \leftarrow \bmu^{\text{init}}_{k}$, probabilities
  $\pi_{kt} \leftarrow 1/K$, and
  $\hat \bSigma_{k} \leftarrow \bSigma_k^{\text{init}} = g\bI_d$ for some constant
  $g>0$.
\item \textbf{M-step} Using
  $\gamma_{itk} = \gamma_{itk}(\hat \alpha, \hat \beta, \hat \Sigma)$, maximize
  \eqref{eq:qfun} with respect to each parameter $\alpha, \beta$ and $\Sigma$:
  \begin{enumerate}
  \item \textbf{Update $\hat \alpha$:} The maximizer of \eqref{eq:qfun} with
    respect to $\alpha$ is
    \begin{align*}
      \hat \alpha \leftarrow   \argmax_{\substack{\{\alpha_{0k}\}_{k=1}^K\\
      \{\balpha_k\}_{k=1}^K}} &\frac{1}{N}\sum_{t=1}^T \left( \sum_{k=1}^K
                                \gamma_{tk} (\alpha_{0k} + {\X^{(t)}}^T
                                \balpha_k) - n_t \log \sum_{l=1}^K
                                \exp(\alpha_{0l} + {\X^{(t)}}^T \balpha_l)
                                \right)\\[-1em]
                              &- \lambda_{\alpha} \sum_{k=1}^K \|\balpha_k\|_1
    \end{align*}
    for sums $\gamma_{tk}=\sum_{i=1}^{n_t} C_i^{(t)}\gamma_{itk}$.
  \item \textbf{Update $\hat \beta$:} Update $\beta$ according to the ADMM
    algorithm described in Section~\ref{sec:admm} and Supplement~B. Since the
    problem decouples across clusters, we solve separately for each $k$:
    \begin{align*} \label{eq:mstep-objective}
      (\hat \bbeta_{0k}, \hat \bbeta_k)
      &\leftarrow \argmin_{\bbeta_{0k}, \bbeta_k}\frac{1}{2N} \sum_{t=1}^T
      \sum_{i=1}^{n_t} C_i^{(t)} \gamma_{itk} (\y_i^{(t)} - \bbeta_{0k} - \bbeta_k^T \X^{(t)}
      )^T \hat \bSigma_k^{-1} ( \y_i^{(t)} - \bbeta_{0k} - \bbeta_k^T
        \X^{(t)})\\[0em]
      & \hspace{17mm} + \lambda_{\beta}\|\bbeta_k\|_1 \\
      &\text{\small\hspace{7mm}subject to\;\;} \| \bbeta_k^T \X^{(t)} \|_2 \le r
        \;\;\forall t=1,\cdots, T. \nonumber
\end{align*}
  \item \textbf{Update $\hat \Sigma$:} The maximizer of \eqref{eq:qfun} with
    respect to $\bSigma_k$ for each $k=1,\dots, K$ is
    \begin{equation*} 
      \hat \bSigma_k \leftarrow \frac{\sum_{t=1}^T\sum_{i=1}^{n_t} C_i^{(t)}\gamma_{itk} \cdot \br_{itk} \br_{itk}^T}{\sum_{t=1}^T \sum_{i=1}^{n_t} C_i^{(t)}\gamma_{itk}}
    \end{equation*}
    for $\br_{itk} = \y_{i}^{(t)} - \hat \bbeta_{0k} - \hat \bbeta_k^T\X^{(t)}$.
  \end{enumerate}
\end{enumerate}
Note, the M-step breaks into a convex problem over $\alpha$ (step 2a) and a
non-convex problem over $(\beta, \Sigma)$ (step 2b and 2c). For the latter part
of the M-step, instead of jointly optimizing over $(\beta, \Sigma)$, we perform
two successive partial optimizations -- first with respect to $\beta$, and next,
with respect to $\Sigma$.

This algorithm is terminated when the penalized log-likelihood has a negligible
relative improvement.  In practice, we run the EM algorithm multiple times and
retain the run with the highest final log-likelihood, for a better chance at
achieving the true optimum.  For $\bmu_k^{\text{init}}$ we randomly choose $K$
out of all $\sum_{t=1}^{T}n_t $ cytogram particles. Initial covariances
$\{\bSigma^{\text{init}}_k\}_{k=1}^K$ are set to have diagonal entries $g$ equal
to $1/K$ times the cytogram range in each dimension. The $\alpha$ part of the
M-step is solved using \texttt{glmnet}, with \texttt{family} set to
\texttt{``multinomial''} \citep{glmnet-paper}. The $\beta$ part of the M-step
requires a custom alternating direction method of multipliers (ADMM) solver,
outlined in the next section.

\subsection{ADMM algorithm in M-step for $\beta$}\label{sec:admm}

The $\beta$ M-step (in step b) is very slow if computed using a non-customized
solver -- for instance, using CVX \citep{cvx}, it is the slowest component of
the EM algorithm by a factor of ten or more. To improve performance, we devise a
customized alternating direction method of multipliers (ADMM) algorithm
\citep{boyd-admm}. We start by observing that this optimization problem
decouples across $k$. Since each $k \in \{1,\cdots, K\}$ can be solved
separately, we will drop the subscript $k$ hereon and write the variables
$\bbeta_{0k}$ and $\bbeta_k$ as $\bbeta_0$ and $\bbeta$, $\gamma_{itk}$ as
$\gamma_{it}$, and $\hat \bSigma_k$ as $\hat \bSigma$ for notational simplicity.

Consider the minimization problem in step b of the M-step of the penalized EM
algorithm. The objective to minimize can be written as
\begin{multline*}
  f(\bbeta_0, \bbeta) = \frac{1}{2N} \sum_{i,t} C_i^{(t)} \gamma_{it}(\y_i^{(t)} - \bbeta_0
  - \bbeta^T \X^{(t)}) ^T \hat \bSigma^{-1} (\y_i^{(t)} - \bbeta_0 - \bbeta^T \X^{(t)}
  ) \\[-1em]
  + \lambda_{\bbeta} \| \bbeta \|_1 + \one_\infty \{ \|\bbeta^T \X^{(t)}\|_2 \le r\}.
\end{multline*}
We can obtain the overall minimizer via partial minimization with respect to
$\bbeta_0$; writing
$\hat \bbeta_0(\bbeta) := \argmin_{\bbeta_0} f(\bbeta_0, \bbeta)$ for this
partial minimizer, setting the gradient to $0$ yields a closed form expression of
$\hat \bbeta_0(\bbeta)= \frac{\sum_{i,t} C_i^{(t)} \gamma_{it} (\y_i^{(t)} -
  \bbeta^T \X^{(t)})}{\sum_{i,t} C_i^{(t)} \gamma_{it}}.$
The objective to minimize with respect to  $\bbeta$ then becomes
\begin{align*}
  f(\hat \bbeta_0 (\bbeta), \bbeta)
  =& \frac{1}{2N} \sum_{i,t} C_i^{(t)} \gamma_{it} (\tilde \y_{i}^{(t)} -  \bbeta^T \tilde \X^{(t)}
    )^T \hat \bSigma^{-1} ( \tilde \y_{i}^{(t)} -  \bbeta^T \tilde \X^{(t)} )\\[-.5em]
& + \lambda_{\bbeta} \|\bbeta\|_1 + \one_\infty \{ \|\bbeta^T \X^{(t)}\|_2 \le r\},
\end{align*}
where $\tilde \y^{(t)}:= \y_i^{(t)} - \bar \y$ and
$\tilde \X^{(t)}:= \X^{(t)} - \bar \X$ are data centered by weighted averages
$\bar \y :={\sum_{i,t} C_i^{(t)} \gamma_{it} \y_i^{(t)}}\; / \; {\sum_{i,t}
  C_i^{(t)} \gamma_{it}}$ and
$\bar \X:= {\sum_{i,t} C_i^{(t)} \gamma_{it} \X^{(t)}}\;/\;{\sum_{i,t} C_i^{(t)}
  \gamma_{it}}$.  Now, introducing augmented variables $\bZ \in \R^{T \times d}$
and $\bW \in \R^{p \times d}$, we can rewrite
$\min_{\bbeta} f(\hat \bbeta_0(\bbeta), \bbeta)$ as:
\begin{mini}{
  \bbeta, \bZ, \bW}{\frac{1}{2N} \sum_{i,t}C_i^{(t)} \gamma_{it} (\tilde
    \y_i^{(t)} - \bbeta^T \tilde \X^{(t)} )^T \hat \bSigma^{-1} (\tilde \y_i^{(t)}
    - \bbeta^T \tilde \X^{(t)}) + \lambda \|\bW\|_1 }{}{}
  \addConstraint{\|\bZ^{(t)}\|_2 \le r}
  \addConstraint{\icol{\X \\ \bI } \bbeta = \icol{\bZ \\ \bW},}
\label{eq:mstep-objective-augmented}
\end{mini}
which can be solved using an ADMM whose full details are deferred to Supplement B. All steps are computationally simple, consisting of least
squares reduced to rapidly solvable Sylvester equations, $\ell_2$ ball projection,
and soft-thresholding. The implementation in the \texttt{flowmix} \texttt{R} package is highly optimized
and faster than any other component of the EM algorithm.

\subsection{Cross-validation for selection of $\lambda_{\alpha}$, $\lambda_{\beta}$}\label{sec:cv}

We choose the regularization parameter values
$(\lambda_{\alpha}, \lambda_{\beta})$ using five-fold cross-validation over a
discrete $2$-dimensional grid of candidate values
$L_{\alpha} \times L_{\beta}$, in which $L_{\alpha}$ and $L_{\beta}$ each
contain logarithmically-spaced positive real numbers. We form the five folds
consisting of every fifth time block containing $20$ consecutive time
points. Denote these five test folds' time points as sets $\{I_o\}_{o=1}^5$, so
that $I_1 = \{1, \cdots, 20,101,\cdots, 120, \cdots\}$,
$I_2 = \{21, \cdots, 40, 121, \cdots, 140, \cdots\}$, and so forth. Writing
$I_{-o} = \{1,\cdots,T\} \backslash I_o$, the test datasets comprise of the
subsetted data $\{(\X_o, \y_o, \C_o)\}_{o=1}^5$ for
$\X_o:=\{\X^{(t)}:t \in I_o\}$, $\y_o:=\{\y^{(t)}: t\in I_o\}$ and
$\C_o:=\{\C^{(t)}: t\in I_o\}$, and the corresponding training dataset comprise
of $\{(\X_{-o}, \y_{-o}, \C_{-o})\}_{o=1}^5$.

The five-fold cross-validation score is calculated as the average of the
out-of-sample negative log-likelihood in \eqref{eq:loglik-mult}:
\begin{equation*}
S(\lambda_{\alpha}, \lambda_{\beta}) = -\frac{1}{5} \sum_{o=1}^5 \ell_{\{n_t: t \in I_o \}} ( \hat \alpha_{-o}, \hat \beta_{-o}, \hat \Sigma_{-o}; \y_o, \X_o, \C_o),
\end{equation*}
where $\hat \alpha_{-o}$, $\hat \beta_{-o}$ and $\hat \Sigma_{-o}$ are the
estimated coefficients from the training data set $(\X_{-o}, \y_{-o},
\C_{-o})$. (We include $\X_o$ in $\ell(\cdot)$ to emphasize which subset of the
covariates the log-likelihood is based on.) The cross-validated regularization
parameter values $\lambda_{\alpha}$ and $\lambda_{\beta}$ are the minimizer of
the cross-validation score:
\begin{equation*}
(\hat \lambda_{\alpha}, \hat \lambda_{\beta}) =   \argmin_{\lambda_{\alpha} \in L_{\alpha}, \lambda_{\beta} \in L_{\beta}}  S(\lambda_{\alpha}, \lambda_{\beta}).
\end{equation*}
A real data example of cross-validation scores in action is shown in
Figures
12
and
13 in the Supplement.
Our scheme of
training/test splits places a strong emphasis on even temporal coverage of the
test data. Since our data are in hourly resolution (equivalent to 20 kilometers
in space) and cross-validation folds are made of $20$-hour-long time blocks, the
temporal closeness of the test time points $I_o$ and the training time points
$I_{-o}$ is negligible. For data with finer time resolution, our recommendation
is to form a time \textit{barrier} between the training and test time points, or
to form larger time blocks for test folds. Also, in this work, we do not discuss
how to select the number of clusters $K$ based on data. In simulation, we
demonstrate that slightly overspecifying the number of clusters results in
equivalent predictive performance as the true number of clusters. See Section
\ref{sec:numerical-numclust} for details.

 \section{Numerical results} \label{sec:numerical}

\subsection{Simulated data}

In order to examine the numerical properties of our proposed method, we apply
our model to simulated data whose setup is closely related to our main flow
cytometry datasets.

\subsubsection{Noisy covariates} \label{sec:numerical-noisy-covariates}

The main source of noise in our data is in the environmental covariates from a
variety of sources -- in-situ and remote-sensing measurements, and oceanographic
model-derived product \citep{world-ocean-atlas}, each with different temporal
and spatial resolution, and varying amounts of uncertainties. In order to
investigate the effect of uncertainty in the covariates, we conduct a simulation
in which synthetic cytograms are generated from a true model and underlying
covariates, and then our model is estimated with access to only artificially
obscured covariates.

\begin{figure}[ht!]
  \centering
  \includegraphics[width=.48\linewidth]{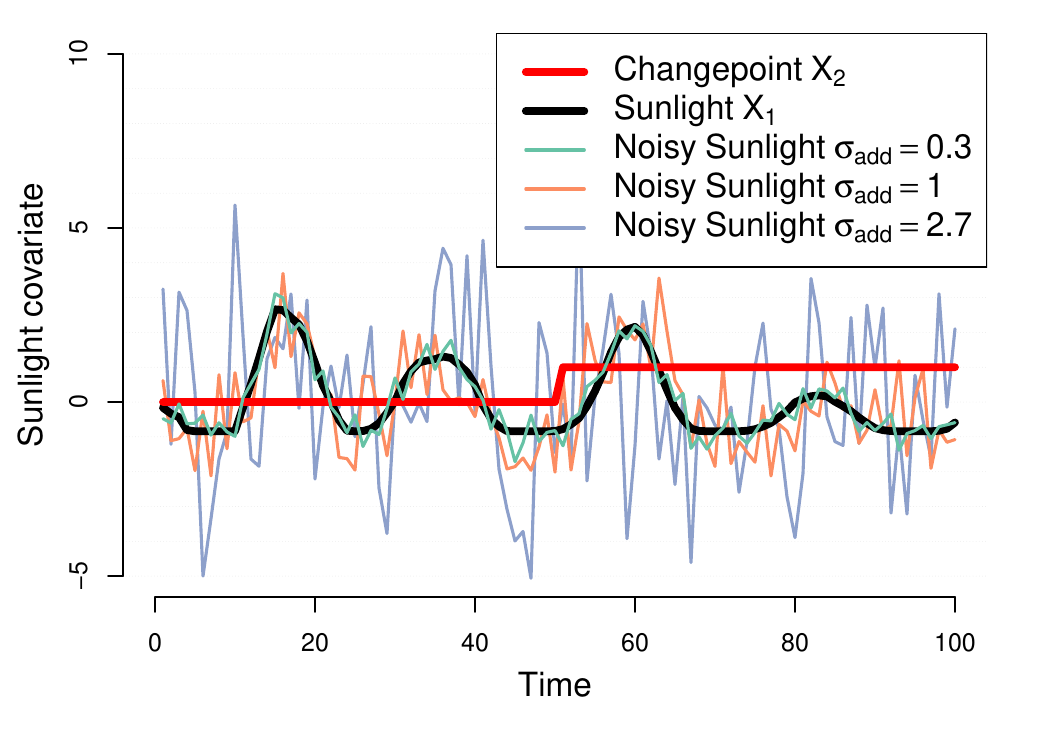} \hspace{0mm} 
  \includegraphics[width=.48\linewidth]{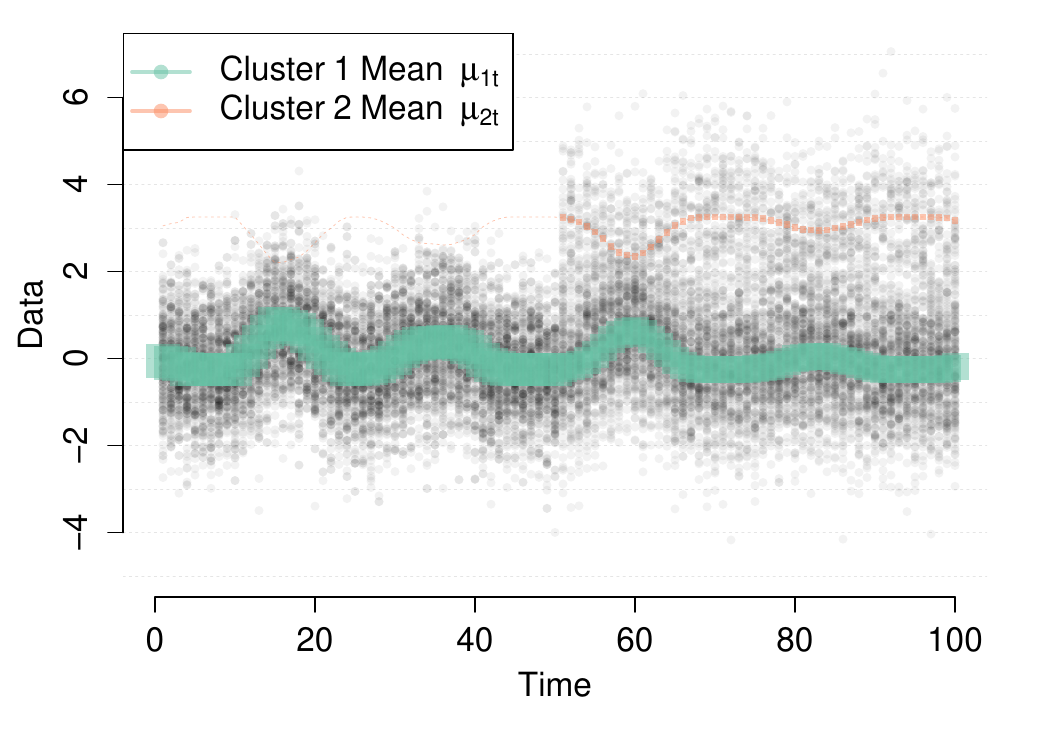} 
  \caption{\it (Left) The thick black line shows the first covariate
    $\X_1 \in \R^T$, which is a smoothed and standardized version of the
    \texttt{par} (sunlight) covariate from Section \ref{sec:1d-real-data}. The
    three thin lines show the obscured sunlight variables for three different
    noise levels $\sigma_{\text{add}}$. The next covariate is a changepoint
    variable $\X_2 \in \R^T$, shown as a thick red line. The remaining $8$
    spurious covariates $\{\X_i\}_{i=3}^{10}$ are generated as $T$
    i.i.d. entries from $\cN(0, 1+\sigma_{\text{add}}^2)$; these are not shown
    here.  (Right) An example of a generated dataset, whose particles are shown
    as grey points in the background. The two true cluster means are plotted as
    colored lines whose thickness is proportional to the cluster
    probabilities. Particles for both clusters are generated as $\cN(0,1)$
    around the cluster means. Cluster $1$ is only present in the second half,
    and has one quarter of the number of particles in cluster $2$ in those time
    points. A thin dashed line is shown in the first half where the cluster
    probability is zero.}
  \label{fig:noisy-covariates-data}
\end{figure}

We generate synthetic data with $T=100$ time points, $K=2$ clusters, and $p=10$
covariates $\{X_i\in\R^T\}_{i=1}^{10}$ as shown in
Figure~\ref{fig:noisy-covariates-data} -- one sunlight variable $\X_1$, one
changepoint variable $\X_2$, and eight spurious covariates
$\{\X_i\}_{i=3}^{10}$.  From these covariates, $T$ 1-dimensional cytograms are
generated from the generative model in Section~\ref{sec:likelihood} with the
true underlying coefficient values, 

\begin{equation}\label{eq:noisy-covariates-coef}
\begin{array}{l}
  \alpha_{0,1} = 0,  \hspace{5mm}
  \alpha_{0,2} = 0, \hspace{5mm}
  \balpha_1 = (0 \;\; 0 \;\; \cdots \;\; 0)^T, \hspace{5mm}
  \balpha_2 = (0 \;\; 8.61 \;\; \cdots \;\; 0)^T,\\
  \beta_{0,1} = 0,  \hspace{5mm}
  \beta_{0,2} = 3, \hspace{5mm}
  \bbeta_1 = (0.3 \;\; 0 \;\;\cdots \;\; 0)^T, \hspace{5mm}
  \bbeta_2 = (-0.3 \;\; 0 \;\; \cdots \;\; 0)^T.
\end{array}
\end{equation}

Both clusters' means follow the sunlight $\X_1$. Cluster $1$ has $n_t=200$
particles for all time points $t=1,\cdots,100$. Cluster $2$ overlaps with
cluster $1$, is present only in the second half of the time range
$t=51,\cdots, 100$, and is 1/4th as populous as cluster 1 at those time
points. Both cluster variances are equal to $1$ so that particles from each
cluster are generated from $\cN(0,1)$ around their respective means, and the
spurious covariates play no role in data generation i.e. all other coefficients
not specified in \eqref{eq:noisy-covariates-coef} are zero.

On each new synthetic dataset, we estimate a cross-validated $2$-cluster model
using radius $r=1.5$, but instead of sunlight covariate
$\X_1$, we use the \textit{obscured}
$\X_1^{\text{noisy}} = \X_1 + \bepsilon, \bepsilon \sim
\cN(0,\sigma_{\text{add}}^2\bI_T)$ for estimation. Also, the eight spurious
covariates $\{\X_i\}_{i=3}^{10}$ are each generated as
$\cN(0, 1+\sigma_{\text{add}}^2)$ to match the magnitude of
$\X_1^{\text{noisy}}$.  We consider a certain range of additive noise
$\sigma_{\text{add}} \in \{0, 0.3, 0.6, \cdots, 2.7\}$, and $100$ synthetic
datasets for each value $\sigma_{\text{add}}$.

\begin{figure}[ht!]
  \centering
  \makebox[\linewidth]{
  \begin{minipage}{0.47\linewidth} 
  \includegraphics[width=\linewidth]{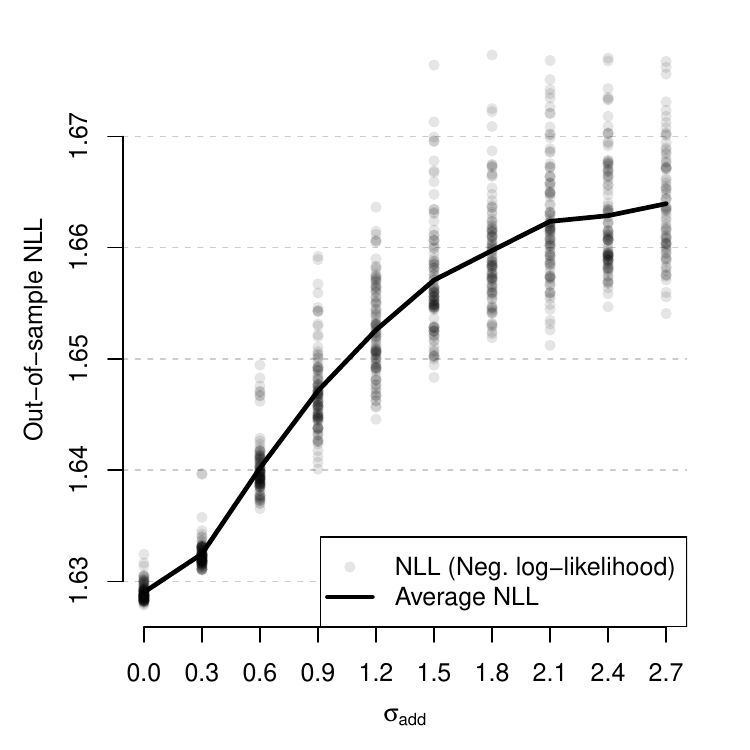}
  \end{minipage}
  \begin{minipage}{0.47\linewidth}
  \includegraphics[width=\linewidth]{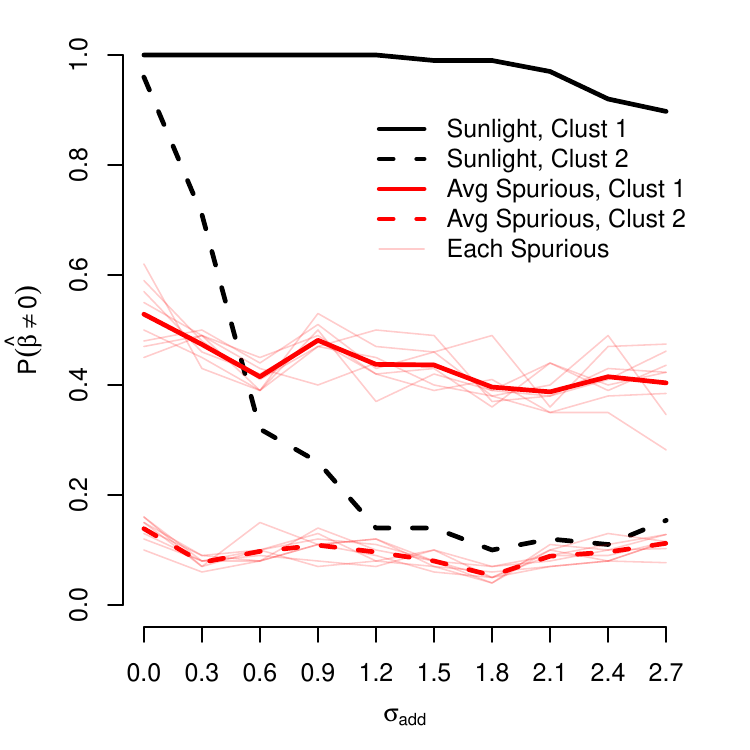} 
    \end{minipage} 
  }
  \caption{\it (Left) Out-of-sample prediction performance using covariates
    obscured by Gaussian noise variance $\sigma_{\text{add}}^2$, for the
    simulation setup described in Section
    \ref{sec:numerical-noisy-covariates}. (Right) The probability of the
    sunlight covariate (the only relevant covariate for cluster means) being
    estimated as nonzero is shown in black lines. The corresponding
    probabilities for the eight spurious covariates are shown in red lines (thin
    red lines are individual covariates, and the thick red line is the average).
    The solid and dashed lines show results from cluster 1 and cluster 2
    respectively. In both clusters, the sunlight variable is more likely to be
    selected than the spurious variables. This advantage is more pronounced for
    cluster $1$ than for cluster $2$, which is only has data in the second half
    of the time range.}
  \label{fig:noisy-covariates-results}
\end{figure}

The left plot of Figure~\ref{fig:noisy-covariates-results} shows the
out-of-sample model prediction performance of $100$ estimated models
for each noise level $\sigma_{\text{add}}$, measured as the negative log
likelihood evaluated on a large independent test dataset. As expected,
out-of-sample prediction gradually worsens with increasing covariate noise
$\sigma_{\text{add}}$, then plateaus at about $\sigma_{\text{add}}=2.7$.

The right plot of Figure~\ref{fig:noisy-covariates-results} demonstrates the
variable selection property of our method, focusing on the $\beta$
coefficients. Focusing on the sunlight variable -- the only true predictor of
mean movement -- we see that it is more likely to be selected than are spurious
covariates, and is less likely to be selected as $\sigma_{\text{add}}$
increases. Additionally, we see that selecting sunlight is possible even when
$\sigma_{\text{add}}$ is high if the cluster has higher relative probability and
has nonzero probability in a longer time range.

\subsubsection{Cluster number misspecification} \label{sec:numerical-numclust}

\begin{figure}[ht!] 
  \centering
  \includegraphics[width=.5\linewidth]{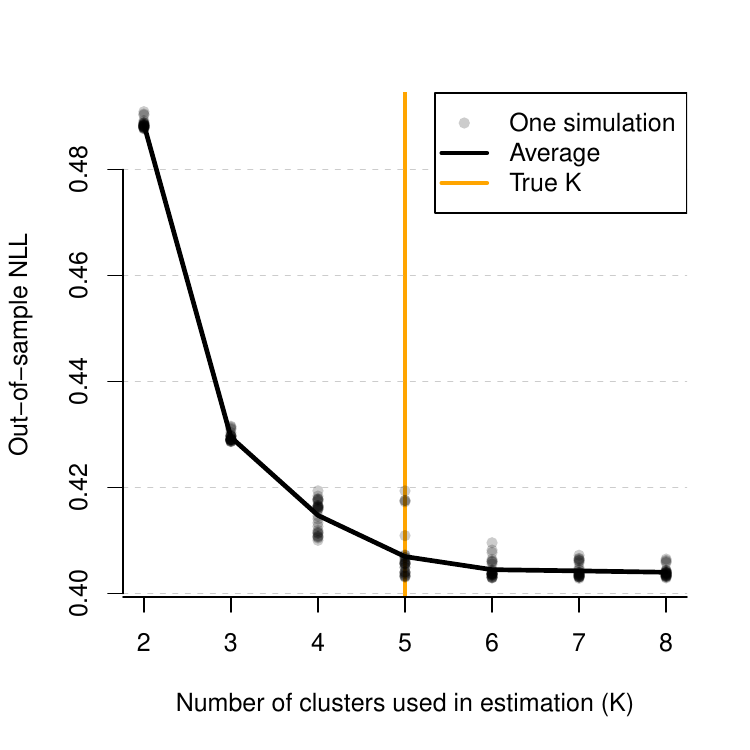}
  \caption{\it Out-of-sample prediction performance for $K$-cluster models
    estimated from 5-cluster pseudo-real datasets (which were each generated
    from a simplified version of a model estimated from real 1-dimensional data,
    in Section \ref{sec:1d-real-data}). Models estimated with fewer than $5$
    clusters have sharply worse out-of-sample prediction performance. On the
    other hand, estimated models with $5$ clusters or more have similar
    out-of-sample prediction performance, because the extra clusters are
    estimated to have zero probability, and play no role in the prediction.}
  \label{fig:numclusts-results}
\end{figure}

In addition to covariate noise, we explore the effect of misspecifying the
number of clusters $K$ in the model. We first form a ground truth model by
taking the five-cluster estimated model from the 1-dimensional $T=296$ data in
Section~\ref{sec:1d-real-data} and Figure~\ref{fig:1d}, and zero-thresholding
the smaller estimated coefficients. We then generate new data $30$ times from
this underlying true model, and estimate a $K$-cluster cross-validated model,
for $K \in \{2,3,4,5,6,7,8\}$. Figure \ref{fig:numclusts-results} shows
out-of-sample prediction performance, measured as the negative log-likelihood on
a large independent test set generated from the true model. We see that models
estimated with $K<5$ clusters have sharply deteriorating out-of-sample
prediction. On the other hand, models estimated with $K>5$ than five clusters
have average out-of-sample prediction performance in the same range as that of
$K=5$ cluster models. A closer examination of the estimated models reveals that,
out of the $K>5$ clusters, five clusters are usually estimated accurately, and
the remaining $K-5$ clusters are estimated with near-zero probability. These
results suggest that one can slightly overspecify the number of clusters for
estimation with little harm to prediction performance. Automatic approaches to
choosing $K$ is an interesting area of future work.

 \section{Application to Seaflow cruise}
\label{sec:real-data}

In this section, we apply our model to data collected on a research cruise in
the North Pacific Ocean, and from the Simons CMAP database
(\url{https://simonscmap.com/}). The MGL1704 cruise traversed two oceanographic
regions over the course of about 2 weeks, between dates 2017-05-28 and
2017-06-13. As seen in Figure~\ref{fig:intro}, the cruise started in the North
Pacific Subtropical Gyre (low latitude, dominated by warm, saltier water),
traveling north to the Subpolar Gyre (high latitude, low-temperature, low-salt,
nutrient-rich water), and returned back south. We first describe the data and
model setup, then discuss the results.

\textbf{Environmental covariates.}  A total of 33 environment covariates (see
Table
1
and
Figure
11
of the Supplement) were \textit{colocalized} with
cytometric data by averaging the environmental data measurements within a
rectangle of every discrete point of the cruise trajectory in space and time,
aggregated to an hourly resolution. These data were processed and downloaded
from the Simons CMAP database \citep{simonscmap-paper} accessed through the
\texttt{CMAP4R} R package \citep{cmap4r}. In addition to these covariates, we
created four new covariates by lagging the sunlight covariate in time by
$\{3,6,9,12\}$ hours. This was motivated by scientific evidence showing that the
peak of phytoplankton cell division is out of phase with sunlight
\citep{Ribalet2015}. We also created two new changepoint variables demarcating
the two crossings events of the cruise through a biological transition line at
latitude $37\textdegree$. These derived covariates play the role of allowing a
more flexible \textit{conditional} representation of the cytograms, using
information from the covariates.  All covariates except for the two changepoint
variables were centered and scaled to have sample variance of 1. Altogether, we
formed a covariate matrix $\X \in \R^{(308-12) \times 39}$. (The first twelve
time points are deleted due to the the lagging of the sunlight variable.)

\textbf{Response data (cytograms).} The response data (cytograms) were collected
on-board using a continuous-time flow cytometer called SeaFlow, which
continuously analyzes sea water through a small opening and measures the optical
properties of individual microscopic particles \citep{seaflow}. The data
consist of measurements of light scatter and fluorescence emissions of
individual particles. Data are organized into files recorded every $3$ minutes,
where each file contains measurements of the cytometric characteristics of
between $1,000$ and $100,000$ particles ranging from $0.5$ to $5$ microns in
diameter. The size of data in any given file depends on the cell abundance of
phytoplankton within the sampled region. Each particle is characterized by two
measures of fluorescence emission (chlorophyll and phycoerythrin), its diameter
(estimated from light scatter measurements by the application of Mie theory for
spherical particles), its carbon content (cell volume is converted to carbon
content) and its label (identified based on a combination of manual gating and a
semi-supervised clustering method), as described in \cite{seaflow-data}. Note
that we use the particle labels only for comparison to our approach in
Section~\ref{sec:gating}. Particles were aggregated by hour, resulting in $T=296$
cytograms for the duration of the cruise, with matching time points as rows of
$\X$.

Lastly, the cytogram data $\{\y_i^{(t)} \in \R^3: i=1,\cdots, n_t\}_t$ were log
transformed due to skewness of the original distributions, augmented with
biomass multiplicity $\{C_i^{(t)}: i=1,\cdots, n_t\}_t$, and binned using $D=40$
equally sized bins in each dimension, as described in Section
\ref{sec:multiplicity}. In the analyses to follow in Sections
\ref{sec:1d-real-data} - \ref{sec:3d-real-data}, we consider two data
representations for analysis: a $d=1$ case with only the binned cell diameter
biomass cytograms, and the full $d=3$ dimensional binned biomass cytograms.

\textbf{Practicalities.} The regularization parameters
$(\lambda_\alpha, \lambda_\beta)$ were chosen using 5-fold cross-validation as
described in Section \ref{sec:cv}.  Every application of the EM algorithm was
repeated $5$ times (for 3-dimensional data) or $10$ times (for 1-dimensional
example).  The model means were restricted using a ball constraint of radius $r$
as described in Section \ref{sec:penalty}. In the 1-dimensional data analysis in
\ref{sec:1d-real-data}, the radius reflects the underlying assumption that
carbon quotas should at most double or halve, peaking during the day due to
carbon fixation via photosynthesis by the cell, and halving due to cell division
(i.e. the mother cell divides into two equal daughter cells). Assuming spherical
particles, this would correspond to a log scale day-night cell diameter
difference of $\log(2)/3 \simeq 0.231$, halved to obtain $r=0.1153$. The
3-dimensional data analysis in Section \ref{sec:3d-real-data} first shifts and
scales the log cell diameter to be in the same range as the other axes, and uses
$r=0.5$, which is similar in scale to the radius used in the 1-dimensional
analysis.

\subsubsection{Application to 1-dimensional cell diameter data} \label{sec:1d-real-data}

In this section, we apply our model to 1-dimensional cytograms at the hourly
time resolution. The 1-dimensional setting is useful for visualization because
single plots can display the entire data and fitted model parameters, displaying
cluster means $\{\bmu_{k \cdot} \in \R^{T}\}_1^K$ as lines and cluster
probabilities $\{\pi_{k \cdot} \in \R^T\}_1^K$ as line thickness, as well as
shaded approximate 95\% conditional density intervals from
$\phi(\cdot, \bmu_{kt}, \bSigma_k)$. The estimated means and
probabilities are shown in Figure \ref{fig:1d}, and the estimated coefficients
can be seen in
Table
2 of the Supplement.

Overall, the estimated model effectively captures the visual patterns in the
cytogram data. Clusters $3$ and $5$ correspond to two well-known populations
called \textit{Synechococcus} and \textit{Prochlorococcus}, respectively.  The
most prominent phenomenon is the daily fluctuation of the mean of cluster $5$,
which is clearly predicted using a combination of time-lagged sunlight and ocean
altimetry. Also notable is change in probability of cluster $3$, which is
predicted well by physical and chemical covariates such as sea surface
temperature and phosphate. The overlapping two clusters $3$ and $4$ are also
accurately captured as separate clusters.

As we will see shortly in the 3-dimensional analysis, introducing the other two
axes of the cytograms (i.e. 1-dimensional cytograms to 3-dimensional cytograms)
clearly helps further distinguish between clusters and identify finer-grain
cluster mean movement. Furthermore, cluster $4$, which has a large variance and
serves as a \textit{catch-all} background cluster, does not appear to represent
a specific cell population, and rather exists to improve the other clusters'
model fits.

We also estimated the \textit{stability} of $\beta$ coefficients of this model,
by calculating the nonzero proportion of each of the estimated coefficients
produced from subsampled datasets. These nonzero proportions are displayed
alongside the original coefficient estimates in
Tables~7
and
8
in the Supplement,
and the entire
procedure is detailed in Supplement~D. The stability estimates seem quite
sensible -- they show high nonzero probability of sunlight variables for
Prochlorococcus (cluster $5$), as well as overall low nonzero probabilities for
the covariates of cluster $4$, the background cluster.

\begin{figure}[ht!]
  \centering
  \makebox[\textwidth]{
    \includegraphics[width=1.2\linewidth]{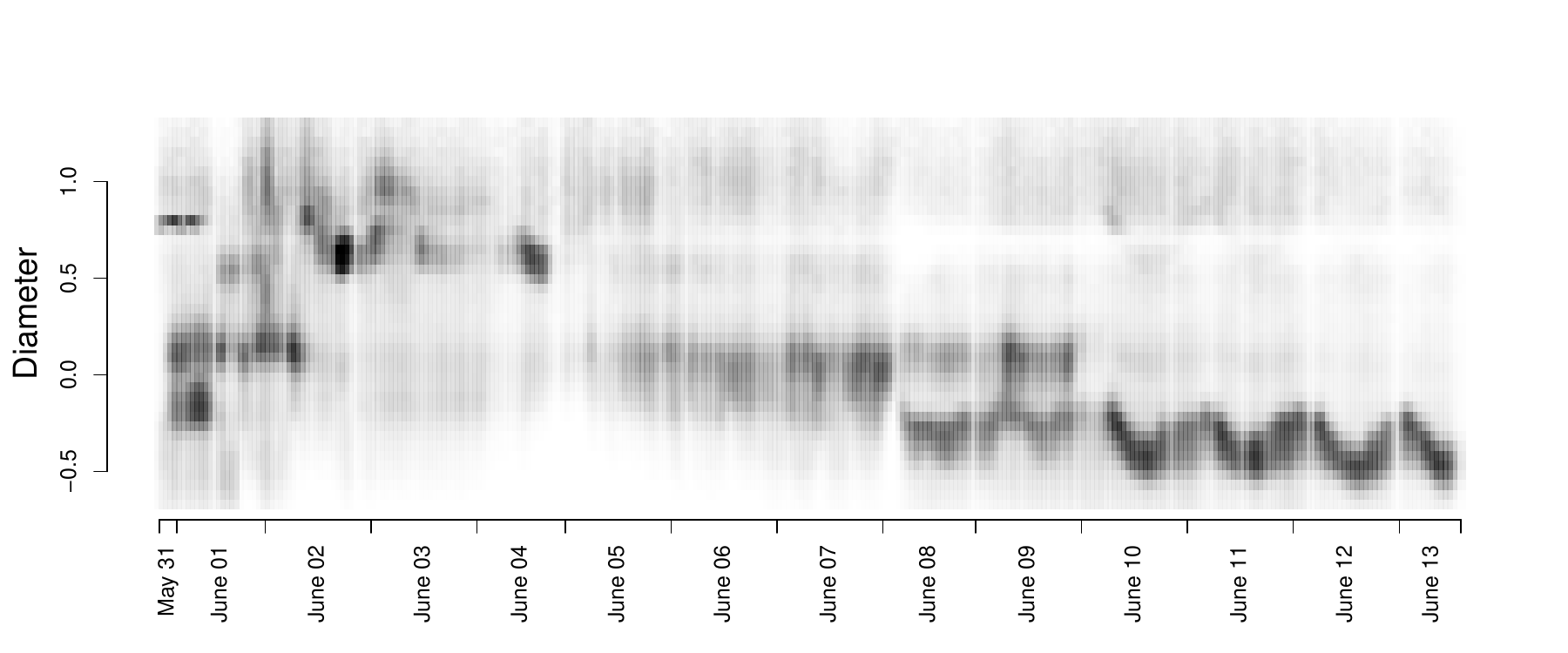}
  }
  \makebox[\textwidth]{
    \vspace{-20mm}
    \includegraphics[width=1.2\linewidth]{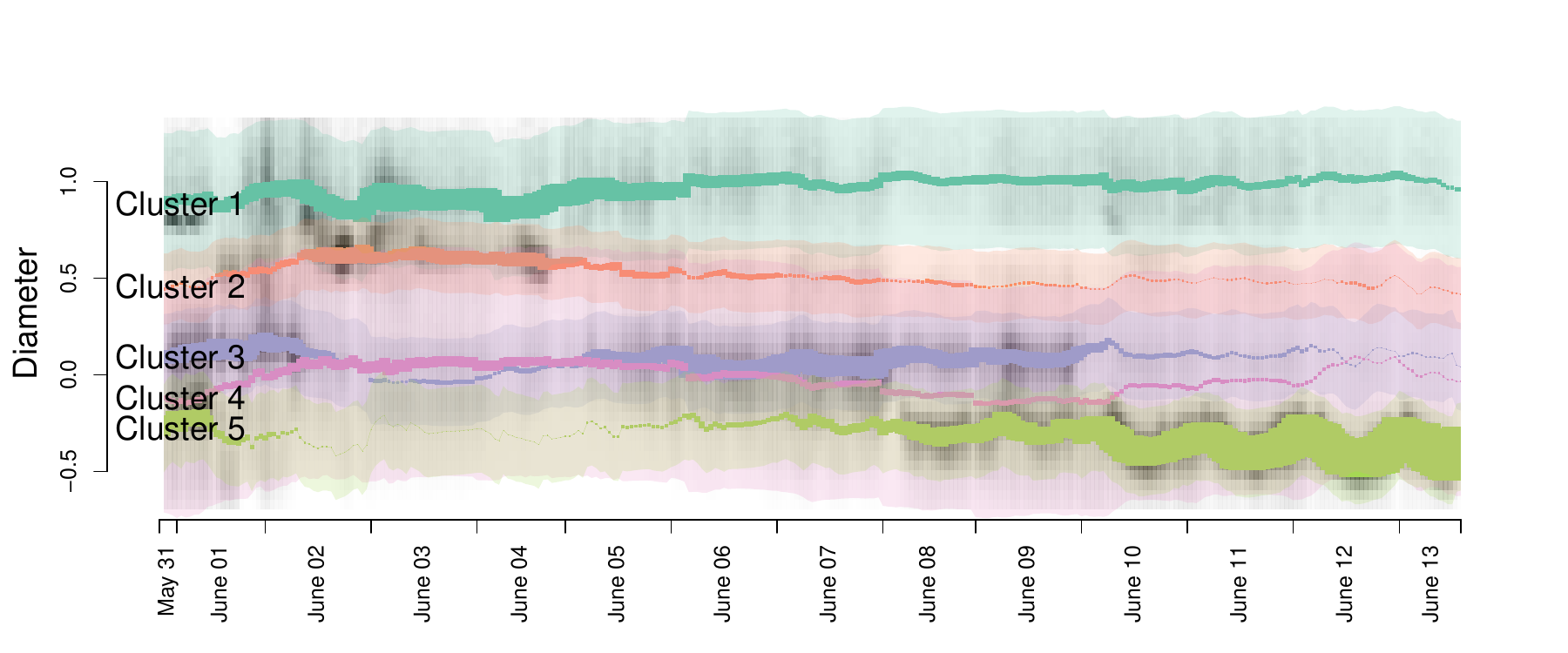}
  }
  \caption{\it (Top) The 1-dimensional cell diameter biomass cytograms (log
    transformed) at an hourly time resolution is shown here. In the background,
    the 1-dimensional biomass distribution of binned cell diameter data is shown
    in greyscale.  (Bottom) The estimated $5$-cluster model is overlaid on the
    same plot; the five solid lines are the five estimated cluster means, whose
    thickness show the values of the $K=5$ cluster probabilities
    $\{\pi_{kt}\}_{k=1}^K$ over time $t=1,\cdots, 296$ (individual hours). The
    shaded region around the solid lines are the estimated $\pm 2$ standard
    deviation around the cluster means.}
  \label{fig:1d}
\end{figure}

\subsubsection{Application to 3-dimensional data}\label{sec:3d-real-data}

\begin{figure}
  \makebox[\textwidth]{
    \includegraphics[width=1.2\linewidth]{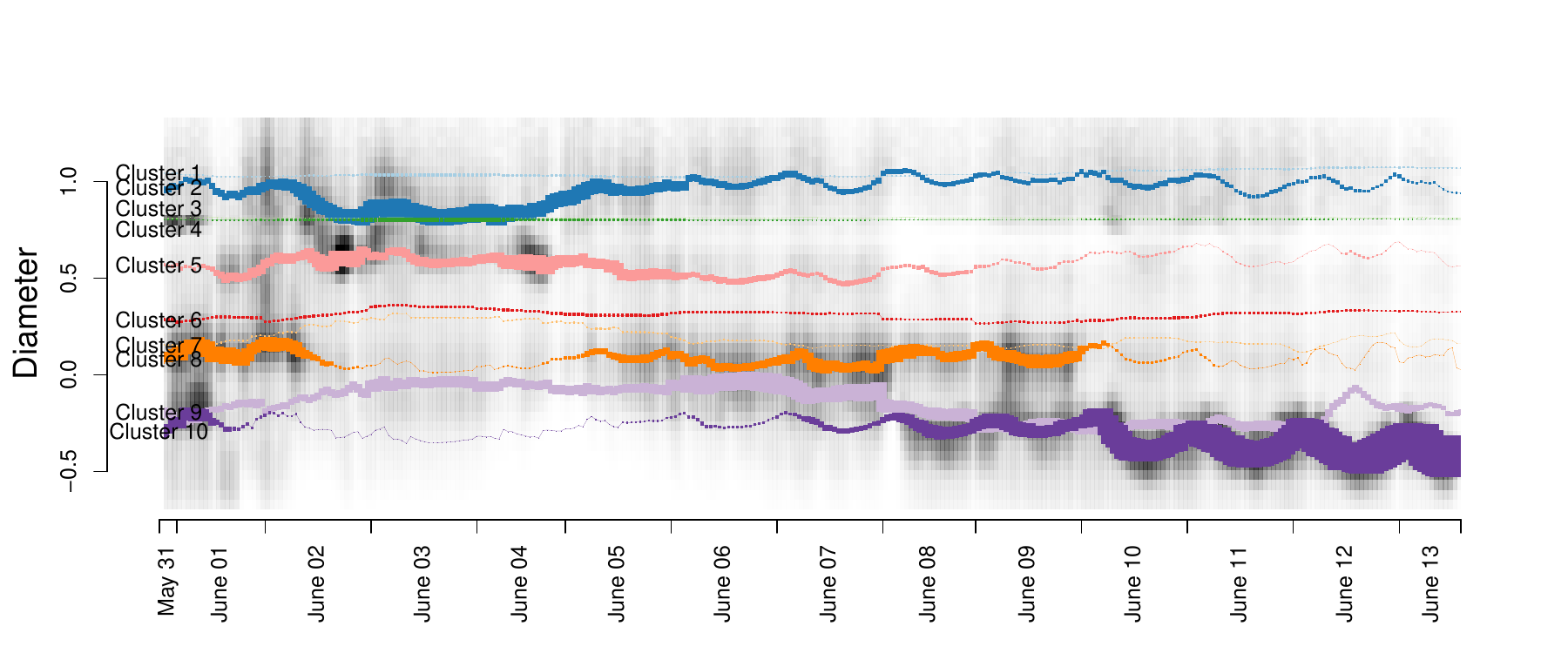}
  }
  \caption{\it A one-dimensional slice of the estimated model of the full
    $3$-dimensional data, showing only the cell diameter axis.  This figure is
    directly comparable to Figure \ref{fig:1d} using only 1-dimensional cell
    diameter data. The colored solid lines track the ten estimated cluster means
    over time, and the line thickness shows the cluster probabilities over
    time. (The shaded 95\% probability regions were omitted for clarity of
    presentation.)  This model on 3-dimensional data suggests finer movement of
    a larger number of cell populations that is not detectable using only the
    1-dimensional data. In particular, a clean separation of the heavily
    overlapping clusters $9$ and $10$ was not possible in the 1-dimensional
    model, but is clear in the 3-dimensional model (also see Figure
    \ref{fig:3d-full} that this separation is made apparent by using the
    additional \texttt{red} axis). }
  \label{fig:3d-cell-diam}
\end{figure}

In this section, we apply our model to the full 3-dimensional data. First, in
Figure~\ref{fig:3d-cell-diam}, we display one dimension (cell diameter) of the
estimated $10$-cluster $3$-dimensional model, as a direct comparison to the
1-dimensional cell diameter analysis in Section \ref{sec:1d-real-data}. Cluster
$10$ is recognized by domain experts to correspond to \textit{Prochlorococcus}. The
separation of the two heavily overlapping clusters $9$ and $10$, and their independent
means' movement, are visually not apparent in the cell diameter data alone;
indeed, the estimated 1-dimensional model in Figure \ref{fig:1d} only captures a
single \textit{Prochlorococcus} cluster $5$.

\begin{figure}
  \centering
\makebox[\linewidth]{\includegraphics[width=1.2\linewidth]{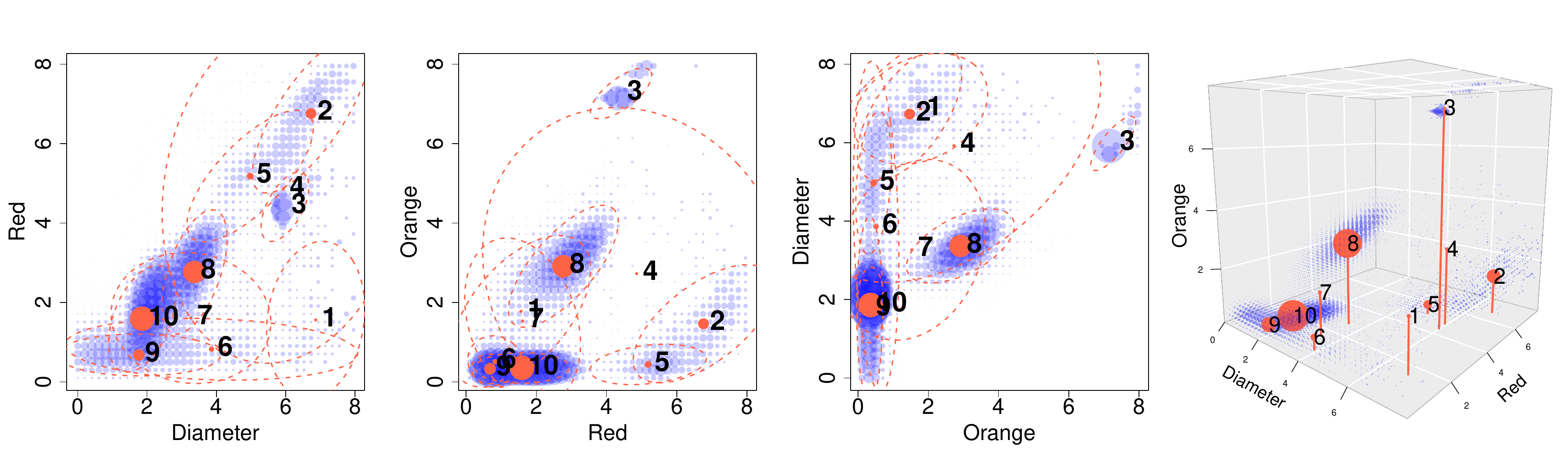}}
  \caption{\it The estimated 3-dimensional 10-cluster model described in
    Section~\ref{sec:3d-real-data}, at one time point.  The size of the blue
    points represents the biomass in each of the $40^3$ bins. The panels show
    various views of the cytograms -- three 2d scatterplots and our estimated
    parameters (means, probabilities, and covariances).  The red dots mark the
    cluster centers at this time point, and the size (radius) of these red dots
    are proportional to the cluster probabilities. The red ellipses in dashed
    lines show the estimated 95\% probability region of the data formed from the
    estimated Gaussian covariance of each cluster.  The $10$ estimated model
    clusters' mean fluctuations and cluster probability dynamics over time can
    be seen in the full video in \url{https://youtu.be/jSxgVvT2wr4} -- a single
    frame of this video is shown in Figure~\ref{fig:3d-full}.}
  \label{fig:3d-full}
\end{figure}

The full $3$-dimensional data and estimated model are challenging to display in
print. A better medium than flat images is a video of $t=1,\cdots,T$ images over
time, which we show in
\url{https://youtu.be/jSxgVvT2wr4}.
Figure~14 of the Supplement
shows one frame
from this video (corresponding to one $t$), which overlays with several plots:
three $2$-dimensional projections of the cytogram, two different angles of the
$3$-dimensional cytograms, the cruise location on a map, the covariates over
time, and the cluster probabilities at each time and as a time series.  The
first four panels of this snapshot are shown in Figure~\ref{fig:3d-full} in
higher resolution.  The mean fluctuations and cluster probability dynamics over
time are clearly captured in the full video, and are explained next, in the
context of covariates.

The estimated mean movement and the $\beta$ coefficients shown in
Tables~4-6 in the Supllement
reveal interesting scientific
insights. The cell diameter of \textit{Prochlorococcus} seems to be well
predicted by sunlight and lagged variants of sunlight. To elaborate, the
estimated entries of $\beta_5$ corresponding to the covariates \texttt{p1},
\texttt{p2} and \texttt{p3} and the cell diameter axis, were estimated as
$0.008$, $0.010$ and $0.013$ -- meaning that the mean cell diameters of
\textit{Prochlorococcus} are predicted to increase by these amounts with a unit
increase in each covariate value. This supports biochemical
intuition about the cell size being directly driven by sunlight. Indeed,
important physiological processes of phytoplankton cells, including growth,
division, and fluorescence (particularly of the pigment chlorophyll-A), are
known to undergo diel variability, i.e. timed with the day-night or light cycle.

Estimated cluster probabilities and the coefficients $\alpha$ shown in
Tables~3
are also quite interpretable. A higher positive
estimated entry of $\alpha_k$ means that a unit increase of that covariate
corresponds to a larger increase of the relative probability of the $k$'th
cluster. The probability of Cluster $8$ (which occupies a region in the orange
fluorescence axis that clearly corresponds to the \textit{Synechococcus}
population) is associated with primary productivity (coefficient value of
$0.19$), oxygen ($0.46$) and nitrate ($-0.35$). Rapid increases in the abundance
and biomass of \textit{Synechococcus} associated with high productivity have
previously been observed over narrow regions of the Pacific at the boundary
between the Subtropical and Subpolar Gyres \citep{Gradoville2020} . High
productivity in the ocean is often linked to high oxygen saturation, a result of
oxygen production during photosynthesis, and low nitrate, as a result of
consumption of this nutrient required for \textit{Synechococcus}'s cell growth
\citep{Moore2002}. Linkages to such biochemical factors unique to this specific
\textit{Synechococcus} cluster are otherwise difficult to identify, but are
clearly identified in our model. In contrast, for cluster $10$
(\textit{Prochlorococcus}), the largest $\alpha$ coefficients correspond to sea
surface temperature ($0.87$) and phosphate ($-0.94$). These results reflect this
organism's observed distribution in the Pacific Ocean; namely its Subtropical
Gyre, where high surface temperatures and low concentrations of phosphate tend
to favor small-celled \textit{Prochlorococcus} leading to higher cluster
probabilities. Interestingly, nitrate was not detected by the model as a
relevant covariate, which is in good agreement with the physiology of
\textit{Prochlorococcus}, which often lack the genes necessary for nitrate
assimilation \citep{Berube2015}.

On the other hand, the large positive $\alpha$ coefficients for cluster $2$
(\textit{Picoeukaryotes}) associated with phosphate ($0.35$) reflects its more
northerly distribution in the North Pacific Subpolar Gyre, a region of the ocean
distinguished by higher surface concentration of nutrients including phosphate
which allow for greater growth of these relatively larger phytoplankton.

Finally, cluster $3$ is particularly interesting as it captures the calibration
beads injected by the instrument as an internal standard. The location of this
cluster is much more apparent in the full 3-dimensional representation in Figure
\ref{fig:3d-full}.  This is the only population whose origin and location is
known a priori, and thus serves as a negative control, which the model is
expected to capture. Indeed, in our estimated $10$-cluster 3-dimensional model,
this bead is clearly captured as a separate population whose mean movement is
minimal over time. Interestingly, 3-dimensional models with fewer than $10$
clusters fail to capture the calibration bead as a separate population.

\subsection{Comparison to gating}
\label{sec:gating}
\begin{figure}
  \centering
  \makebox[\linewidth]{\includegraphics[width=\linewidth]{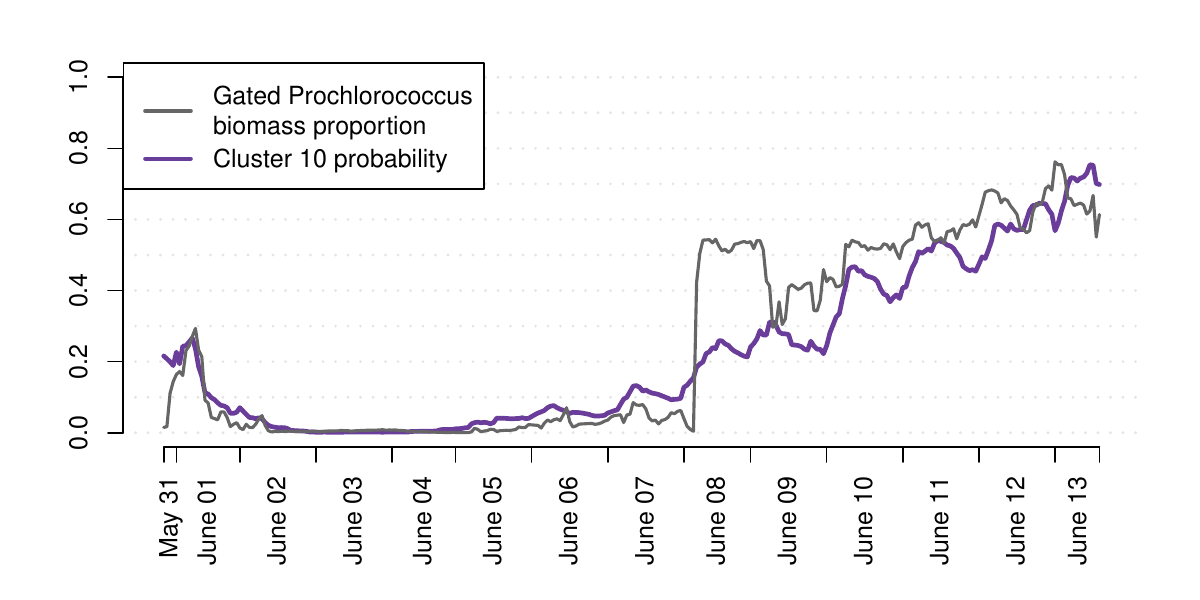}}
  \caption{\it This figure shows the relative biomass of
    \textit{Prochlorococcus}, measured in two ways -- using traditional gating
    (black line), and using the estimated cluster probability of cluster $10$
    (purple) in the 3-dimensional data in Section \ref{sec:3d-real-data} and
    Figure \ref{fig:3d-full}. One noticeable discrepancy is on June 8th and
    9th. The gating (black line) abruptly jumps from $0$ to $0.5$ due to flaws
    in automatic gating, while our model (purple) suggests a gradual increase on
    June 8th and onwards. Visual inspection and expert annotation of this
    cluster in the cytogram suggests that our model cluster $10$ is correctly
    tracking \textit{Prochlorococcus}.}
  \label{fig:gating-prochloro}
\end{figure}

In Figure \ref{fig:gating-prochloro}, we compare the relative biomass of
\textit{Prochlorococcus}, measured in two ways. The dark grey line shows the
relative biomass of \textit{Prochloroccocus}, gated in \citet{seaflow-data}
using \texttt{flowDensity} bioconductor package \citep{Malek2015a}, applied
semi-automatically to individual 3-dimensional cytograms recorded roughly every
$3$ minutes, then aggregated to an hourly level. There is a noticeable
discrepancy between the two methods on June 8th and 9th. The dark grey line
abruptly rises from near $0$ to about $0.5$, while the purple line follows a
gradual increase from June 8th onwards. The reasons for this discrepancy are
apparent from visual examination of the gated cytograms.  First, the gating
results have no continuity or smoothness over time, having been applied to
individual cytograms. More importantly, while our model consistently tracks the
\textit{Prochlorococcus} cluster as a single ellipsoidal cluster $10$, the
semi-automatic gating function erroneously includes external particles -- many
from our model's cluster $9$, which domain experts would not consider to be
\textit{Prochlorococcus}.

 \section{Conclusion}

In this work, we propose a novel sparse mixture of multivariate regressions
model for modeling flow cytometry data. We devise a penalized
expectation-maximization algorithm with parameter constraints and implement a
specific ADMM solver, which is called in the M-step. Our simulations and
application results in Section \ref{sec:numerical} and \ref{sec:real-data}
demonstrate that our proposed model can reveal interpretable insights from flow
cytometry data, and help scientists identify how environmental conditions
influence the dynamics of phytoplankton populations.

Our method provides scientists with a rich description of the association
between environmental factors and phytoplankton cell populations.  It leverages
covariates and all cytograms to identify cell populations. This means two cell
populations that might be indistinguishable in a single cytogram could be
differentiated if their dynamics (i.e. dependence on covariates) are distinct
from each other. Thus, even when one is not interested in the covariates
themselves but only the estimation of cell populations (as in gating) this
method still may be the best choice. In applying the method, we recover some
known associations, such as \textit{Prochlorococcus} and light (positive
controls), we did not identify some known non-associations (negative controls),
and also produced some new associations that can be studied. Also, in
investigating a discrepancy between our method and a pre-existing gating
approach, we uncovered some undesirable behavior of the pre-existing approach,
and showcased our method's ability to perform the difficult task of automatic
and consistent gating of overlapping clusters in cytograms over time.

While the motivation from this methodology comes from oceanography, the flow
cytometry technology is important to many other areas, including biomarker
detection \citep{biomarker-flow-cytometry}, diagnosis of human diseases such as
tumors \citep{clinical-flow-cytometry}, and ecological studies
\citep{ecology-flow-cytometry}.  For instance, in a biomedical application,
covariates can be patient attributes, and the response can be cytograms obtained
from patient blood samples. In fact, the statistical methodology developed here
can be applied to any context in which modeling cytograms in terms of features
is reasonable -- the time ordering of the data is not required for application.
We therefore expect it to be valuable in a wide range of fields.

Our model diagnostics in Supplement~E indicate some leftover time dependence in
the data residuals from our model. To remedy this within the framework of our
model, one might add time-lagged versions of the covariates, or even summaries
from cytograms $\y^{(t)}$, to directly incorporate time-space autocorrelation in
our model. Alternatively, one could also extend the $d$-by-$d$ cluster
covariance $\bSigma_k \in \R^{d \times d}$ to be a time-varying matrix
$\bSigma_{kt}$. This covariance matrix can take time structure that is not
driven by covariates $X^{(t)}$, but has dependence (e.g. time autocorrelation)
or smoothness that is learned directly from the data. 
However, a time series extension also complicates our existing cross-validation
strategy for tuning $\lambda_\alpha$ and $\lambda_\beta$, and constitutes a
significant departure from our current proposed model. We view a time-series
extension of our model to be an excellent methodology direction to pursue next.

The methodology has several exciting directions for future work. Our mixture
model methodology would greatly benefit from a principled, automatic choice of
the number of $K$ based on the data.  It would be also be interesting to see how
relaxing the Gaussian cluster assumption to different distributions --
e.g. skewed, multivariate $t$ distributions -- helps improve the flexibility of
our approach.  A model with feature-dependent covariances
$\{\bSigma_k\}_{k=1,\cdots, K}$ could enable more flexible prediction as
well. Also promising are the extension and comparison to more non-parametric
approaches to the conditional distribution of cytograms, or to the entire joint
model of cytograms and environmental covariates.

On the application side, it would be interesting to compare estimated models on
data from other oceanographic cruises traversing the same trajectory or
different areas, and see to what extent the estimated relationship between
cytograms and environmental covariates can be replicated.

\section*{Acknowledgments}

The authors acknowledge the Center for Advanced Research Computing (CARC) at the
University of Southern California for providing computing resources that have
contributed to the research results reported within this
publication. \url{https://carc.usc.edu}.

This work was supported by grants by the Simons Collaboration on Computational
Biogeochemical Modeling of Marine Ecosystems/CBIOMES (Grant ID: 549939 to JB,
Microbial Oceanography Project Award ID 574495 to FR). Dr. Jacob Bien was also
supported in part by NIH Grant R01GM123993 and NSF CAREER Award DMS-1653017. We
thank Dr. E. Virginia Armbrust for supporting SeaFlow deployment on the cruise
in the North Pacific funded by the Simons Foundation grant (SCOPE Award ID
426570SP to EVA).  We also thank Chris Berthiaume and Dr. Annette Hynes for
their help in processing and curating SeaFlow data.

\bibliographystyle{unsrtnat}
\bibliography{flowmix}
\end{document}


\maketitle

\section*{Supplement A: Proof of Proposition 1}
\setcounter{equation}{13}
\setcounter{figure}{10}

\begin{proof}
  Let us denote $ \theta := (\alpha, \beta, \Sigma)$ and write $g(\theta)$ in
  place of $g(\alpha, \beta)$ for brevity. Also recall our shorthand for
  particles $\y := \{\y_i^{(t)}\}_{i,t}$. First write the penalized likelihood in
  (3)
  (and the objective of
  (7)) as:
  \begin{equation*}
    f(\theta,\y):=-(1/N)\log \cL(\alpha, \beta, \Sigma; \{\y_i^{(t)}\}_{i,t}) + g(\theta),
  \end{equation*}
  to emphasize it is a bivariate function of $\theta$ and $\y$. The latter $\y$ is
  taken to be a $\R^{d\sum_tn_t}$ vector, having vectorized entries of
  $\{\y_i^{(t)}\}_{i,t}$.   Now, define the particle mapping $h_B: \R^d \to \R^d$,
  \begin{equation*}
    h_B(\y_i^{(t)}) = \sum_{b=1}^B  \tilde \y_b \one \{ \y_i^{(t)} \in E_b\},
  \end{equation*}
which maps a point to the center of the bin containing it. Writing
  $h_B(\y) = \{h_B(\y_i^{(t)})\}_{i,t}$, we can succinctly express the objective
  of
  (6), $f(\theta, h_B(\y))$.

Next, we want to see that $f(\theta, \y)$ as a function of $\y$ is Lipschitz
  over $\theta$ i.e. for any datasets $\y$, $\y'$ in the data domain
  $\R^{d\sum_t n_t}$, there exists a finite constant $L$ such that,
  \begin{equation}\label{eq:unif-lipschitz}
    \max_{\theta\in\Theta} |f(\theta, \y) - f(\theta, \y')| \le L \cdot \|\y - \y'\|_2.
  \end{equation}
By the mean value theorem and Cauchy-Schwarz, there exists
  $\tilde \y \in \mathcal{Y}$ such that:
  \begin{equation*}
    |f(\theta, \y ) - f(\theta, \y')| \le \|\nabla_\y f (\theta, \tilde \y)\|_2 \cdot \|\y-\y'\|_2
  \end{equation*}
  The gradient of $f(\theta,\y)$ has subvectors of the form:
  \begin{equation*}
     \nabla_{\y_{i}^{(t)}} f(\theta,\y) = \frac{\sum_{k=1}^K  \pi_{kt}(\alpha) \cdot \phi(\y_i^{(t)};\bmu_{kt}(\beta), \bSigma_k )\cdot (-1/2) \cdot  \bSigma_k^{-1}\cdot (\y_i^{(t)} - \bmu_{kt}(\beta))}{\sum_{k=1}^K \pi_{kt}(\alpha) \cdot\phi(\y_i^{(t)}; \bmu_{kt}(\beta),
      \bSigma_k)}.
\end{equation*}
Given that $\bSigma_k \ge c \bI_d$ for all $k$, this is a continuous function on
  the compact domain $\Theta \times \mathcal{Y}$, so by Weierstrass's extreme
  value theorem, it attains a finite maximum, and so
  \eqref{eq:unif-lipschitz}
  holds with
  $L := \max_{(\theta, \y) \in \Theta \times \mathcal{Y}} \|\nabla_\y f(\theta,
  \y)\|_2 $.

  It follows that:
  \begin{equation}\label{eq:unif-conv}
    \max_{\theta\in\Theta} | f (\theta, h_B(\y)) - f (\theta, \y) |  \le L \cdot \|h_B(\y) - \y\|_2
    \le L \cdot \sqrt{\sum_{t=1}^T n_t} \cdot  R \sqrt{d} {B}^{-\frac{1}{d}} \;\; \xrightarrow{B \to \infty} 0,
  \end{equation}
  using that the largest distance between any $\y_i^{(t)}$ and $h_B(\y_i^{(t)})$,
  both $d$-vectors, is smaller than the length
  $\sqrt{d} R B^{-\frac{1}{d}}$ of the main diagonal of a
  $d$-dimensional hypercube. This establishes that $f(\theta, \y)$ and
  $f(\theta, h_B(\y))$ are arbitrarily close \textit{uniformly} in $\theta$, as
  $B$ increases.

  Now, further denote $f(\theta, \y)$ as $f(\theta)$, and $f(\theta, h_B(\y))$
  as $f_B(\theta)$, whose subscript $B$ emphasizes the dependence on the number
  of bins $B$. For any sequence $\tilde \theta_B$ of elements taken from the set
  sequence $\tilde \Theta_B$, there exists a convergent subsequence
  $\tilde \theta_{s_B}$ for some sequence $s_B$, by the Bolzano-Weierstrass
  theorem. Now, we proceed to show that
  $\lim_{B \to \infty} \tilde \theta_{s_B}$ is in $\hat \Theta$. For any
  $\hat \theta \in \hat \Theta$ and $\tilde \theta_{s_B}$:
  \begin{equation}\label{eq:lemma-proof-1}
    f_{s_B}(\tilde \theta_{s_B}) \ge f_{s_B}(\hat \theta).
  \end{equation}
  Taking the lim inf of both sides and using, by \eqref{eq:unif-conv}, that
  $\lim_{B \to \infty} f_{s_B}(\hat \theta) = f(\hat \theta)$,
  \begin{equation}\label{eq:lemma-proof-2}
\liminf_{B \to \infty} f_{s_B}(\tilde \theta_{s_B}) \ge f(\hat \theta).
  \end{equation}
  Now, bound $f(\hat \theta) - f(\tilde \theta_{s_B})$ from above as follows:
  \begin{align*}
    \limsup_{B \to \infty} \left[ f(\hat \theta) - f(\tilde \theta_{s_B})\right]
    &= 
      \limsup_{B \to \infty} \left[ f(\hat \theta) - f_{s_B}(\tilde \theta_{s_B}) + f_{s_B} (\tilde \theta_{s_B}) - f(\tilde \theta_{s_B})\right] \\
    &\le 0 + \limsup_{B \to \infty} \left[ f_{s_B}(\tilde \theta_{s_B}) - f(\tilde \theta_{s_B})\right]\;\;\text{by}\;\; \eqref{eq:lemma-proof-2} \\
    &\le \limsup_{B \to \infty} \left[\sup_\theta | f_{s_B}(\theta) - f_{s_B}( \theta)|\right]\\
    &=0 \;\;\text{by}\;\; \eqref{eq:unif-conv}.
  \end{align*}
  Replacing the lim sup with the limit, we have
  \begin{equation}\label{eq:lim-of-f}
    \lim_{B \to \infty} f(\tilde \theta_{s_B}) = f(\hat \theta)
  \end{equation}
Since $f$ is continuous on $\Theta$, we have: 
  \begin{equation}\label{eq:f-of-lim}
    \lim_{B \to \infty} f(\tilde \theta_{s_B}) = f(\lim_{B \to \infty} \tilde \theta_{s_B}) =   f(\hat \theta),
  \end{equation}
which proves the limit $\lim_{B\to\infty} \tilde
  \theta_{s_B}$ of any convergent subsequence $\tilde
  \theta_{s_B}$ is a minimizer of the function
  $f$. This proves our original statement
  (8).
\end{proof}

\section*{Supplement B: ADMM details}

Continuing directly from the end of
Section 2.5, we describe the
ADMM algorithm for solving
(12)
in full
detail. (As a reminder, the subscript $k$ has been dropped for notational
simplicity.) Using augmented variables $\bU_{\bZ} \in \R^{T \times d}$ and
$\bU_{\bW} \in \R^{p \times d}$ (combined as
$\bU = \icol{\bU_Z\\\bU_{\bW}} \in \R^{(p+T) \times d}$) and penalty parameter
$\rho \in \R$, the augmented Lagrangian is:
\footnote{ The variables
  and data are of dimension:
  $ \bZ^{(t)} \in \R^d, \; \bW,\bbeta \in \R^{p \times d}, \; \X \in \R^{T \times p},
  \; \X^{(t)} \in \R^{p}, \; \bZ \in \R^{T \times d}$.}
\begin{align}\label{eq:lagrangian}
  L_\rho =& \frac{1}{2N} \sum_{i,t} C_i^{(t)} \gamma_{it} (\tilde \y_i^{(t)} -
            \bbeta^T \tilde \X^{(t)} )^T \hat \bSigma^{-1} (\tilde \y_i^{(t)} - \bbeta^T\tilde \X^{(t)}) +
  \langle \bU,\icol{\X \\ \bI} \bbeta - \icol{\bZ \\ \bW}   \rangle  \\
  &+\frac{\rho}{2} \| \icol{\X \\ \bI} \bbeta - \icol{\bZ \\ \bW} \|_F^2
  + \lambda_{\bbeta} \|\bW\|_1 + \sum_{t=1}^T \one_{\infty}(\|\bZ^{(t)}\|_2 \le
  r). \nonumber
\end{align}
Since $L_\rho$ is a quadratic function of $\bbeta$, we can find the solution by
setting the gradient of $L_\rho$ with respect to $\bbeta^T$ equal to zero:
\begin{align}
  \nabla_{\bbeta^T} L_\rho
  &= \frac{1}{N} \sum_{i,t} C_i^{(t)} \gamma_{it} \hat \bSigma^{-1}(\bbeta^T
    \tilde \X^{(t)} - \tilde \y_{i}^{(t)} ) {\tilde \X^{(t)}}^T + \bU^T \icol{ \X \\
  \bI} + \rho \left[ \bbeta^T \icol{\X \\ \bI}^T - \icol{\bZ \\ \bW}^T \right]\icol{\X\\\bI}
 = \bzero.  \label{eq:cow}
\end{align}
The first term in \eqref{eq:cow} can be simplified using
$\dbtilde \y^{(t)}:= \frac{1}{N} \sum_i C_i^{(t)} \gamma_{it} \tilde
\y_{i}^{(t)}$ and
$\bD:=\text{diag}( \{\frac{1}{N}\gamma_{t}/\sum_t \gamma_{t} \}_t)$,
(recalling that $\gamma_{t}:=\sum_{i=1}^{n_t} C_i^{(t)}\gamma_{it}$.)
as follows:
\begin{align*}
  \frac{1}{N} &\sum_{i,t} C_i^{(t)} \gamma_{it} \hat \bSigma^{-1} (\bbeta^T \tilde \X^{(t)} -  \tilde \y_{i}^{(t)}) {\tilde \X^{(t)}}^T \\
              &=  \hat \bSigma^{-1} \bbeta^T \sum_{t} \left(\frac{1}{N}\sum_{i} C_i^{(t)} \gamma_{it}\right) \tilde \X^{(t)} {\tilde \X^{(t)}}^T - \hat \bSigma^{-1} \sum_{t}
                \underbrace{ \frac{1}{N} \sum_i C_i^{(t)} \gamma_{it} \tilde \y_{i}^{(t)}}_{\dbtilde \y^{(t)}} {\tilde \X^{(t)}}^T\\
              &= \hat \bSigma^{-1} \bbeta^T \tilde \X^T \underbrace{\bD}_{\text{diag}( \{\frac{1}{N}\gamma_{ t}/\sum_t \gamma_{ t} \}_t) } \tilde \X \hspace{4mm}-\hspace{4mm} \hat \bSigma^{-1} \dbtilde \y^T \tilde \X.
\end{align*}
Using this, \eqref{eq:cow} can be rewritten as:
\begin{align} \nabla_{\bbeta^T} L_\rho &= \hat \bSigma^{-1} \bbeta^T \tilde \X^T \bD \tilde \X - \hat \bSigma^{-1} \dbtilde \y^T \tilde \X + \bU^T \icol{\X \\ \bI} + \rho \bbeta^T(\X^T\X + \bI) - \rho(\bZ^T \X + \bW^T) =\bzero,\label{eq:horse}
\end{align}
which can be further simplified to
\begin{align*}
  \hat \bSigma^{-1} \bbeta^T \tilde \X^T \bD\tilde \X + \rho \bbeta^T (\tilde \X^T\tilde \X + \bI) + \bE = \bzero,
\end{align*}
for $\bE := -\hat \bSigma^{-1}\dbtilde \y^T \tilde \X +  \bU^T \icol{\X \\ \bI} - \rho (\bZ^T \X + \bW^T)$
and $\dbtilde \y$ whose rows are $\dbtilde \y^{(t)}$. Lastly, if we multiply
$(\X^T\X+\bI)^{-1}$ on the right and $\hat \bSigma$ on the left, and take the
transpose, we get:
\begin{equation}
  \label{eq:bartels-form}
  (\X^T\X + \bI)^{-1} \tilde \X^T \bD\tilde \X \bbeta  +
 \bbeta \cdot \rho \hat \bSigma  + (\X^T\X + \bI)^{-1}\hat  \bE^T \bSigma = \bzero,
\end{equation}
which is of the form ${\bA}\bbeta + \bbeta {\bB} + \bC=\bzero$ for the matrix-valued
variable $\bbeta$ and square matrices $\bA$ and $\bB$. This is the famed
Sylvester equation that is typically solved the Bartels-Stewart algorithm
\citep{bartels-stewart}, an algorithm that is well studied and implemented in
major software packages (e.g. C++ Armadillo \cite{Sanderson2016}).

However, invoking Bartels-Stewart to solve \eqref{eq:bartels-form} at every ADMM
iteration is unnecessarily expensive, since the algorithm always starts with
two Schur decompositions of $\bA = \rho \hat \bSigma $ and
${\bB} = \tilde \X^T \bD\tilde \X (\X^T\X + \bI)^{-1}$ that do not
need to be repeated (i.e. ${\bA}$ and ${\bB}$ are the same for all ADMM iterations in
this EM algorithm iteration).

Instead, if we perform the Schur decompositions once before the first ADMM iteration
to obtain $\bA=\bU_{\bA} \bT_{\bA} \bU_{\bA}^T$ and
${\bB}=\bU_{\bB}\bT_{\bB}\bU_{\bB}^T$ (producing orthonormal matrices
$\bU_{\bA}, \bU_{\bB}$ and upper-triangular matrices $\bT_{\bA}, \bT_{\bB}$),
the equation ${\bA}\bbeta + \bbeta {\bB} + {\bC}=\bzero$ to solve at every subsequent ADMM iteration is much faster to solve, since it can be written
as:
\begin{align}
  \bU_{\bA}\bT_{\bA}\bU_{\bA}^T\bbeta + \bbeta \bU_{\bB}\bT_{\bB}\bU_{\bB}^T + {\bC} = \bzero.
\end{align}
which is equivalent to:
\begin{align}
  \bT_{\bA}\left(\bU_{\bA}^T\bbeta \bU_{\bB}\right) + \left(\bU_{\bA}^T \bbeta \bU_{\bB}\right)\bT_{\bB} + \bU_{\bA}^T {\bC} \bU_{\bB}= \bzero,
\end{align}
which is a special case of a Sylvester equation -- in the variable
$\bbeta' = \bU_{\bA}^T \bbeta \bU_{\bB}$ -- that is extremely fast to solve since the
coefficients $\bT_{\bA}$ and $\bT_{\bB}$ are triangular matrices.
The ADMM iterations are run until a particular convergence check is met,
described in the next subsection.  In addition, in the \texttt{flowmix} R
package, we use the approach of \cite{laadmm} to adaptively change the penalty
parameter $\rho$, which makes the algorithm's convergence more robust compared
to when using a fixed choice of $\rho$.

\subsubsection*{Convergence criterion}

Following Section 3.3 in \cite{boyd-admm}, a convergence criterion is
established by limiting the \textit{primal} and \textit{dual residuals} of our
problem. This can derived by first framing our augmented problem in
(12)
in terms of the canonical ADMM problem
(using block matrix notation):

\begin{mini}{\bbeta, \bZ, \bW}{
    f(\bbeta) + g(\bZ, \bW)
    }{}{}
    \addConstraint{\bF\bbeta + \bG \left(\begin{matrix} \bW \\ \bZ\end{matrix}\right) =
      \left( \begin{matrix} \bzero \\ \bzero \end{matrix} \right) \in \R^{(T + p)\times d}}
\end{mini}
for matrices $\bF$ and $\bG$ defined as:
\begin{equation}
  \bF = \left(\begin{array}{c}  {\bI}_p \\
                \X\end{array}\right),
            \bG = \left(\begin{matrix} - {\bI}_p & \bzero \\
                                       \bzero      & {\bI}_T \end{matrix}\right),
\end{equation}
where the left-most column of $\bF$ is a single zero vector, and $\bI_p \in
\R^{p \times p}$ and
$\bI_T \in \R^{T \times T}$ are identity matrices. Using this notation,
the primal and dual residuals are written as $\br$ and $\bs$ respectively:
\begin{align*}
  \br &= \bF  \bbeta + \bG \begin{pmatrix} \bW \\ \bZ \end{pmatrix} \in \R^{(p+T) \times d} \\
  \bs &= \rho \bF^T \bG \left[ \begin{pmatrix} \bW\\\bZ \end{pmatrix} - \begin{pmatrix} \bW^{\text{prev}}\\\bZ^{\text{prev}} \end{pmatrix}\right] \in \R^{p \times d},
\end{align*}
where the $\bW^{\text{prev}}$ denotes the value of $\bW$ from the previous
iteration.  Since $\br$ and $\bs$ are matrices, we use the Frobenius norm in
controlling their entrywise size as the stopping rule. The stopping criterion is
then
\begin{equation*}
\|\br\|_F \le \epsilon^{\text{pri}} \text{ \;\; and \;\; } \|\bs\|_F \le \epsilon^{\text{dual}},
\end{equation*}
for a stopping tolerance value
\begin{align*}
\epsilon^{\text{pri}} &= \epsilon^{\text{rel}}\max\{\|\bF \bbeta \|_F, \left\|\bG\begin{pmatrix}\bW \\ \bZ\end{pmatrix}\right\|_F \},  \\
\epsilon^{\text{dual}} &= \epsilon^{\text{rel}}\|\bF^T\begin{pmatrix}\bU_{\bW} \\ \bU_{\bZ}\end{pmatrix} \|_F,  
\end{align*}
for a relative tolerance threshold value $\epsilon^{\text{rel}} = 10^{-3}$.

 \section*{Supplement C: Additional data analysis results}

\begin{figure}[ht!]
\makebox[\linewidth][c]{\subfloat{\includegraphics[width=.6\linewidth]{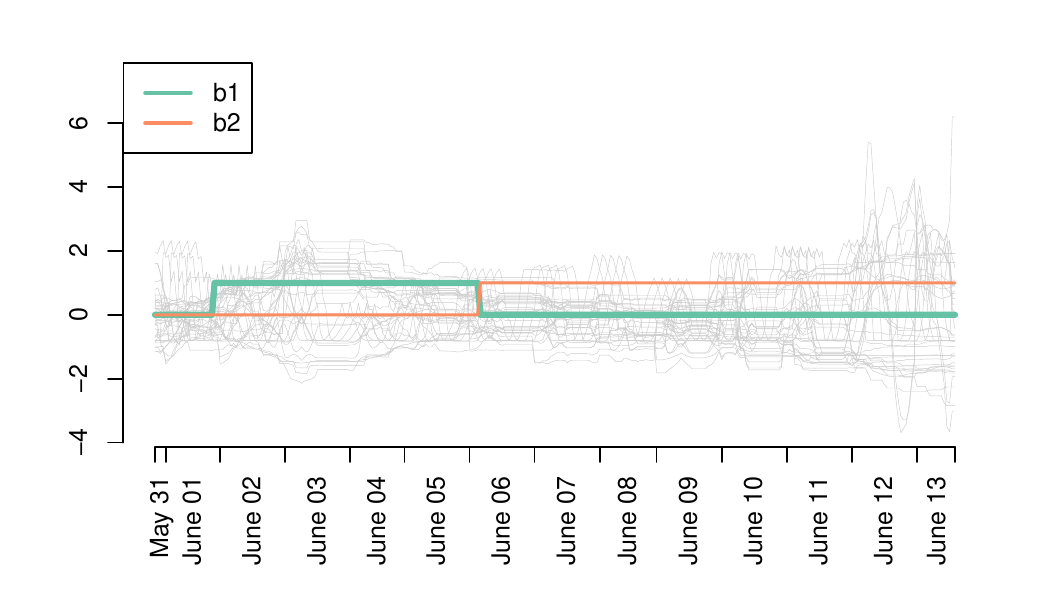}}
  \hskip -4ex
  \subfloat{\includegraphics[width=.6\linewidth]{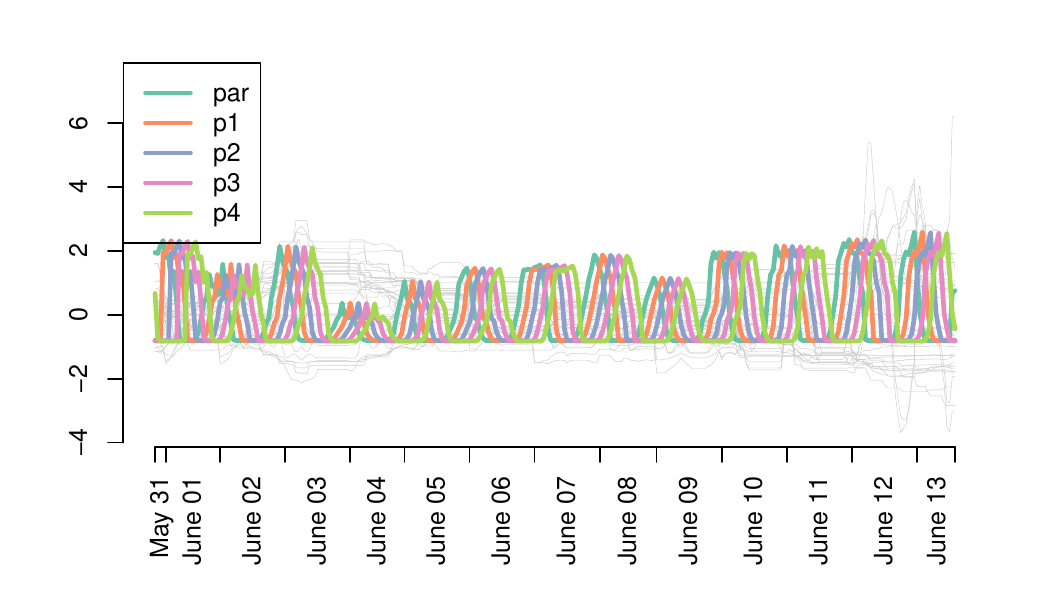}}
}\\[-5ex] 
\makebox[\linewidth][c]{\subfloat{\includegraphics[width=.6\linewidth]{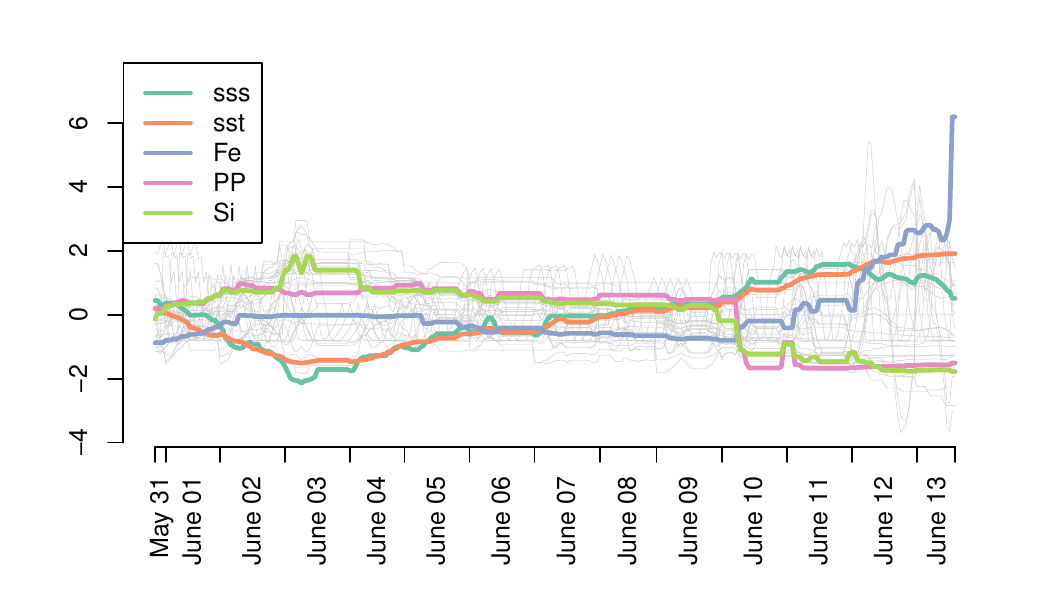}}
  \hskip -4ex
  \subfloat{\includegraphics[width=.6\linewidth]{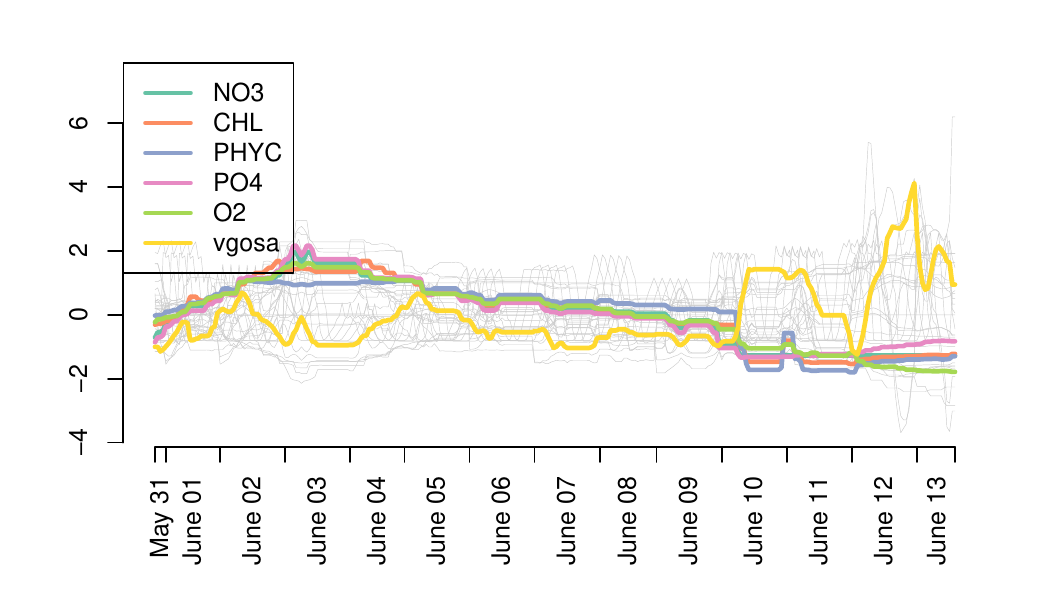}}
}\\[-5ex] 
\makebox[\linewidth][c]{\subfloat{\includegraphics[width=.6\linewidth]{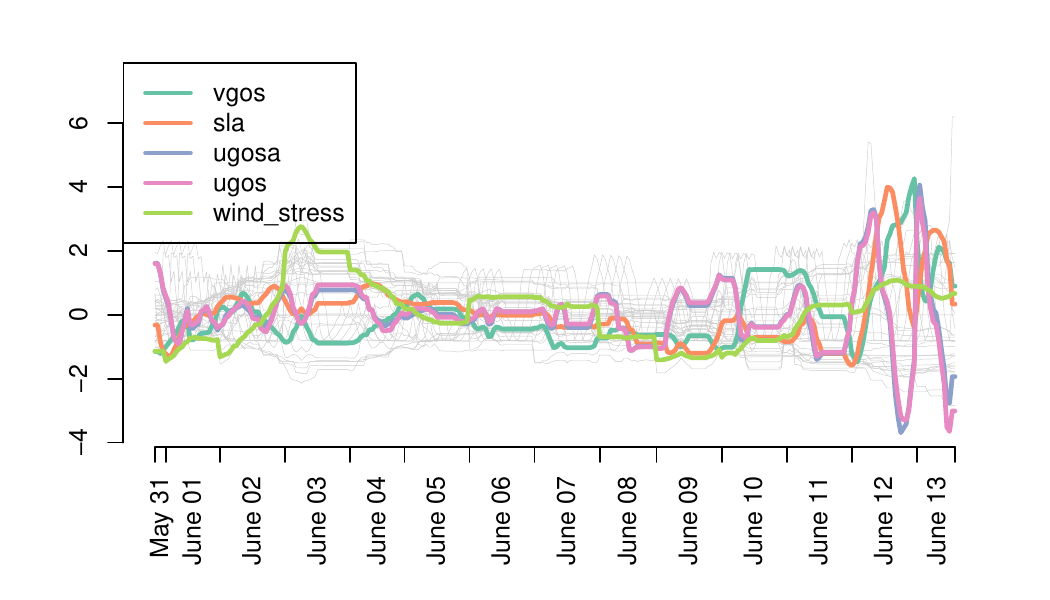}}
  \hskip -4ex
  \subfloat{\includegraphics[width=.6\linewidth]{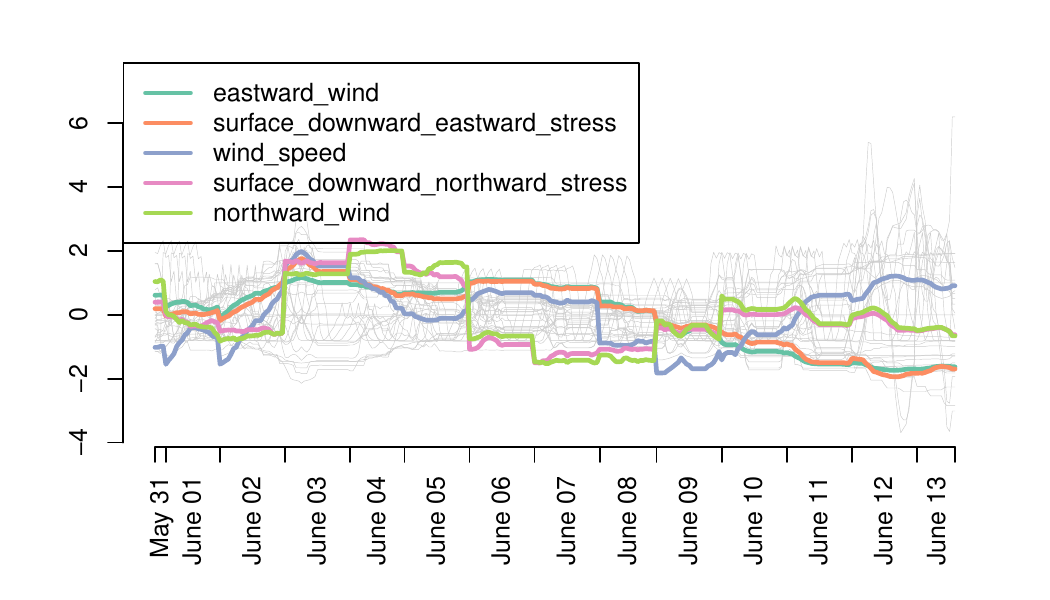}}
}\\[-5ex] 
\makebox[\linewidth][c]{\subfloat{\includegraphics[width=.6\linewidth]{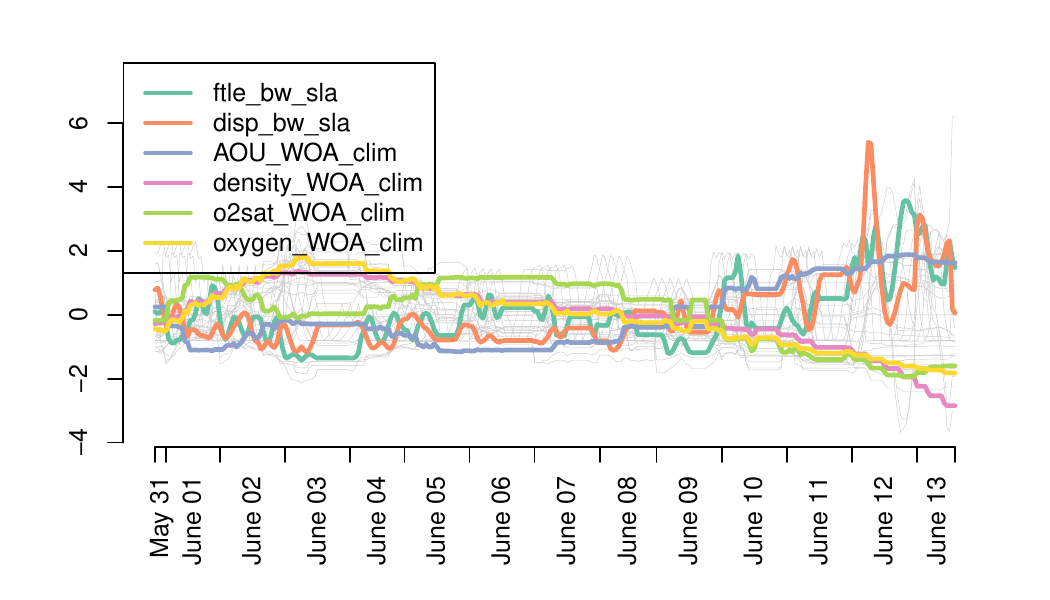}}
  \hskip -4ex
  \subfloat{\includegraphics[width=.6\linewidth]{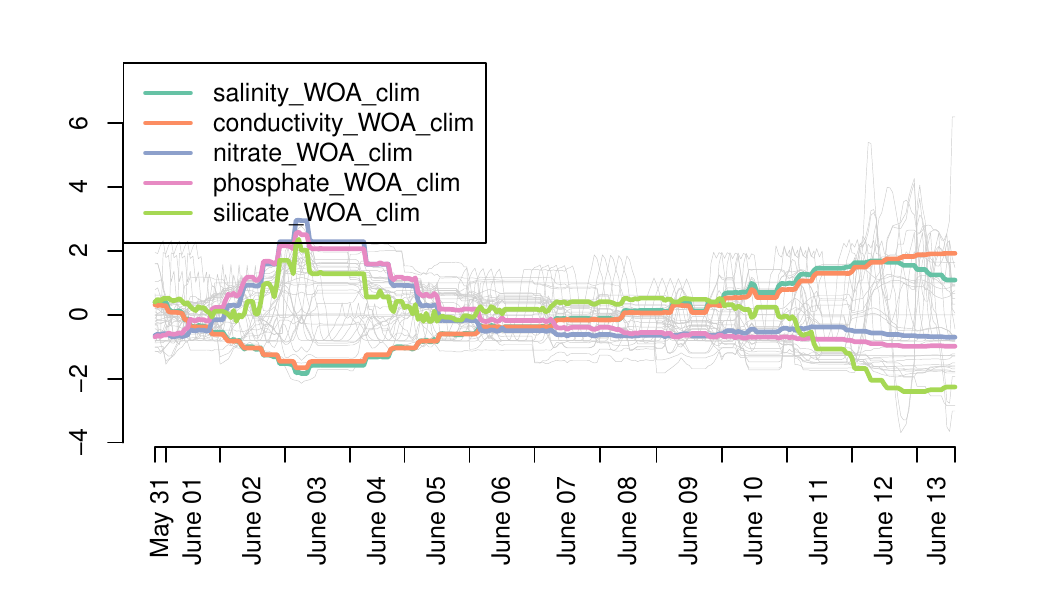}}
}
\caption{\it Eight panels showing the four or five environmental covariates at a
  time. The first two figures show manually created covariates. In the first
  figure, \texttt{b1} and \texttt{b2} are indicator variables for regions
  crossings of the cruise across an important ecological transition zone. In the
  second figure, \texttt{p1}, \texttt{p2}, \texttt{p3}, \texttt{p4} are the
  sunlight variable \texttt{par} lagged by 3,6,9 and 12 hours. The rest of the
  covariates are described briefly in Table
  \ref{tab:environmental-covariates-labels}. All covariates except for
  \texttt{b1} and \texttt{b2} were standardized to have mean 0 and sample
  standard deviation 1.}
\label{fig:environmental-covariates}
\end{figure}

\begin{figure}[ht!]
  \centering
  \makebox[\textwidth]{\includegraphics[width=.8\linewidth]{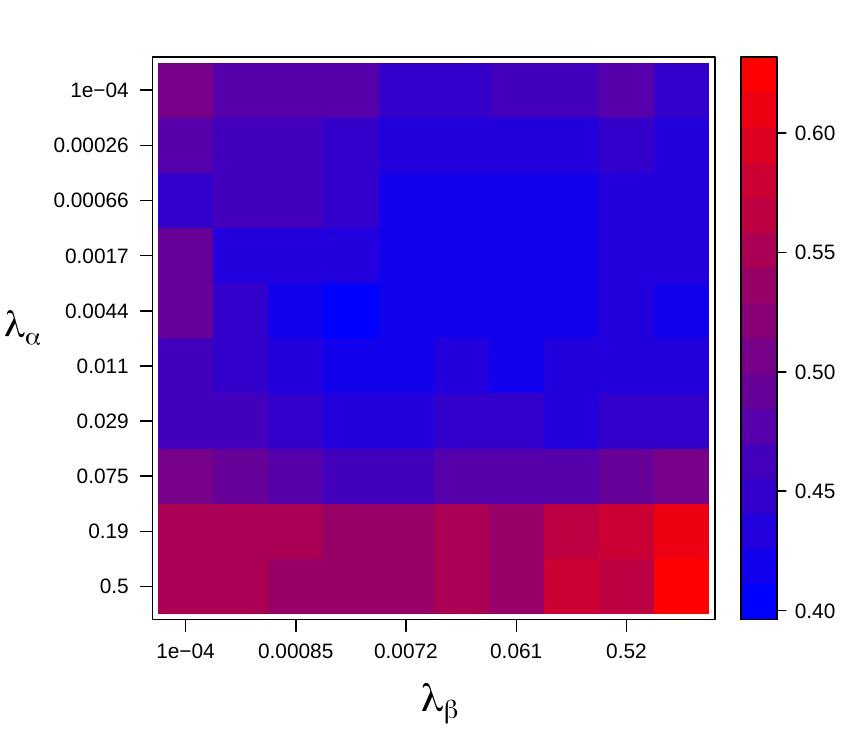}}
  \caption{A $10 \times 10$ cross-validation (CV) score matrix from the
    1-dimensional data analysis in
    Figure~7
    visualized as a
    2-dimensional heatmap. Blue shows a low average out-of-sample negative log
    likelihood across the five CV folds, and red shows high. In this case, the
    couplet at row 5 and column 4, $\lambda_\alpha=0.0044$ and
    $\lambda_\beta=0.0025$ was chosen. (To be clear, smaller CV score (blue)
    means a better model according to averaged out-of-sample prediction
    performance, measured by the negative log-likelihood.)}
  \label{fig:1d-cvscoremat}
\end{figure}

\begin{figure}[ht!]
  \centering
  \makebox[\textwidth]{\includegraphics[width=1.2\linewidth]{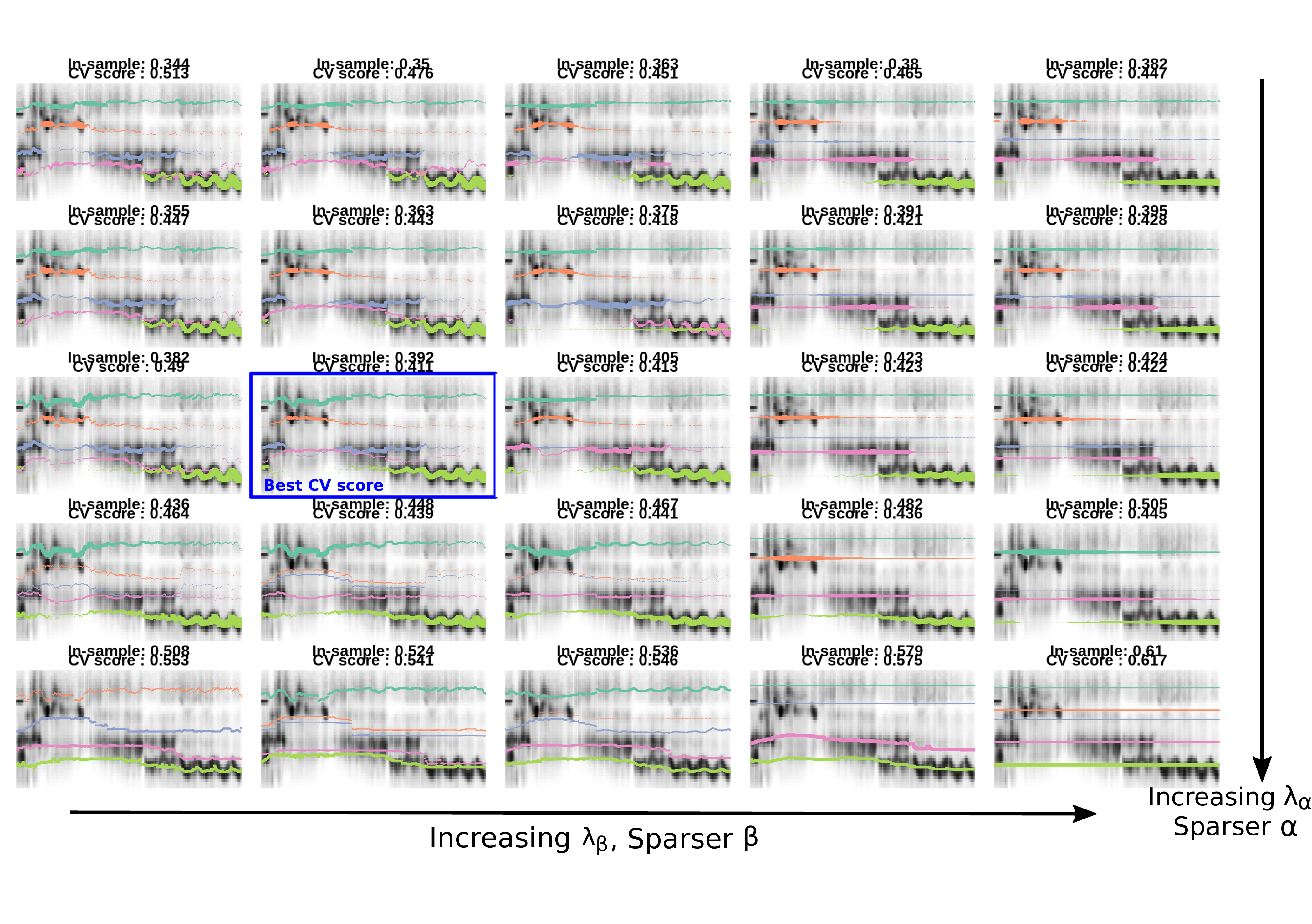}} 
  \caption{\it This figure shows $25 = 5 \times 5$ models each from different
    pairs of $(\lambda_\alpha, \lambda_\beta)$ values, from the 1-dimensional
    data analysis in
    Figure~7.
    The top row shows the models with the
    lowest $\lambda_\alpha$ in each column, and the left-most column in each row
    shows the model with the lowest $\lambda_\beta$ value in each row. The
    top-left figure shows the most complex, least regularized model with all
    non-zero coefficients -- overfitting to the data -- and the bottom right
    shows the simplest, most regularized model with every clusters' mean and
    probability constant over time. The blue box highlights the final estimated
    model with best cross-validated out-of-sample likelihood, as shown in
    Figure~7.
    The plot titles show two measures -- in-sample
    objective value, and the average out-of-sample negative log-likelihood
    across 5 cross-validation test folds; these show that the complex models
    perform well in-sample, but the cross-validation score suggests the best
    model is in the middle. The actual cross validation for the analysis was
    done using a $10 \times 10$ 2d grid of $(\lambda_\alpha, \lambda_\beta)$
    values, but this plot only shows a subset of the rows and columns, for
    illustration purposes. Figure~\ref{fig:1d-cvscoremat} shows the full
    $10 \times 10$ cross-validation score matrix, visualized as a heatmap.}
  \label{fig:1d-allmodels}
\end{figure}

\begin{figure}
  \centering
  \makebox[\linewidth]{\includegraphics[width=1.2\linewidth]{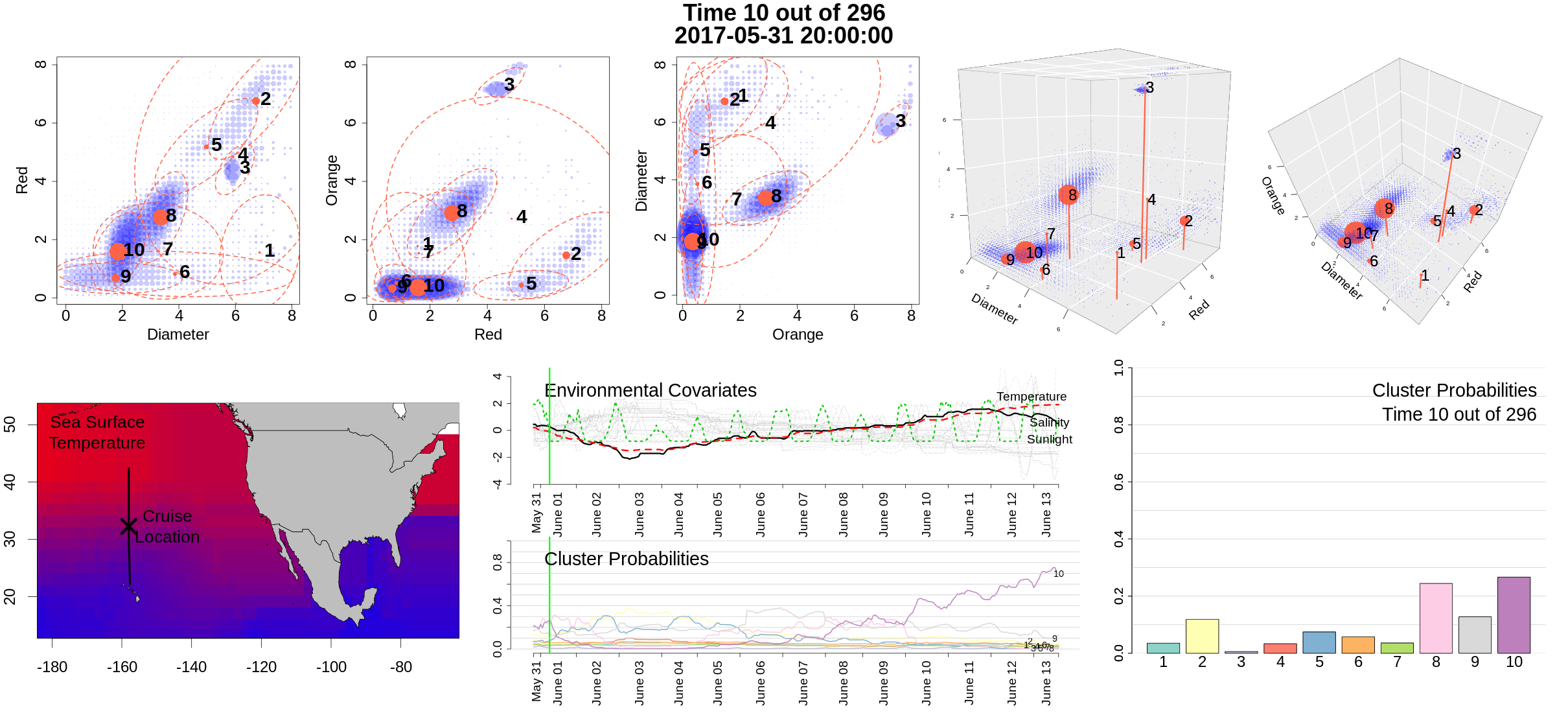}}
  \caption{\it A frame of a video (\url{https://youtu.be/jSxgVvT2wr4}) showing
    the estimated 3-dimensional 10-cluster model described in
    Section~4.0.2
    and
    Figure~9.
    The size of the blue points represents the
    biomass in each of the $40^3$ bins. The top panel shows various views of the
    cytograms and our estimated parameters (means, probabilities, and
    covariances). The lower panel shows the cruise location on a map, covariates
    over time, and finally cluster probabilities at each time and as a time
    series. The $10$ estimated model clusters' mean fluctuations and cluster
    probability dynamics over time can be seen in the full video.}
  \label{fig:3d-video}
\end{figure}

This section contains additional figures and tables for the data analysis from
Section~4.
Here is a summary of the material:
\begin{enumerate}
\item Figure~\ref{fig:environmental-covariates} and
  Table~\ref{tab:environmental-covariates-labels} show plots of all
  environmental covariates, and longer names of the covariates.
\item Figure~\ref{fig:1d-cvscoremat} and Figure~\ref{fig:1d-allmodels} show more
  detailed results from the 1d data application in
  Section~4.0.1.
\item Figure~\ref{fig:1d-cvscoremat} shows a $10$ by $10$ heatmap of the
  cross-validation scores from the 2d grid of candidate
  $(\lambda_\alpha, \lambda_\beta)$ values, and Figure~\ref{fig:1d-allmodels}
  shows the estimated models.
\item The estimated coefficients from the 1-dimensional analysis in
  Section~4.0.1
  are shown as tables in
  Table~\ref{tab:1d-coef-table}. Additionally, stability estimates are shown in
  \ref{tab:stability-beta-1} and \ref{tab:stability-beta-2}, whose details are
  described in Supplement D.
\item The estimated coefficients from the 3-dimensional analysis in
  Section~4.0.2
  are shown in Table~\ref{tab:3d-alpha} through
  Table~\ref{tab:3d-beta-3}.
\item A frame of a video showing a complete visualization of the 3-dimensional
  model in
  Section~4.0.2
  is shown in
  Figure~\ref{fig:3d-video}.
\end{enumerate}

\begin{table}[ht!]
\centering
\begin{tabular}{p{3.5cm}l}
\toprule
 Covariate Name & Long Name\\ 
\toprule
 \texttt{sss} & Sea surface salinity \\ 
\hline
  \texttt{sst} & Sea surface temperature \\ 
\hline
  \texttt{Fe} & Mole concentration of dissolved iron in sea water \\ 
\hline
  \texttt{PP} & Net primary productivity of Carbon per unit volume \\ 
\hline
  \texttt{Si} & Mole concentration of Silicate in sea water \\ 
\hline
  \texttt{NO3} & Mole concentration of Nitrate in sea water \\ 
\hline
  \texttt{CHL} & Mass concentration of Chlorophyll in sea water \\ 
\hline
  \texttt{PHYC} & Mole concentration of Phytoplankton expressed as carbon in sea water \\ 
\hline
  \texttt{PO4} & Mole concentration of Phosphate in sea water \\ 
\hline
  \texttt{O2} & Mole Concentration of dissolved Oxygen in sea water \\ 
\hline
  \texttt{vgosa} & Geostrophic velocity anomalies: meridian component \\ 
\hline
  \texttt{vgos} & Absolute geostrophic velocity: meridian component \\ 
\hline
  \texttt{sla} & Sea level anomaly \\ 
\hline
  \texttt{ugosa} & Geostrophic velocity anomalies: zonal component \\ 
\hline
  \texttt{ugos} & Absolute geostrophic velocity: zonal component \\ 
\hline
  \texttt{wind\_stress} & Wind stress \\ 
\hline
  \texttt{eastward\_wind} & Eastward wind speed \\ 
\hline
  \texttt{\parbox[][22pt][c]{3cm}{\linespread{0.5} surface\_downward \_eastward\_stress}} & Eastward wind stress \\ 
\hline
  \texttt{wind\_speed} & Wind speed \\ 
\hline
  \texttt{\parbox[][22pt][c]{3cm}{\linespread{0.5} surface\_downward \_northward\_stress}} & Northward wind stress \\ 
\hline
  \texttt{northward\_wind} & Northward wind speed \\ 
\hline
  \texttt{ftle\_bw\_sla} & FTLE backward-in-time using geostrophic velocity anomaly \\ 
\hline
  \texttt{disp\_bw\_sla} & Displacement backward-in-time using geostrophic velocity anomaly \\ 
\hline
  \texttt{AOU\_WOA\_clim} & Objectively analyzed climatology for apparent oxygen utilization\\ 
\hline
  \texttt{density\_woa\_clim} & Objectively analyzed climatology for density\\ 
\hline
  \texttt{o2sat\_woa\_clim} & Objectively analyzed climatology for percent oxygen saturation \\ 
\hline
  \texttt{oxygen\_woa\_clim} & Objectively analyzed climatology for dissolved oxygen\\ 
\hline
  \texttt{salinity\_woa\_clim} & Objectively analyzed climatology for salinity\\ 
\hline
  \texttt{conductivity\_woa\_clim} & Objectively analyzed climatology for conductivity\\ 
\hline
  \texttt{nitrate\_woa\_clim} & Objectively analyzed climatology for nitrate\\ 
\hline
  \texttt{phosphate\_woa\_clim} & Objectively analyzed climatology for phosphate\\ 
\hline
  \texttt{silicate\_woa\_clim} & Objectively analyzed climatology for silicate\\        
\hline
  \texttt{par} & Photosynthetically active radiation \\ 
\hline
  \texttt{p1, p2, p3, p4} & 3-, 6-, 9-, and 12- hour lagged {\tt par}.\\
\hline
  \texttt{b1, b2} & Indicator variables for ecological regions.\\
\hline
\end{tabular}
\caption{\it The environmental covariates used in our 1d and 3d analysis in
  Section~4
  were retrieved from the Simon's CMAP database
  using a process called ``colocalization'', which is to take average of moving
  time/space boxes. The covariate names, except for those in the last two rows,
  can be used to query data from the Simons CMAP database. For ease of
  presentation, in Tables~\ref{tab:1d-coef-table} through \ref{tab:3d-beta-3}, the
  short hand of \texttt{sdns} and \texttt{sdes} is used for
  \texttt{surface\_downward\_northward\_stress} and
  \texttt{surface\_downward\_northward\_stress}, and the suffixes
  \texttt{\_WOA\_clim} are omitted.}
\label{tab:environmental-covariates-labels}
\end{table}

\begin{table}[ht!]

\centering
\begin{tabular}{rlrrrrr}
  \toprule
Covariate name & type & Cluster 1 & Cluster 2 & Cluster 3 & Cluster 4 & Cluster 5 \\ 
  \toprule
  \texttt{p2} & light & $\cdot$ & -0.231 & -0.005 & $\cdot$ & 0.006 \\ 
  \texttt{p3} & light & $\cdot$ & -0.007 & $\cdot$ & $\cdot$ & $\cdot$ \\ 
  \texttt{par} & light & 0.018 & $\cdot$ & $\cdot$ & $\cdot$ & -0.027 \\ 
  \texttt{sst} & phys & $\cdot$ & $\cdot$ & $\cdot$ & $\cdot$ & 0.913 \\ 
  \texttt{vgosa} & phys & $\cdot$ & $\cdot$ & -0.116 & $\cdot$ & $\cdot$ \\ 
  \texttt{vgos} & phys & 0.133 & $\cdot$ & $\cdot$ & $\cdot$ & $\cdot$ \\ 
  \texttt{sla} & phys & $\cdot$ & 0.185 & $\cdot$ & $\cdot$ & $\cdot$ \\ 
  \texttt{ugos} & phys & $\cdot$ & -0.134 & 0.022 & $\cdot$ & -0.086 \\ 
  \texttt{wind\_stress} & phys & $\cdot$ & $\cdot$ & $\cdot$ & 0.011 & $\cdot$ \\ 
  \texttt{wind\_speed} & phys & $\cdot$ & $\cdot$ & $\cdot$ & $\cdot$ & -0.165 \\ 
  \texttt{northward\_wind} & phys & 0.138 & $\cdot$ & $\cdot$ & $\cdot$ & -0.063 \\ 
  \texttt{ftle\_bw\_sla} & phys & 0.032 & $\cdot$ & -0.094 & $\cdot$ & $\cdot$ \\ 
  \texttt{disp\_bw\_sla} & phys & $\cdot$ & $\cdot$ & $\cdot$ & $\cdot$ & 0.018 \\ 
  \texttt{PP} & bio & $\cdot$ & $\cdot$ & 0.088 & $\cdot$ & $\cdot$ \\ 
  \texttt{CHL} & bio & $\cdot$ & 0.134 & $\cdot$ & $\cdot$ & -0.045 \\ 
  \texttt{o2sat} & bio & $\cdot$ & $\cdot$ & 0.280 & $\cdot$ & $\cdot$ \\ 
  \texttt{nitrate} & chem & $\cdot$ & $\cdot$ & -0.636 & $\cdot$ & $\cdot$ \\ 
  \texttt{phosphate} & chem & $\cdot$ & 0.158 & $\cdot$ & $\cdot$ & -0.860 \\ 
  \texttt{silicate} & chem & -0.015 & $\cdot$ & $\cdot$ & $\cdot$ & $\cdot$ \\ 
   \hline
\end{tabular}

\bigskip
\begin{tabular}{rlrrrrr}
  \toprule
 Covariate name & type & Cluster 1 & Cluster 2 & Cluster 3 & Cluster 4 & Cluster 5 \\ 
  \toprule
  \texttt{p1} & light & 0.002 & 0.004 & 0.000 & -0.002 & 0.008 \\ 
  \texttt{p2} & light & $\cdot$ & -0.004 & 0.006 & -0.006 & 0.013 \\ 
  \texttt{p3} & light & 0.007 & $\cdot$ & 0.005 & -0.004 & 0.010 \\ 
  \texttt{p4} & light & -0.004 & 0.010 & 0.002 & -0.002 & -0.006 \\ 
  \texttt{par} & light & 0.002 & $\cdot$ & 0.004 & -0.002 & 0.002 \\ 
  \texttt{sst} & phys & $\cdot$ & -0.028 & -0.004 & $\cdot$ & -0.070 \\ 
  \texttt{vgosa} & phys & $\cdot$ & 0.014 & $\cdot$ & 0.025 & $\cdot$ \\ 
  \texttt{vgos} & phys & 0.012 & $\cdot$ & $\cdot$ & $\cdot$ & -0.006 \\ 
  \texttt{sla} & phys & 0.004 & -0.012 & 0.001 & 0.002 & -0.022 \\ 
  \texttt{ugosa} & phys & $\cdot$ & $\cdot$ & -0.005 & 0.007 & -0.050 \\ 
  \texttt{ugos} & phys & 0.005 & 0.007 & $\cdot$ & $\cdot$ & 0.059 \\ 
  \texttt{wind\_stress} & phys & 0.011 & $\cdot$ & -0.029 & $\cdot$ & 0.025 \\ 
  \texttt{eastward\_wind} & phys & -0.023 & $\cdot$ & -0.020 & $\cdot$ & 0.021 \\ 
  \texttt{sdes} & phys & $\cdot$ & $\cdot$ & -0.011 & $\cdot$ & $\cdot$ \\ 
  \texttt{wind\_speed} & phys & $\cdot$ & $\cdot$ & $\cdot$ & 0.033 & $\cdot$ \\ 
  \texttt{sdns} & phys & $\cdot$ & -0.002 & $\cdot$ & -0.007 & 0.011 \\ 
  \texttt{northward\_wind} & phys & $\cdot$ & -0.009 & -0.008 & $\cdot$ & $\cdot$ \\ 
  \texttt{ftle\_bw\_sla} & phys & 0.005 & $\cdot$ & 0.025 & 0.000 & -0.011 \\ 
  \texttt{disp\_bw\_sla} & phys & 0.001 & $\cdot$ & 0.004 & $\cdot$ & -0.003 \\ 
  \texttt{density} & phys & $\cdot$ & 0.026 & $\cdot$ & 0.051 & -0.038 \\ 
  \texttt{PP} & bio & 0.066 & $\cdot$ & -0.008 & $\cdot$ & $\cdot$ \\ 
  \texttt{CHL} & bio & -0.141 & $\cdot$ & $\cdot$ & $\cdot$ & $\cdot$ \\ 
  \texttt{o2sat} & bio & 0.010 & -0.004 & 0.017 & $\cdot$ & -0.022 \\ 
  \texttt{Si} & chem & 0.014 & -0.021 & $\cdot$ & $\cdot$ & -0.005 \\ 
  \texttt{phosphate} & chem & $\cdot$ & 0.052 & $\cdot$ & 0.012 & -0.132 \\ 
  \texttt{silicate} & chem & -0.016 & -0.022 & $\cdot$ & -0.045 & -0.001 \\ 
   \hline
\end{tabular}

\caption{\it Estimated $\alpha$ coefficients (top) $\beta$ coefficients (bottom)
  for the 5-cluster, 1-dimensional model in
  Figure~7.
  Some of the
  names of the covariates are abbreviated from the full versions in
  Table~\ref{tab:environmental-covariates-labels}. The rows (covariates) whose
  coefficients were all estimated to be zero were omitted. Additionally,
  stability estimates for the $\beta$ coefficients are shown in
  Tables~\ref{tab:stability-beta-1} and \ref{tab:stability-beta-2}.}
\label{tab:1d-coef-table}
\end{table}

\begin{table}[ht]
\end{table}

\begin{sidewaystable}[ht]
\centering
\begin{tabular}{rl|rrrrrrrrrr}
  \toprule
Covariate  & type & Cluster 1 & Cluster 2 & Cluster 3 & Cluster 4 & Cluster 5 & Cluster 6 & Cluster 7 & Cluster 8 & Cluster 9 & Cluster 10 \\ 
  \hline
  \texttt{p1} & {light} & $\cdot$ & $\cdot$ & 0.086 & $\cdot$ & $\cdot$ & $\cdot$ & $\cdot$ & $\cdot$ & 0.012 & $\cdot$ \\ 
  \texttt{p2} & {light} & $\cdot$ & $\cdot$ & $\cdot$ & $\cdot$ & -0.130 & $\cdot$ & $\cdot$ & $\cdot$ & $\cdot$ & 0.058 \\ 
  \texttt{p3} & {light} & $\cdot$ & $\cdot$ & $\cdot$ & $\cdot$ & $\cdot$ & $\cdot$ & $\cdot$ & 0.047 & -0.038 & 0.018 \\ 
  \texttt{p4} & {light} & $\cdot$ & $\cdot$ & $\cdot$ & $\cdot$ & $\cdot$ & $\cdot$ & $\cdot$ & $\cdot$ & -0.023 & $\cdot$ \\ 
  \texttt{par} & {light} & $\cdot$ & $\cdot$ & 0.431 & $\cdot$ & $\cdot$ & $\cdot$ & $\cdot$ & $\cdot$ & 0.013 & -0.079 \\ 
  \texttt{sst} & {phys} & $\cdot$ & $\cdot$ & $\cdot$ & $\cdot$ & $\cdot$ & $\cdot$ & $\cdot$ & $\cdot$ & $\cdot$ & 0.868 \\ 
  \texttt{vgosa} & {phys} & $\cdot$ & $\cdot$ & $\cdot$ & $\cdot$ & $\cdot$ & $\cdot$ & $\cdot$ & $\cdot$ & -0.061 & $\cdot$ \\ 
  \texttt{vgos} & {phys} & $\cdot$ & 0.079 & -0.002 & $\cdot$ & $\cdot$ & $\cdot$ & $\cdot$ & -0.088 & $\cdot$ & $\cdot$ \\ 
  \texttt{sla} & {phys} & $\cdot$ & $\cdot$ & $\cdot$ & 0.025 & 0.201 & $\cdot$ & $\cdot$ & -0.056 & $\cdot$ & $\cdot$ \\ 
  \texttt{ugosa} & {phys} & $\cdot$ & 0.025 & $\cdot$ & $\cdot$ & $\cdot$ & $\cdot$ & $\cdot$ & $\cdot$ & $\cdot$ & -0.034 \\ 
  \texttt{ugos} & {phys} & $\cdot$ & $\cdot$ & $\cdot$ & $\cdot$ & -0.188 & $\cdot$ & $\cdot$ & 0.175 & $\cdot$ & -0.060 \\ 
  \texttt{wind\_stress} & {phys} & 0.160 & $\cdot$ & -0.436 & 0.036 & -0.085 & $\cdot$ & $\cdot$ & -0.371 & 0.108 & $\cdot$ \\ 
  \texttt{sdes} & {phys} & $\cdot$ & $\cdot$ & $\cdot$ & $\cdot$ & $\cdot$ & $\cdot$ & $\cdot$ & $\cdot$ & 0.140 & $\cdot$ \\ 
  \texttt{wind\_speed} & {phys} & $\cdot$ & $\cdot$ & $\cdot$ & $\cdot$ & $\cdot$ & $\cdot$ & $\cdot$ & $\cdot$ & $\cdot$ & -0.255 \\ 
  \texttt{northward\_wind} & {phys} & 0.019 & $\cdot$ & $\cdot$ & $\cdot$ & $\cdot$ & $\cdot$ & $\cdot$ & -0.181 & $\cdot$ & -0.105 \\ 
  \texttt{ftle\_bw\_sla} & {phys} & $\cdot$ & $\cdot$ & $\cdot$ & $\cdot$ & $\cdot$ & $\cdot$ & $\cdot$ & -0.180 & $\cdot$ & $\cdot$ \\ 
  \texttt{disp\_bw\_sla} & {phys} & $\cdot$ & $\cdot$ & $\cdot$ & $\cdot$ & -0.161 & $\cdot$ & $\cdot$ & $\cdot$ & $\cdot$ & 0.011 \\ 
  \texttt{density} & {phys} & $\cdot$ & 0.111 & $\cdot$ & $\cdot$ & 0.429 & $\cdot$ & $\cdot$ & $\cdot$ & $\cdot$ & $\cdot$ \\ 
  \texttt{PP} & {bio} & $\cdot$ & $\cdot$ & $\cdot$ & $\cdot$ & $\cdot$ & $\cdot$ & $\cdot$ & 0.195 & $\cdot$ & $\cdot$ \\ 
  \texttt{CHL} & {bio} & $\cdot$ & $\cdot$ & $\cdot$ & $\cdot$ & 0.214 & $\cdot$ & $\cdot$ & $\cdot$ & $\cdot$ & -0.061 \\ 
  \texttt{o2sat} & {bio} & -0.143 & $\cdot$ & $\cdot$ & $\cdot$ & $\cdot$ & -0.049 & $\cdot$ & 0.455 & 0.003 & $\cdot$ \\ 
  \texttt{AOU} & {chem} & $\cdot$ & $\cdot$ & $\cdot$ & $\cdot$ & -0.020 & 0.032 & $\cdot$ & $\cdot$ & $\cdot$ & $\cdot$ \\ 
  \texttt{nitrate} & {chem} & 0.015 & $\cdot$ & $\cdot$ & 0.298 & $\cdot$ & $\cdot$ & $\cdot$ & -0.347 & -0.005 & $\cdot$ \\ 
  \texttt{phosphate} & {chem} & $\cdot$ & 0.348 & $\cdot$ & $\cdot$ & $\cdot$ & $\cdot$ & $\cdot$ & $\cdot$ & $\cdot$ & -0.937 \\ 
   \hline
\end{tabular}
\caption{\it Estimated $\alpha$ coefficients for 10-cluster, 3-dimensional model
  from
  Section~3.
  Rows whose coefficients are all zero have
  been omitted.}
\label{tab:3d-alpha}
\end{sidewaystable}

\begin{sidewaystable}[ht]
\centering
\begin{tabular}{rl|lll|lll|lll|lll}
  \toprule
   Covariate & Covariate & Clust 1 & Clust 1 & Clust 1 & Clust 2 & Clust 2 & Clust 2 & Clust 3 & Clust 3 & Clust 3 & Clust 4 & Clust 4 & Clust 4 \\ 
   name & type &  Diam &  Red &  Orange &  Diam &  Red &  Orange &  Diam &  Red &  Orange &  Diam &  Red &  Orange \\ 
  \toprule
  \texttt{p1} & light & $\cdot$ & $\cdot$ & $\cdot$ & 0.031 & $\cdot$ & $\cdot$ & $\cdot$ & 0.088 & $\cdot$ & -0.000 & $\cdot$ & $\cdot$ \\ 
  \texttt{p2} & light & $\cdot$ & $\cdot$ & $\cdot$ & 0.044 & -0.004 & -0.006 & 0.003 & $\cdot$ & $\cdot$ & $\cdot$ & $\cdot$ & $\cdot$ \\ 
  \texttt{p3} & light & $\cdot$ & $\cdot$ & $\cdot$ & 0.042 & $\cdot$ & -0.001 & 0.002 & 0.005 & -0.005 & $\cdot$ & $\cdot$ & $\cdot$ \\ 
  \texttt{p4} & light & $\cdot$ & $\cdot$ & $\cdot$ & 0.009 & $\cdot$ & -0.011 & $\cdot$ & -0.012 & 0.023 & $\cdot$ & $\cdot$ & $\cdot$ \\ 
  \texttt{par} & light & $\cdot$ & -0.015 & -0.007 & 0.040 & -0.015 & 0.056 & $\cdot$ & $\cdot$ & $\cdot$ & $\cdot$ & $\cdot$ & $\cdot$ \\ 
  \texttt{sst} & phys & 0.011 & $\cdot$ & $\cdot$ & $\cdot$ & $\cdot$ & $\cdot$ & $\cdot$ & $\cdot$ & $\cdot$ & $\cdot$ & $\cdot$ & $\cdot$ \\ 
  \texttt{vgosa} & phys & $\cdot$ & $\cdot$ & $\cdot$ & 0.004 & $\cdot$ & $\cdot$ & $\cdot$ & $\cdot$ & $\cdot$ & $\cdot$ & $\cdot$ & $\cdot$ \\ 
  \texttt{vgos} & phys & $\cdot$ & $\cdot$ & $\cdot$ & 0.060 & $\cdot$ & $\cdot$ & $\cdot$ & $\cdot$ & $\cdot$ & $\cdot$ & $\cdot$ & $\cdot$ \\ 
  \texttt{sla} & phys & $\cdot$ & $\cdot$ & $\cdot$ & $\cdot$ & -0.040 & $\cdot$ & $\cdot$ & -0.002 & 0.026 & $\cdot$ & $\cdot$ & 0.048 \\ 
  \texttt{ugosa} & phys & $\cdot$ & $\cdot$ & $\cdot$ & $\cdot$ & $\cdot$ & $\cdot$ & $\cdot$ & -0.044 & $\cdot$ & $\cdot$ & $\cdot$ & $\cdot$ \\ 
  \texttt{ugos} & phys & $\cdot$ & $\cdot$ & $\cdot$ & 0.014 & $\cdot$ & $\cdot$ & $\cdot$ & $\cdot$ & $\cdot$ & $\cdot$ & $\cdot$ & $\cdot$ \\ 
  \texttt{wind\_stress} & phys & $\cdot$ & $\cdot$ & $\cdot$ & $\cdot$ & 0.002 & 0.004 & $\cdot$ & $\cdot$ & $\cdot$ & $\cdot$ & $\cdot$ & $\cdot$ \\ 
  \texttt{eastward\_wind} & phys & -0.010 & $\cdot$ & $\cdot$ & $\cdot$ & $\cdot$ & 0.068 & $\cdot$ & $\cdot$ & $\cdot$ & $\cdot$ & $\cdot$ & $\cdot$ \\ 
  \texttt{sdes} & phys & $\cdot$ & $\cdot$ & $\cdot$ & $\cdot$ & 0.014 & $\cdot$ & $\cdot$ & $\cdot$ & $\cdot$ & $\cdot$ & $\cdot$ & $\cdot$ \\ 
  \texttt{wind\_speed} & phys & $\cdot$ & $\cdot$ & $\cdot$ & -0.022 & 0.024 & $\cdot$ & $\cdot$ & $\cdot$ & $\cdot$ & $\cdot$ & $\cdot$ & $\cdot$ \\ 
  \texttt{sdns} & phys & $\cdot$ & $\cdot$ & $\cdot$ & $\cdot$ & $\cdot$ & $\cdot$ & $\cdot$ & $\cdot$ & $\cdot$ & $\cdot$ & $\cdot$ & $\cdot$ \\ 
  \texttt{northward\_wind} & phys & $\cdot$ & $\cdot$ & $\cdot$ & 0.031 & -0.019 & $\cdot$ & $\cdot$ & $\cdot$ & $\cdot$ & $\cdot$ & $\cdot$ & $\cdot$ \\ 
  \texttt{ftle\_bw\_sla} & phys & $\cdot$ & $\cdot$ & $\cdot$ & 0.004 & $\cdot$ & $\cdot$ & $\cdot$ & $\cdot$ & $\cdot$ & $\cdot$ & $\cdot$ & $\cdot$ \\ 
  \texttt{disp\_bw\_sla} & phys & 0.003 & -0.015 & $\cdot$ & 0.007 & $\cdot$ & $\cdot$ & $\cdot$ & $\cdot$ & $\cdot$ & $\cdot$ & $\cdot$ & $\cdot$ \\ 
  \texttt{density} & phys & $\cdot$ & $\cdot$ & $\cdot$ & $\cdot$ & $\cdot$ & $\cdot$ & $\cdot$ & $\cdot$ & $\cdot$ & $\cdot$ & $\cdot$ & $\cdot$ \\ 
  \texttt{PP} & bio & $\cdot$ & $\cdot$ & 0.078 & 0.270 & -0.053 & 0.023 & $\cdot$ & $\cdot$ & $\cdot$ & $\cdot$ & $\cdot$ & $\cdot$ \\ 
  \texttt{CHL} & bio & $\cdot$ & $\cdot$ & $\cdot$ & -0.393 & $\cdot$ & $\cdot$ & $\cdot$ & $\cdot$ & $\cdot$ & $\cdot$ & $\cdot$ & $\cdot$ \\ 
  \texttt{o2sat} & bio & -0.047 & 0.015 & $\cdot$ & 0.118 & 0.137 & $\cdot$ & $\cdot$ & $\cdot$ & $\cdot$ & -0.005 & $\cdot$ & $\cdot$ \\ 
  \texttt{Si} & chem & $\cdot$ & $\cdot$ & $\cdot$ & $\cdot$ & 0.030 & $\cdot$ & $\cdot$ & $\cdot$ & $\cdot$ & $\cdot$ & $\cdot$ & $\cdot$ \\ 
  \texttt{AOU} & chem & $\cdot$ & $\cdot$ & $\cdot$ & $\cdot$ & $\cdot$ & $\cdot$ & $\cdot$ & $\cdot$ & $\cdot$ & $\cdot$ & $\cdot$ & $\cdot$ \\ 
  \texttt{nitrate} & chem & $\cdot$ & $\cdot$ & $\cdot$ & -0.009 & $\cdot$ & -0.051 & $\cdot$ & $\cdot$ & $\cdot$ & $\cdot$ & $\cdot$ & $\cdot$ \\ 
  \texttt{phosphate} & chem & $\cdot$ & $\cdot$ & $\cdot$ & $\cdot$ & $\cdot$ & $\cdot$ & $\cdot$ & $\cdot$ & $\cdot$ & $\cdot$ & $\cdot$ & $\cdot$ \\ 
  \texttt{silicate} & chem & $\cdot$ & 0.023 & $\cdot$ & -0.032 & 0.009 & -0.018 & $\cdot$ & $\cdot$ & $\cdot$ & $\cdot$ & $\cdot$ & -0.016 \\ 
   \hline
\end{tabular}
\caption{Estimated $\beta$ coefficients from the 10-cluster, 3-dimensional model
  in
  Section~3.
  (part 1 of 3). The column names Diam, Red and
  Orange refer to the names of the 3-dimensional cytogram axes.}
\label{tab:3d-beta-1}
\end{sidewaystable}
\begin{sidewaystable}[ht]
\centering
\begin{tabular}{rl|lll|lll|lll|lll}
  \toprule
   Covariate& Covariate & Clust 5 & Clust 5 & Clust 5 & Clust 6 & Clust 6 & Clust 6 & Clust 7 & Clust 7 & Clust 7 & Clust 8 & Clust 8 & Clust 8 \\ 
   name & type &  Diam &  Red &  Orange &  Diam &  Red &  Orange &  Diam &  Red &  Orange &  Diam &  Red &  Orange \\ 
  \toprule
  \texttt{p1} & light & $\cdot$ & -0.004 & $\cdot$ & $\cdot$ & $\cdot$ & -0.005 & $\cdot$ & $\cdot$ & $\cdot$ & 0.029 & -0.016 & $\cdot$ \\ 
  \texttt{p2} & light & 0.042 & -0.002 & -0.002 & $\cdot$ & $\cdot$ & -0.000 & -0.028 & $\cdot$ & $\cdot$ & 0.034 & -0.005 & 0.018 \\ 
  \texttt{p3} & light & 0.036 & 0.035 & -0.006 & $\cdot$ & $\cdot$ & -0.005 & $\cdot$ & $\cdot$ & $\cdot$ & 0.040 & 0.022 & $\cdot$ \\ 
  \texttt{p4} & light & $\cdot$ & 0.065 & -0.003 & $\cdot$ & $\cdot$ & -0.005 & $\cdot$ & $\cdot$ & $\cdot$ & $\cdot$ & 0.018 & 0.025 \\ 
  \texttt{par} & light & 0.035 & -0.016 & 0.001 & $\cdot$ & $\cdot$ & -0.002 & $\cdot$ & $\cdot$ & $\cdot$ & 0.020 & $\cdot$ & 0.039 \\ 
  \texttt{sst} & phys & $\cdot$ & $\cdot$ & $\cdot$ & $\cdot$ & $\cdot$ & $\cdot$ & $\cdot$ & $\cdot$ & $\cdot$ & $\cdot$ & $\cdot$ & -0.001 \\ 
  \texttt{vgosa} & phys & 0.094 & $\cdot$ & 0.011 & $\cdot$ & $\cdot$ & $\cdot$ & 0.071 & $\cdot$ & $\cdot$ & $\cdot$ & -0.020 & $\cdot$ \\ 
  \texttt{vgos} & phys & $\cdot$ & $\cdot$ & $\cdot$ & $\cdot$ & $\cdot$ & $\cdot$ & $\cdot$ & $\cdot$ & $\cdot$ & 0.044 & $\cdot$ & $\cdot$ \\ 
  \texttt{sla} & phys & $\cdot$ & $\cdot$ & $\cdot$ & $\cdot$ & $\cdot$ & $\cdot$ & 0.002 & $\cdot$ & $\cdot$ & 0.049 & -0.011 & 0.014 \\ 
  \texttt{ugosa} & phys & 0.040 & $\cdot$ & 0.005 & $\cdot$ & -0.006 & $\cdot$ & $\cdot$ & $\cdot$ & $\cdot$ & $\cdot$ & $\cdot$ & $\cdot$ \\ 
  \texttt{ugos} & phys & $\cdot$ & 0.014 & $\cdot$ & $\cdot$ & $\cdot$ & $\cdot$ & $\cdot$ & $\cdot$ & $\cdot$ & -0.074 & $\cdot$ & 0.063 \\ 
  \texttt{wind\_stress} & phys & $\cdot$ & -0.062 & $\cdot$ & $\cdot$ & $\cdot$ & $\cdot$ & $\cdot$ & $\cdot$ & $\cdot$ & $\cdot$ & -0.096 & -0.056 \\ 
  \texttt{eastward\_wind} & phys & $\cdot$ & $\cdot$ & 0.018 & $\cdot$ & 0.044 & $\cdot$ & $\cdot$ & 0.044 & $\cdot$ & -0.016 & $\cdot$ & $\cdot$ \\ 
  \texttt{sdes} & phys & $\cdot$ & $\cdot$ & $\cdot$ & $\cdot$ & $\cdot$ & 0.016 & $\cdot$ & $\cdot$ & $\cdot$ & $\cdot$ & -0.051 & $\cdot$ \\ 
  \texttt{wind\_speed} & phys & -0.092 & $\cdot$ & 0.010 & 0.090 & $\cdot$ & $\cdot$ & $\cdot$ & $\cdot$ & $\cdot$ & -0.112 & 0.078 & 0.097 \\ 
  \texttt{sdns} & phys & 0.002 & -0.073 & $\cdot$ & $\cdot$ & $\cdot$ & $\cdot$ & $\cdot$ & $\cdot$ & $\cdot$ & 0.022 & $\cdot$ & $\cdot$ \\ 
  \texttt{northward\_wind} & phys & $\cdot$ & $\cdot$ & $\cdot$ & $\cdot$ & -0.001 & $\cdot$ & 0.025 & $\cdot$ & $\cdot$ & $\cdot$ & 0.015 & -0.002 \\ 
  \texttt{ftle\_bw\_sla} & phys & $\cdot$ & 0.056 & $\cdot$ & $\cdot$ & $\cdot$ & $\cdot$ & $\cdot$ & $\cdot$ & $\cdot$ & 0.099 & 0.075 & 0.006 \\ 
  \texttt{disp\_bw\_sla} & phys & $\cdot$ & $\cdot$ & $\cdot$ & $\cdot$ & $\cdot$ & $\cdot$ & $\cdot$ & $\cdot$ & $\cdot$ & 0.066 & $\cdot$ & -0.015 \\ 
  \texttt{density} & phys & 0.046 & $\cdot$ & $\cdot$ & $\cdot$ & $\cdot$ & $\cdot$ & $\cdot$ & $\cdot$ & $\cdot$ & $\cdot$ & $\cdot$ & 0.020 \\ 
  \texttt{PP} & bio & $\cdot$ & -0.396 & $\cdot$ & $\cdot$ & $\cdot$ & 0.007 & $\cdot$ & $\cdot$ & $\cdot$ & $\cdot$ & -0.114 & -0.099 \\ 
  \texttt{CHL} & bio & $\cdot$ & 0.215 & 0.039 & $\cdot$ & $\cdot$ & $\cdot$ & $\cdot$ & $\cdot$ & $\cdot$ & -0.137 & 0.185 & 0.081 \\ 
  \texttt{o2sat} & bio & -0.175 & $\cdot$ & -0.028 & $\cdot$ & $\cdot$ & $\cdot$ & $\cdot$ & $\cdot$ & $\cdot$ & 0.002 & 0.135 & $\cdot$ \\ 
  \texttt{Si} & chem & $\cdot$ & $\cdot$ & $\cdot$ & $\cdot$ & $\cdot$ & $\cdot$ & $\cdot$ & $\cdot$ & $\cdot$ & 0.266 & $\cdot$ & $\cdot$ \\ 
  \texttt{AOU} & chem & $\cdot$ & $\cdot$ & $\cdot$ & $\cdot$ & $\cdot$ & $\cdot$ & $\cdot$ & $\cdot$ & $\cdot$ & $\cdot$ & $\cdot$ & -0.168 \\ 
  \texttt{nitrate} & chem & $\cdot$ & 0.049 & $\cdot$ & $\cdot$ & $\cdot$ & 0.045 & 0.005 & $\cdot$ & $\cdot$ & $\cdot$ & -0.017 & $\cdot$ \\ 
  \texttt{phosphate} & chem & 0.128 & 0.096 & 0.028 & 0.013 & $\cdot$ & $\cdot$ & 0.197 & $\cdot$ & $\cdot$ & $\cdot$ & $\cdot$ & 0.042 \\ 
  \texttt{silicate} & chem & $\cdot$ & 0.095 & -0.002 & $\cdot$ & $\cdot$ & $\cdot$ & $\cdot$ & $\cdot$ & $\cdot$ & $\cdot$ & $\cdot$ & -0.101 \\ 
   \hline
\end{tabular}
\caption{Estimated $\beta$ coefficients from the 10-cluster, 3-dimensional model
  in
  Section~3.
  (part 2 of 3). The column names Diam, Red and
  Orange refer to the names of the 3-dimensional cytogram axes.}
\label{tab:3d-beta-2}
\end{sidewaystable}
\begin{sidewaystable}[ht]
\centering
\begin{tabular}{rl|lll|lll}
  \toprule
   Covariate& Covariate  & Clust 9 & Clust 9 & Clust 9 & Clust 10 & Clust 10 & Clust 10 \\ 
   name& type &  Diam &  Red &  Orange &  Diam &  Red &  Orange \\ 
  \toprule
  \texttt{p1} & light & -0.017 & 0.012 & -0.002 & 0.040 & -0.020 & 0.001 \\ 
  \texttt{p2} & light & -0.008 & 0.000 & -0.001 & 0.073 & 0.013 & $\cdot$ \\ 
  \texttt{p3} & light & -0.026 & $\cdot$ & -0.003 & 0.022 & -0.028 & -3.659e-05 \\ 
  \texttt{p4} & light & -0.006 & $\cdot$ & -0.000 & -0.003 & $\cdot$ & $\cdot$ \\ 
  \texttt{par} & light & $\cdot$ & 0.001 & -0.000 & 0.024 & -0.059 & $\cdot$ \\ 
  \texttt{sst} & phys & $\cdot$ & -0.036 & $\cdot$ & -0.315 & $\cdot$ & $\cdot$ \\ 
  \texttt{vgosa} & phys & 0.061 & $\cdot$ & $\cdot$ & $\cdot$ & -0.024 & $\cdot$ \\ 
  \texttt{vgos} & phys & $\cdot$ & 0.002 & 0.002 & -0.020 & $\cdot$ & -0.000 \\ 
  \texttt{sla} & phys & 0.074 & -0.002 & $\cdot$ & -0.028 & -0.014 & $\cdot$ \\ 
  \texttt{ugosa} & phys & $\cdot$ & $\cdot$ & 0.003 & -0.046 & -0.016 & 0.000 \\ 
  \texttt{ugos} & phys & 0.034 & $\cdot$ & $\cdot$ & 0.057 & $\cdot$ & $\cdot$ \\ 
  \texttt{wind\_stress} & phys & $\cdot$ & -0.032 & $\cdot$ & -0.018 & $\cdot$ & $\cdot$ \\ 
  \texttt{eastward\_wind} & phys & 0.114 & 0.010 & 0.011 & $\cdot$ & 0.048 & 0.012 \\ 
  \texttt{sdes} & phys & 0.139 & $\cdot$ & $\cdot$ & -0.029 & -0.083 & $\cdot$ \\ 
  \texttt{wind\_speed} & phys & 0.136 & 0.043 & 0.005 & $\cdot$ & 0.003 & 0.007 \\ 
  \texttt{sdns} & phys & 0.006 & $\cdot$ & 0.004 & $\cdot$ & $\cdot$ & $\cdot$ \\ 
  \texttt{northward\_wind} & phys & $\cdot$ & -0.012 & $\cdot$ & 0.001 & $\cdot$ & $\cdot$ \\ 
  \texttt{ftle\_bw\_sla} & phys & $\cdot$ & $\cdot$ & 0.002 & 0.003 & -0.021 & $\cdot$ \\ 
  \texttt{disp\_bw\_sla} & phys & -0.056 & $\cdot$ & $\cdot$ & -0.018 & -0.008 & -0.001 \\ 
  \texttt{density} & phys & 0.031 & $\cdot$ & $\cdot$ & -0.060 & 0.020 & $\cdot$ \\ 
  \texttt{PP} & bio & $\cdot$ & -0.016 & $\cdot$ & 0.077 & $\cdot$ & $\cdot$ \\ 
  \texttt{CHL} & bio & $\cdot$ & $\cdot$ & 0.001 & $\cdot$ & 0.098 & $\cdot$ \\ 
  \texttt{o2sat} & bio & 0.038 & 0.025 & -0.000 & -0.041 & $\cdot$ & $\cdot$ \\ 
  \texttt{Si} & chem & -0.010 & $\cdot$ & -0.002 & $\cdot$ & 0.014 & $\cdot$ \\ 
  \texttt{AOU} & chem & $\cdot$ & $\cdot$ & $\cdot$ & $\cdot$ & -0.029 & $\cdot$ \\ 
  \texttt{nitrate} & chem & $\cdot$ & $\cdot$ & $\cdot$ & -0.182 & -0.224 & $\cdot$ \\ 
  \texttt{phosphate} & chem & $\cdot$ & $\cdot$ & 0.019 & -0.086 & $\cdot$ & $\cdot$ \\ 
  \texttt{silicate} & chem & -0.121 & -0.003 & -0.005 & -0.040 & -0.004 & $\cdot$ \\ 
   \hline
\end{tabular}
\caption{Estimated $\beta$ coefficients from the 10-cluster, 3-dimensional model
  in
  Section~3
  (part 3 of 3). The column names Diam, Red and
  Orange refer to the names of the 3-dimensional cytogram axes.}
\label{tab:3d-beta-3}
\end{sidewaystable}

 \section*{Supplement D: Variable Selection Stability}

In order to quantify uncertainty of our model estimates, we use the subsampling
bootstrap \citep{politis-romano-wolf-1999} to estimate stability of $\beta$
coefficients of our model estimated on the 1-dimensional data in
Figure~7.
This closely follows the idea of
\textit{stability selection} in \cite{meinshausen2010stability} originally
designed for regression. To measure stability, we calculate the nonzero
proportion of each of the estimated coefficients ($\beta$), from $100$
subsampled datasets of size $150$ from the original dataset of size
$T=296$. These proportions are displayed alongside the original coefficient
estimates in Tables~\ref{tab:stability-beta-1} and
\ref{tab:stability-beta-2}. We describe the entire procedure next.

Recall we are denoting model estimates in the full data, after cross-validation,
as $\{\hat \bbeta_{k} \in \R^p\}_{k=1,\cdots,K}$. We are not concerned with
the intercept coefficient $\hat \bbeta_{k0}$, so we omit it from the procedure
hereon. We take the following steps.

\begin{enumerate}
\item For $b = 1,\cdots, n_{\text{boot}}$, repeat the following steps:
  \begin{enumerate}
  \item Randomly draw $B=150$ time indices out of $\{1,\cdots, 296\}$, and call
    this set $I_b$.
  \item Take a subsample of the data $\{\y^{(t)}, \X^{(t)}\}_{t \in I_b}$.
  \item On this data, estimate the coefficients $\{\hat \bbeta^{b,B}_{k}\}_{k=1}^K$, with
    cross-validation. (The $b$ and $B$ in the superscript emphasize that it is
    from the $b$'th subsampled dataset of size $B$.)
  \item Use the matching procedure in Algorithm~\ref{alg:match-clusters}, to
    find the optimal cluster permutations $\tilde \pi$ to most closely match the
    original clusters from the full data, i.e. the ones in
    $\{\hat \bbeta_{k} \}_{k=1}^{K}$.
  \item Reorder cluster assignments of the subsampled coefficient estimates
    $\{\hat \bbeta^{b,B}_{k}\}_{k=1}^K$ by assigning new cluster labels
    $\{\tilde \pi(k)\}_{k=1}^{K}$ to existing clusters $k=1,\cdots, K$.
  \end{enumerate}
\item From all models' coefficient estimates
  $\{\hat \bbeta_{k}^{b,B}\}_{b=1}^{n_{\text{boot}}}$, calculate stability
  estimates for each $k=1,\cdots,K$ as:
  $$\frac{1}{n_{\text{boot}}}\sum_{b=1}^{n_{\text{boot}}} \one \{ \be_j^T \hat \bbeta_k^{b,B} \neq 0 \} \text{ for } j = 1,\cdots, p.$$
\end{enumerate}
We use the subsampling bootstrap because the normal bootstrap uses repeated
sampling of time points, which complicates our cross-validation of time points
by often having data shared in both the training folds and test fold. The
subsampling bootstrap, which does not take repeated time points, eschews this
issue.

\begin{sidewaystable}[ht]
\centering
\begin{tabular}{rrr|rrr|rrr}
  \toprule
Cluster 1 & Cluster 1 & Cluster 1 & Cluster 2 & Cluster 2 & Cluster 2 & Cluster 3 & Cluster 3 & Cluster 3 \\ 
Variable & Estim. & P(nonzero) & Variable & Estim. & P(nonzero) & Variable & Estim. & P(nonzero) \\ 
  \toprule
  \texttt{b2} & 0.077 & 0.87 & \texttt{phosphate} & 0.052 & 0.89 & \texttt{ftle\_bw\_sla} & 0.024 & 0.68 \\ 
  \texttt{silicate} & -0.016 & 0.72 & \texttt{p4} & 0.010 & 0.40 & \texttt{wind\_stress} & -0.029 & 0.66 \\ 
  \texttt{ftle\_bw\_sla} & 0.005 & 0.34 & \texttt{p2} & -0.004 & 0.30 & \texttt{p2} & 0.006 & 0.61 \\ 
  \texttt{p4} & -0.004 & 0.32 & \texttt{density} & 0.026 & 0.27 & \texttt{sla} & 0.001 & 0.54 \\ 
  \texttt{PP} & 0.066 & 0.30 & \texttt{sdns} & -0.002 & 0.22 & \texttt{b2} & -0.023 & 0.50 \\ 
  \texttt{sla} & 0.004 & 0.28 & \texttt{CHL} & $\cdot$ & 0.20 & \texttt{p3} & 0.005 & 0.46 \\ 
  \texttt{p2} & $\cdot$ & 0.25 & \texttt{wind\_stress} & $\cdot$ & 0.17 & \texttt{northward\_wind} & -0.008 & 0.36 \\ 
  \texttt{ugos} & 0.005 & 0.24 & \texttt{Si} & -0.021 & 0.15 & \texttt{eastward\_wind} & -0.020 & 0.35 \\ 
  \texttt{par} & 0.002 & 0.24 & \texttt{p1} & 0.004 & 0.14 & \texttt{p1} & 0.000 & 0.32 \\ 
  \texttt{nitrate} & $\cdot$ & 0.22 & \texttt{p3} & $\cdot$ & 0.14 & \texttt{b1} & 0.032 & 0.30 \\ 
  \texttt{b1} & 0.014 & 0.21 & \texttt{vgos} & $\cdot$ & 0.14 & \texttt{ugosa} & -0.005 & 0.29 \\ 
  \texttt{o2sat} & 0.010 & 0.21 & \texttt{northward\_wind} & -0.009 & 0.14 & \texttt{p4} & 0.002 & 0.27 \\ 
  \texttt{eastward\_wind} & -0.023 & 0.20 & \texttt{b2} & -0.000 & 0.13 & \texttt{vgosa} & $\cdot$ & 0.23 \\ 
  \texttt{p1} & 0.002 & 0.18 & \texttt{ugos} & 0.007 & 0.13 & \texttt{par} & 0.004 & 0.23 \\ 
  \texttt{vgosa} & $\cdot$ & 0.15 & \texttt{silicate} & -0.022 & 0.13 & \texttt{vgos} & $\cdot$ & 0.20 \\ 
  \texttt{vgos} & 0.012 & 0.14 & \texttt{par} & $\cdot$ & 0.12 & \texttt{sdns} & -0.010 & 0.19 \\ 
  \texttt{northward\_wind} & $\cdot$ & 0.13 & \texttt{sla} & -0.012 & 0.09 & \texttt{density} & $\cdot$ & 0.19 \\ 
  \texttt{CHL} & -0.141 & 0.12 & \texttt{Fe} & $\cdot$ & 0.08 & \texttt{disp\_bw\_sla} & 0.004 & 0.16 \\ 
  \texttt{AOU} & $\cdot$ & 0.12 & \texttt{nitrate} & $\cdot$ & 0.08 & \texttt{wind\_speed} & $\cdot$ & 0.15 \\ 
  \texttt{p3} & 0.007 & 0.11 & \texttt{sst} & -0.028 & 0.07 & \texttt{phosphate} & $\cdot$ & 0.14 \\ 
  \texttt{PO4} & 0.024 & 0.11 & \texttt{PP} & $\cdot$ & 0.07 & \texttt{AOU} & $\cdot$ & 0.12 \\ 
  \texttt{disp\_bw\_sla} & 0.001 & 0.11 & \texttt{vgosa} & 0.014 & 0.07 & \texttt{silicate} & $\cdot$ & 0.12 \\ 
  \texttt{Si} & 0.014 & 0.09 & \texttt{ftle\_bw\_sla} & $\cdot$ & 0.07 & \texttt{o2sat} & 0.017 & 0.09 \\ 
  \texttt{sdns} & $\cdot$ & 0.09 & \texttt{disp\_bw\_sla} & $\cdot$ & 0.07 & \texttt{sdns} & $\cdot$ & 0.07 \\ 
  \texttt{wind\_speed} & $\cdot$ & 0.08 & \texttt{ugosa} & $\cdot$ & 0.06 & \texttt{ugos} & $\cdot$ & 0.06 \\ 
  \texttt{ugosa} & $\cdot$ & 0.07 & \texttt{wind\_speed} & $\cdot$ & 0.05 & \texttt{sst} & -0.004 & 0.04 \\ 
  \texttt{density} & $\cdot$ & 0.07 & \texttt{eastward\_wind} & $\cdot$ & 0.04 & \texttt{PP} & -0.008 & 0.04 \\ 
  \texttt{O2} & 0.071 & 0.06 & \texttt{o2sat} & -0.004 & 0.04 & \texttt{sss} & $\cdot$ & 0.03 \\ 
  \texttt{wind\_stress} & 0.011 & 0.06 & \texttt{O2} & $\cdot$ & 0.03 & \texttt{PO4} & -0.001 & 0.03 \\ 
  \texttt{sss} & $\cdot$ & 0.03 & \texttt{b1} & $\cdot$ & 0.02 & \texttt{Fe} & -0.011 & 0.02 \\ 
  \texttt{NO3} & $\cdot$ & 0.03 & \texttt{AOU} & $\cdot$ & 0.02 & \texttt{CHL} & $\cdot$ & 0.02 \\ 
  \texttt{sdns} & $\cdot$ & 0.03 & \texttt{PO4} & $\cdot$ & 0.01 & \texttt{O2} & $\cdot$ & 0.02 \\ 
  \texttt{oxygen} & $\cdot$ & 0.03 & \texttt{oxygen} & $\cdot$ & 0.01 & \texttt{nitrate} & $\cdot$ & 0.02 \\ 
  \texttt{phosphate} & $\cdot$ & 0.03 & \texttt{salinity} & $\cdot$ & 0.01 & \texttt{Si} & $\cdot$ & 0.01 \\ 
  \texttt{Fe} & $\cdot$ & 0.01 & \texttt{sss} & $\cdot$ & $\cdot$ & \texttt{NO3} & $\cdot$ & $\cdot$ \\ 
  \texttt{PHYC} & $\cdot$ & 0.01 & \texttt{NO3} & $\cdot$ & $\cdot$ & \texttt{PHYC} & $\cdot$ & $\cdot$ \\ 
  \texttt{salinity} & $\cdot$ & 0.01 & \texttt{PHYC} & $\cdot$ & $\cdot$ & \texttt{oxygen} & $\cdot$ & $\cdot$ \\ 
  \texttt{sst} & $\cdot$ & $\cdot$ & \texttt{sdns} & $\cdot$ & $\cdot$ & \texttt{salinity} & $\cdot$ & $\cdot$ \\ 
  \texttt{conductivity} & $\cdot$ & $\cdot$ & \texttt{conductivity} & $\cdot$ & $\cdot$ & \texttt{conductivity} & $\cdot$ & $\cdot$ \\ 
   \toprule
\end{tabular}
\caption{(Part 1 of 2) Stability of $\beta$ coefficients (measured by how
  frequently nonzero it was estimated in subsamples) in clusters 1 through 3,
  shown along with original coefficient estimates in the full data. Within each
  column, variables are sorted by selection probability. Zeros are shown as
  $\cdot$, to distinguish with $0.000$ or $-0.000$ which are small but nonzero
  numbers. As was mentioned in Table~\ref{tab:environmental-covariates-labels},
  the short hand of \texttt{sdns} and \texttt{sdes} is used for
  \texttt{surface\_downward\_northward\_stress} and
  \texttt{surface\_downward\_northward\_stress}, and the suffixes
  \texttt{\_WOA\_clim} are omitted.}
\label{tab:stability-beta-1}
\end{sidewaystable}

\begin{table}[ht]
\centering
\begin{tabular}{rrr|rrr}
  \toprule
Cluster 4 & Cluster 4 & Cluster 4 & Cluster 5 & Cluster 5 & Cluster 5 \\ 
Variable & Estim. & P(nonzero) & Variable & Estim. & P(nonzero) \\ 
  \toprule
  \texttt{sla} & 0.002 & 0.28 & \texttt{p2} & 0.013 & 0.96 \\ 
  \texttt{wind\_speed} & 0.033 & 0.25 & \texttt{sla} & -0.022 & 0.93 \\ 
  \texttt{p3} & -0.003 & 0.23 & \texttt{p4} & -0.006 & 0.91 \\ 
  \texttt{p4} & -0.002 & 0.23 & \texttt{p3} & 0.010 & 0.83 \\ 
  \texttt{ftle\_bw\_sla} & 0.000 & 0.23 & \texttt{PP} & $\cdot$ & 0.78 \\ 
  \texttt{wind\_stress} & $\cdot$ & 0.22 & \texttt{ugos} & 0.058 & 0.74 \\ 
  \texttt{silicate} & -0.045 & 0.22 & \texttt{p1} & 0.008 & 0.70 \\ 
  \texttt{b2} & $\cdot$ & 0.21 & \texttt{disp\_bw\_sla} & -0.003 & 0.63 \\ 
  \texttt{p1} & -0.002 & 0.20 & \texttt{eastward\_wind} & 0.021 & 0.47 \\ 
  \texttt{p2} & -0.006 & 0.20 & \texttt{par} & 0.002 & 0.33 \\ 
  \texttt{par} & -0.002 & 0.19 & \texttt{ftle\_bw\_sla} & -0.011 & 0.32 \\ 
  \texttt{vgos} & $\cdot$ & 0.18 & \texttt{b2} & $\cdot$ & 0.31 \\ 
  \texttt{eastward\_wind} & $\cdot$ & 0.18 & \texttt{ugosa} & -0.050 & 0.31 \\ 
  \texttt{vgosa} & 0.025 & 0.14 & \texttt{vgos} & -0.006 & 0.30 \\ 
  \texttt{disp\_bw\_sla} & $\cdot$ & 0.14 & \texttt{nitrate} & $\cdot$ & 0.30 \\ 
  \texttt{sdns} & $\cdot$ & 0.11 & \texttt{sdns} & $\cdot$ & 0.27 \\ 
  \texttt{ugosa} & 0.007 & 0.10 & \texttt{sss} & -0.067 & 0.26 \\ 
  \texttt{northward\_wind} & $\cdot$ & 0.10 & \texttt{Fe} & 0.001 & 0.26 \\ 
  \texttt{sss} & $\cdot$ & 0.08 & \texttt{vgosa} & $\cdot$ & 0.23 \\ 
  \texttt{ugos} & $\cdot$ & 0.08 & \texttt{northward\_wind} & $\cdot$ & 0.22 \\ 
  \texttt{b1} & 0.038 & 0.07 & \texttt{PHYC} & 0.010 & 0.21 \\ 
  \texttt{Si} & $\cdot$ & 0.07 & \texttt{wind\_speed} & $\cdot$ & 0.19 \\ 
  \texttt{AOU} & $\cdot$ & 0.07 & \texttt{silicate} & -0.001 & 0.19 \\ 
  \texttt{Fe} & $\cdot$ & 0.06 & \texttt{density} & -0.038 & 0.18 \\ 
  \texttt{sdns} & -0.007 & 0.06 & \texttt{sst} & -0.070 & 0.17 \\ 
  \texttt{density} & 0.051 & 0.06 & \texttt{wind\_stress} & 0.025 & 0.15 \\ 
  \texttt{o2sat} & $\cdot$ & 0.06 & \texttt{sdns} & 0.011 & 0.15 \\ 
  \texttt{phosphate} & 0.012 & 0.05 & \texttt{o2sat} & -0.022 & 0.14 \\ 
  \texttt{nitrate} & $\cdot$ & 0.04 & \texttt{phosphate} & -0.132 & 0.12 \\ 
  \texttt{sst} & $\cdot$ & 0.03 & \texttt{b1} & 0.046 & 0.11 \\ 
  \texttt{PP} & $\cdot$ & 0.03 & \texttt{AOU} & $\cdot$ & 0.07 \\ 
  \texttt{PHYC} & $\cdot$ & 0.03 & \texttt{CHL} & $\cdot$ & 0.05 \\ 
  \texttt{PO4} & $\cdot$ & 0.03 & \texttt{Si} & -0.006 & 0.04 \\ 
  \texttt{CHL} & $\cdot$ & 0.01 & \texttt{oxygen} & $\cdot$ & 0.04 \\ 
  \texttt{O2} & $\cdot$ & 0.01 & \texttt{PO4} & $\cdot$ & 0.03 \\ 
  \texttt{salinity} & $\cdot$ & 0.01 & \texttt{O2} & $\cdot$ & 0.01 \\ 
  \texttt{NO3} & $\cdot$ & $\cdot$ & \texttt{NO3} & $\cdot$ & $\cdot$ \\ 
  \texttt{oxygen} & $\cdot$ & $\cdot$ & \texttt{salinity} & $\cdot$ & $\cdot$ \\ 
  \texttt{conductivity} & $\cdot$ & $\cdot$ & \texttt{conductivity} & $\cdot$ & $\cdot$ \\ 
   \toprule
\end{tabular}
\caption{(Part 2 of 2) Continuing from table~\ref{tab:stability-beta-1}, we show
  the stability of $\beta$ coefficients in clusters 4 through 5, along with
  original coefficient estimates in the full data.}
\label{tab:stability-beta-2}
\end{table}

\section*{Supplement E: Probabilistic gating and residual analysis}

First, we describe the procedure of drawing cluster membership of a given
particle $\y_{i}^{(t)}$. This procedure can be thought of as
\textit{probabilistic gating} (classification) of particles using our mixture-of-regression
model. We make use of the posterior membership probability
$\gamma_{itk}(\hat \alpha, \hat \beta, \hat \Sigma)$ -- responsibilities,
defined as in
(11)
-- which can be calculated from the
estimated model parameters $(\hat \alpha, \hat \beta, \hat \Sigma)$. Then, from
each particle $y_{i}^{(t)}$, a random draw of particle membership $Z_{i}^{(t)}$
can be made as follows:
$$ Z_{i}^{(t)} = k \text{ with probability }\gamma_{itk}(\hat \alpha, \hat \beta, \hat \Sigma), \text{ for } k=1,\cdots, K. $$
Given the randomly drawn latent membership $Z_i^{(t)}$ of each $\y_i^{(t)}$, we
can calculate the residuals by subtracting from $\y_i^{(t)}$ the assigned
cluster's mean. The set of residuals produced are written as,
$$\{\br_{i,t,k} \in \mathbb{R}^{3}\}_{i=1,\cdots, n_{tk}},$$
where $n_{tk}$ is the number of particles that were classified as cluster
$k$.  (Note, the results in the remainder of this section are based on a newly
fit 10-cluster 3-dimensional \textit{flowmix} model using binned MGL1704 cruise
data with finer bins compared to the model fit in
Section~4.0.2,
in order to calculate higher-resolution binned
residuals.)

\begin{figure}
  \centering
  \includegraphics[width = \linewidth]{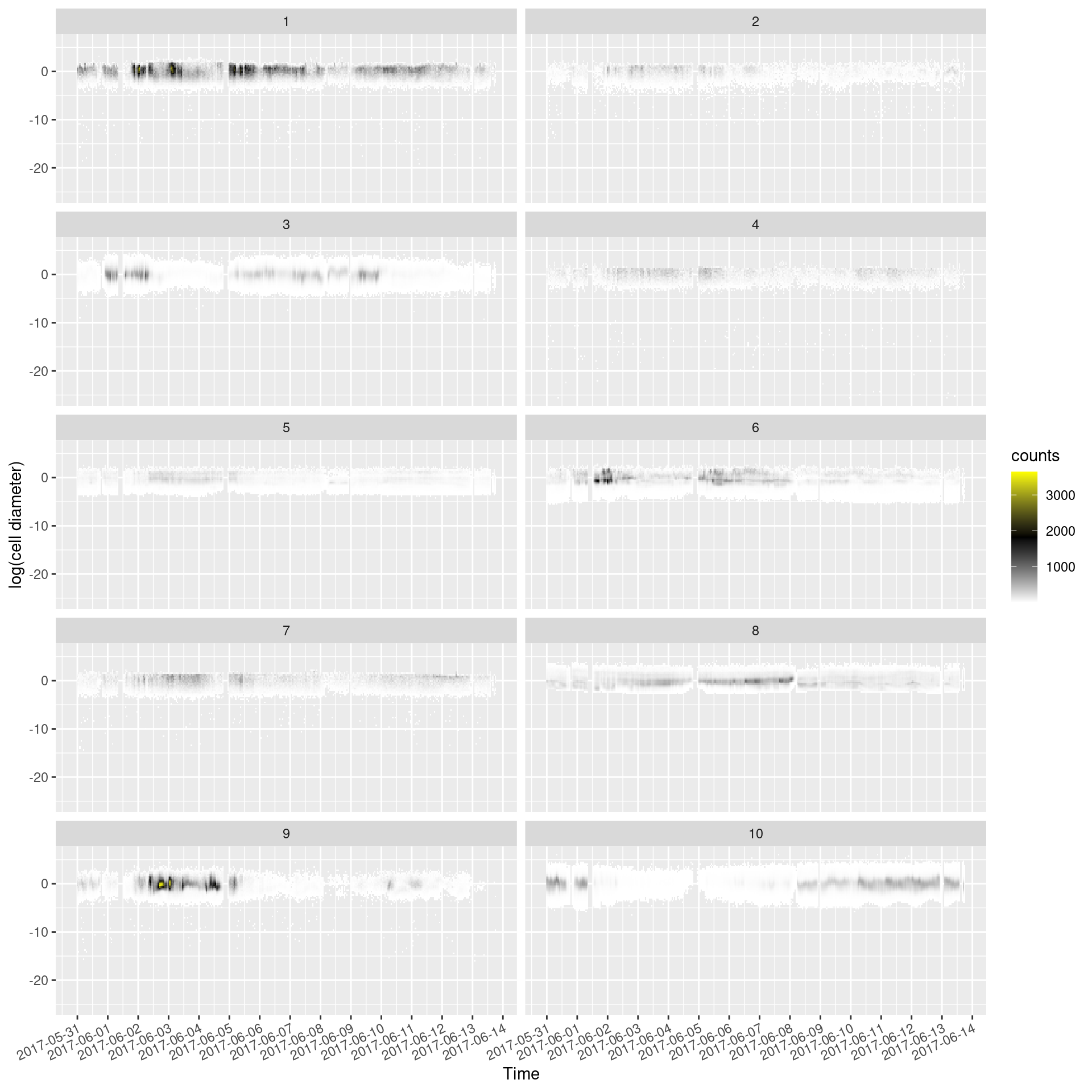}
  \caption{\it Each of the ten panels show the binned, 1-dimensional sphered
    residuals from a 10-cluster 3d model. The dimension (out of three) that is
    shown is the log cell diameter (called Diam elsewhere). For full details about the sphered
    residuals, see Supplement E.}
  \label{fig:3d-residuals}
\end{figure}

Figure~\ref{fig:3d-residuals} shows the binned residuals for one dimension --
cell diameter -- with separate panel for the ten clusters. The color of each bin
represents the total biomass in that bin. From simple visual inspection of this
plot, it is unclear if any time dependence persists in the residuals. In order
to go beyond visual inspection, we devise an approach to quantitatively measure
time-dependency based on Wasserstein distance of sphered residual distributions,
presented next.

\begin{figure}
  \centering
  \includegraphics[width = \linewidth]{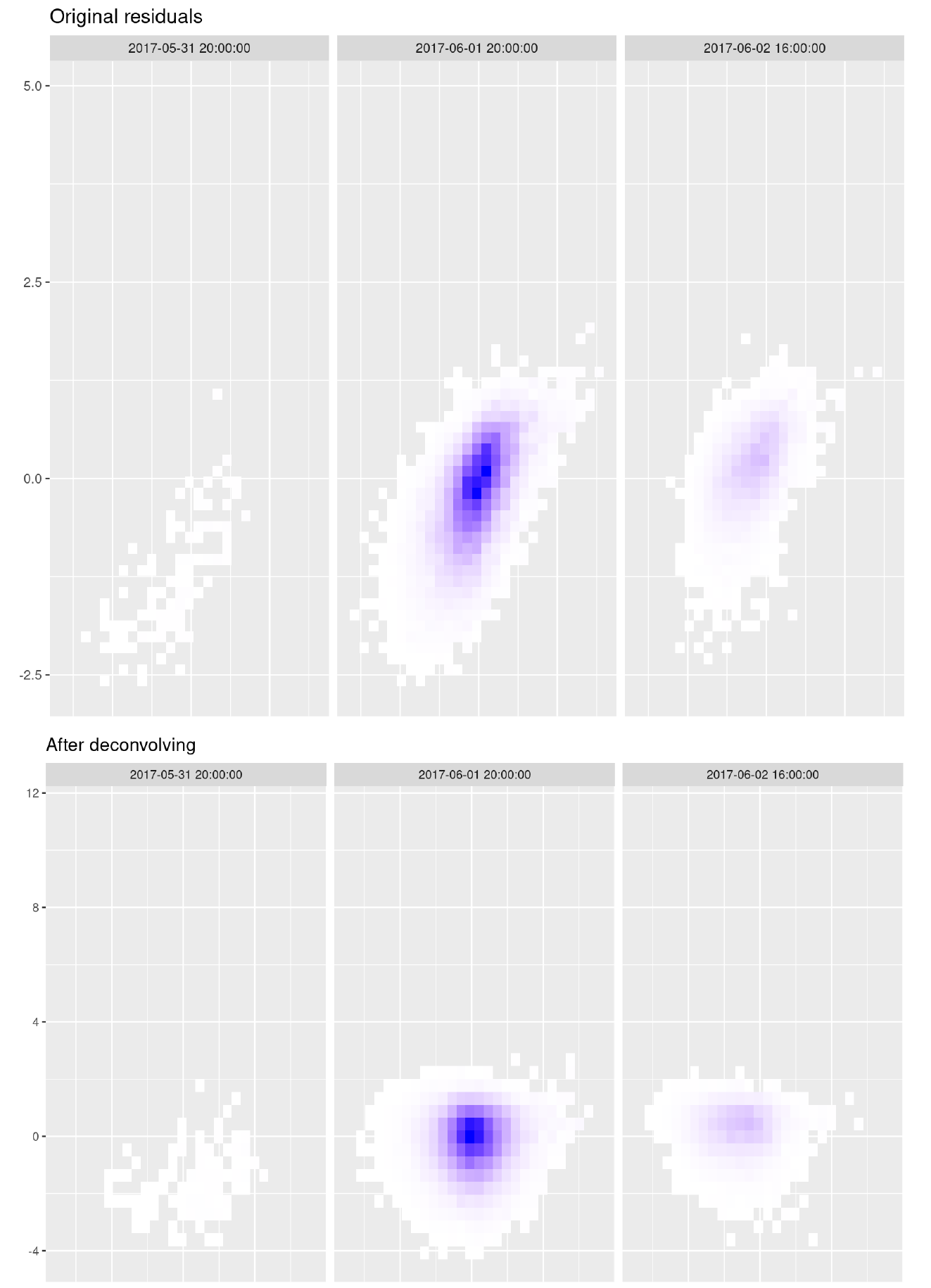}
  \caption{\it The top row shows, at three time points, two dimensions (diam and
    chl) of binned residuals from one cluster. The bottom row shows the
    sphered version of those same residuals. The sphering is done by
    left-multiplication of the square root of the estimated covariance matrix
    ($\bSigma_k^{-1/2}$) on the residuals. For full details, see Supplement E.}
  \label{fig:3d-residuals-sphered}
\end{figure}

In particular, we investigate whether the $k$'th cluster's 3-dimensional
residuals can be deemed white noise if their distributions $F_t$ over time
$t=1,\cdots, T$ do not have time dependence. In our setting, we use
\textit{sphered} residuals, and use Wasserstein distance as a distance
metric for measuring time dependence.

The residuals $\br_{i,t,k}$ are sphered by left-multiplication with the
square-root of the estimated covariance matrix of the $k$'th population
$\bSigma_k^{-1/2} \in \mathbb{R}^{3 \times 3}$. The sphered residuals are
$\tilde \br_{i,t,k} = \bSigma_k^{-1/2} \br_{i,t,k}$, whose distribution has been
standardized to be close to $\mathcal{N}(0, \bI_3)$ for $3$-by-$3$ identity matrix
$\bI_3$.  The sphered residuals are comparable in size across clusters, which
is important in forming a distance measure that is overall consistent in
scale. Several examples of this step are shown in
Figure~\ref{fig:3d-residuals-sphered}.

\begin{figure}
  \centering
\includegraphics[width = .8\linewidth]{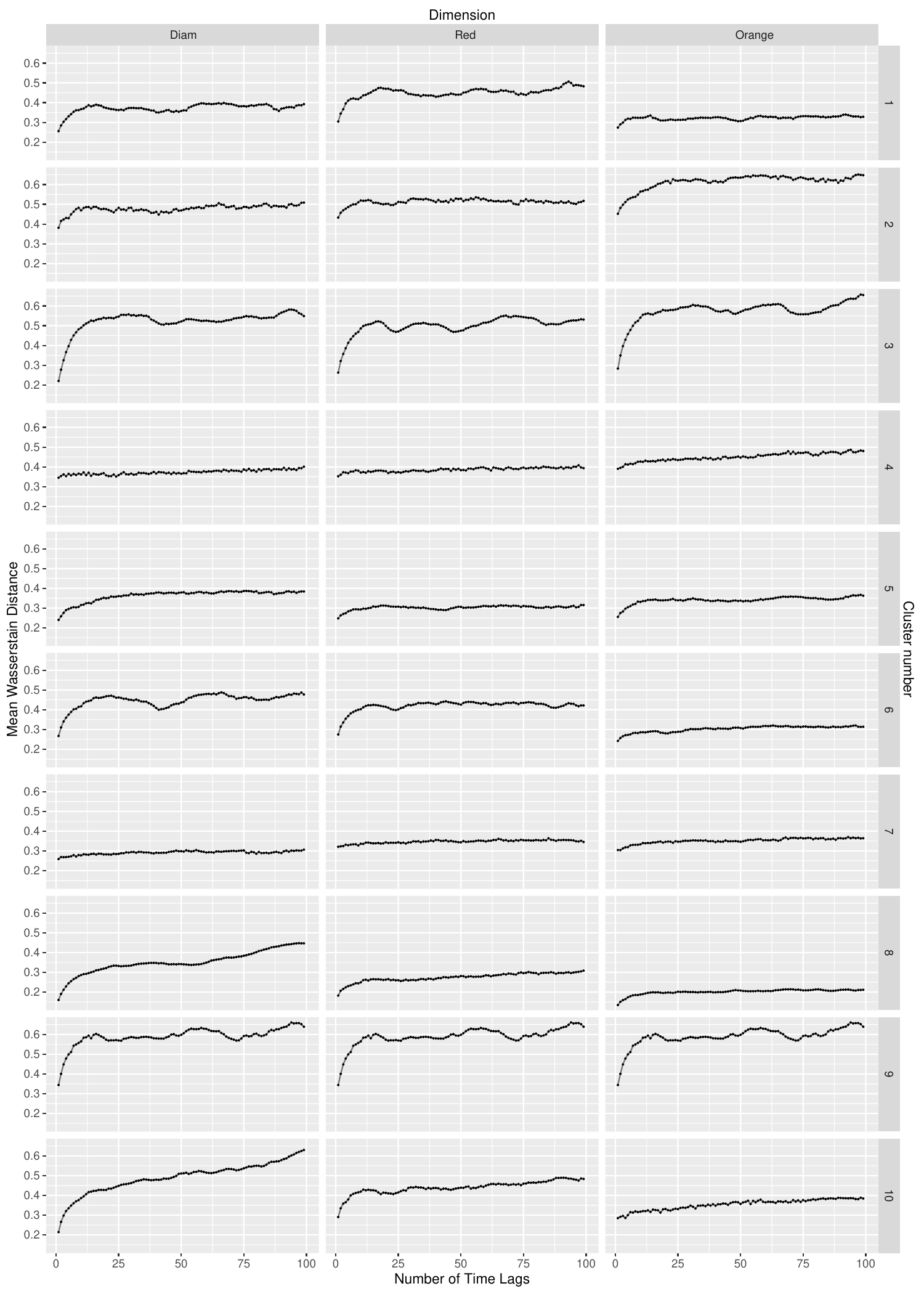}
  \caption{\it Each column is a dimension $m=1,2,3$ corresponding to the
    cytogram axes Diam, Red, and Orange. Each row is a cluster $k$, ten in total. In
    each panel, the y-axis is the average $l$-lagged Wasserstein distance
    for cluster $k$ in dimension $m$, $D_{m,k,l}$, as defined in \eqref{eq:wasserstein} in
    Supplement E, and the x axis is lag $l$. }
  \label{fig:wasserstein}
\end{figure}

Next, we operate on the 1-dimensional sphered residuals
$\{\be_m^T\tilde \br_{i,t,k}\}_d$ for each cluster $k\in\{1,\cdots, K\}$ and
dimension $m\in\{1,2,3\}$. From each time point $t$, we obtained the density
estimate of one dimension by fitting a 100-bin one-dimensional histogram
$h(\be_m^T \tilde \br_{i,t,k}) \in \mathbb{R}^{100}$ for each time point $t$, in
$100$ fixed bins shared across all time points). Then, we calculated the average
of 2-Wasserstein distances between
$\{e_m^T\tilde r_{i,t,k}: t=1,\cdots, T-l\}$ and the $l$-lagged version of it
$\{e_m^T\tilde r_{i,t,k}: t=l+1,\cdots, T\}$:
\begin{equation}\label{eq:wasserstein}
  D_{m,k,l} = \frac{1}{T-l} \sum_{t=1}^{T-l} D_2\left( (h(\be_m^T\tilde
    \br_{i,t,k}), h(\be_m^T \tilde \br_{i,t+l, k}) \right)
\end{equation}
where $D_2(\ba, \bb)$ is a Wasserstein's distance of vector $\ba$ and $\bb$. This is a measure of $l$-lag time dependence since it quantifies the average
displacement in the space of data between the (i) original sphered residual
distributions and (ii) the \textit{time-lagged} version of them. All values of
$D_{m,k,l}$ for each value of dimension $m$ and cluster $k$ are plotted in
Figure~\ref{fig:wasserstein}, in which each panel is a line plot over $l$.

We can see in each panel of Figure~\ref{fig:wasserstein} that all values of
$D_{m, k,l}$ have an increasing trend over $l$. On the right end of each curve,
there is a flat plateau of $\{D_{m,k,l}\}_l$ at larger values of $l$ in each
panel -- this is because a large time-lag explicitly breaks any time-dependence
and effectively guarantees that the distributions under comparison are unrelated. On the
left end of each curve, flatness in $D_{m,k,l}$ indicates $l$-lagged
time-independence for small values of $l$, while a dip in $D_{m, k, l}$
signifies a nonzero time-dependence.  Certain panels of
Figure~\ref{fig:wasserstein} flatness in $D_{m,k,l}$ on the left part of each
curve, indicating non-appreciable $l$-lagged time-dependence for small values of
$l$. On the other hand, certain clusters -- especially $k \in \{3,6,9,10\}$ --
have especially sharp dips for small $l$ which indicates leftover time
dependence in the residuals at short lags.

We suggest two possible explanations for the leftover time dependence: (i) There
are missing environmental covariates that affect the cytograms but are not
accessible to us; and (ii), using smoothly time-varying covariates directly
induces some time dependence in the residuals. Our estimated model supports the
latter explanation -- these clusters ($k \in \{3,6,9,10\}$) correspond to
specific well-known marine phytoplankton populations, and have many nonzero
large $\alpha$ and $\beta$ coefficients predicting their mean and relative
abundance.  The other clusters -- especially $k \in \{4,5\}$ -- are non-specific
populations and have the smallest coefficient values out of all clusters.

\section*{Supplement F: Effect of binning on estimation}

\begin{figure}[ht!]
\includegraphics[width = .5\linewidth]{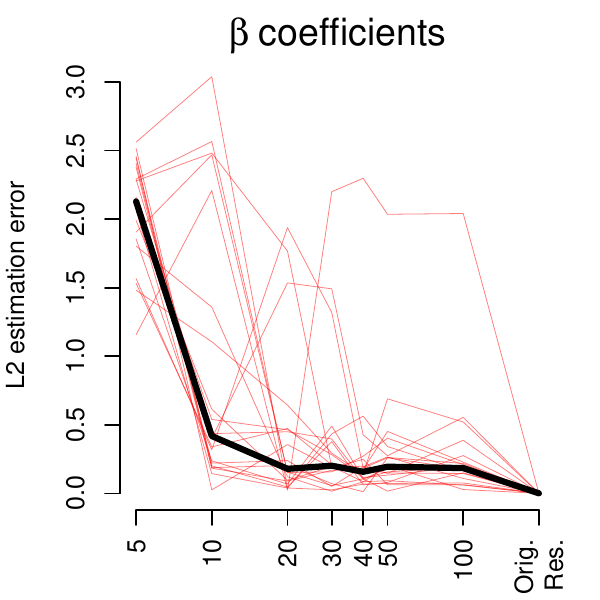}
  \hspace{-5mm}
\includegraphics[width = .5\linewidth]{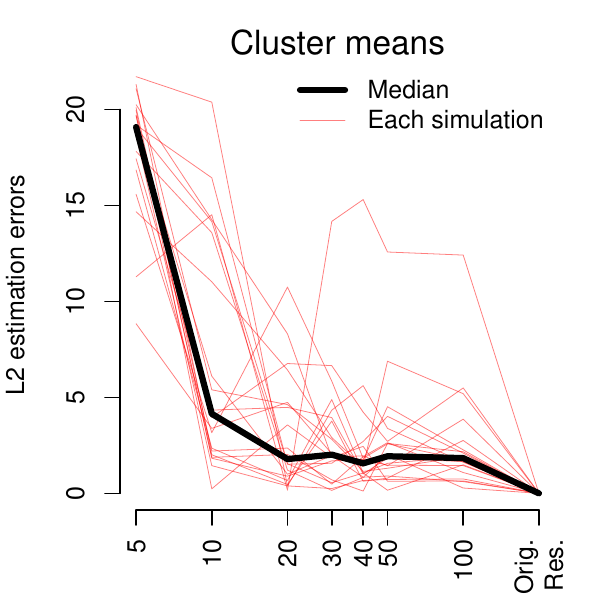}
  \caption{L2 estimation errors of the estimated models from binning, compared
    to the original data, over a varying number of bins (shown in the $x$
    axis). The left panel shows entry-wise L2 estimation error of the $\beta$
    coefficients, and the right panel shows that of the cluster means. (The
    right-most value is of every line of both plots equals zero, and the $x$
    axis is logarithmically spaced.)  These plots show that the estimation error
    incurred by binning decreases rapidly, then plateaus at a low point for a
    relatively coarse bin resolution, of about $B=20$.}
  \label{fig:bin-l2-error}
\end{figure}

Using a simulation, we quantify how much model estimation accuracy suffers as a
result of binning the data. We first generated data according to the model
described in
Figure~4
twenty times. Then, for each
dataset, we binned the data using bin sizes
$B \in \{5, 10, 20, 30, 40, 50, 100\}$, to create coarsely to finely binned
datasets. We estimated models from each dataset, as well as one model from the
original unbinned dataset, and used Algorithm~\ref{alg:match-clusters} to
permute these models' cluster labels to closely match those of a large external
dataset. We then computed the entry-wise L2 estimation error of the estimated
coefficients $\hat \beta$ and cluster means $\{\bmu_{kt}(\hat \beta)\}_{k,t}$ from the
binned data to those from the original data. We then plotted this estimation
error over the number of bins $B$, as shown in
Figure~\ref{fig:bin-l2-error}. The thin red lines show the estimation errors
from each of the twenty simulations, and the black line shows the median of
those values, at each value of $B$. We can see that the estimation error
plateaus to the minimum at a relatively small value of $B$, at about $B=20$.

\begin{algorithm}
\caption{Cluster matching algorithm for two models}\label{alg:match-clusters}
\begin{algorithmic}[1]
\Procedure{Match Clusters}{$(\alpha^1, \beta^1, \bSigma^1), ( \alpha^2, \beta^2, \bSigma^2), \{\y_{it}\}_{i,t} $}\Comment{Two models, and one dataset.}

\State Calculate $\{\gamma^1_{itk}\}_{i,t,k}, \{\gamma^2_{itk}\}_{i,t,k}$ from
each model using
  \begin{equation*}
    \gamma_{itk}( \alpha,  \beta,  \bSigma) = \frac{\phi\left(\y_i^{(t)}; \bmu_{kt}( \beta),
         \bSigma_k\right) \cdot \pi_{kt}( \alpha) }{\sum_{l=1}^L \phi\left(\y_i^{(t)} ;
         \bmu_{lt}( \beta),  \bSigma_l\right) \cdot
      \pi_{lt}( \alpha) },
  \end{equation*}
  \State Form $\bGamma^1, \bGamma^2 \in \R^{(\sum_{t=1}^T n_t) \times K}$ whose rows are
  $\gamma^1_{it\cdot} \in \R^K$ and $\gamma^2_{it\cdot} \in \R^K$,
  \State \hspace{3mm} for $i=1,\cdots, n_t$, and $t=1,\cdots, T$.
  \State Denote $\Pi$ as the set of all permutation maps of $K$ elements
  \Comment{Note that $|\Pi| = K!$.}  \State
  \hspace{4mm}$\pi:\{1, \dots, K\} \to \{1,\cdots, K\}$.  \State Denote
  permutation matrix $\bP_\pi$ of the map $\pi$.  \State Calculate
  $\tilde \pi = \argmax_{\pi \in \Pi} KL( \bGamma^1 \bP_\pi, \bGamma^2)$.  \State
  \Comment{$KL(\bP_1, \bP_2):= \sum_{i,j}\bQ_{i,j}$ where $\bQ=\bP_1 \log(\bP_1/\bP_2)$ is
    calculated entrywise for probability matrices $\bP_1, \bP_2$.}  \State
  \textbf{return} $\{\tilde \pi(k)\}_{k=1,\cdots,K}$ as the cluster labels of
  model 1 closest to model 2 clusters $1,\cdots,K$.  \EndProcedure
\end{algorithmic}
\end{algorithm}

\section*{Supplement G: Model performance with non-Gaussian data}

In order to investigate the model performance when data deviate from mixtures of
Gaussians, we consider two such scenarios in a simulation study. The first is
when the within-cluster data are more heavy-tailed than Gaussian and the other
is when the data are more skewed.  In the data setup described in
Figure~4
and
Section~3.1.1
(with $\sigma_{\text{add}}=0$ since
we are not concerned about covariate noise here) we replace the $\cN(0,1)$
distribution for generating data in each cluster at each time point, with either
a t-distribution or a skewed-Normal distribution. These two replacements are
described in detail next.

\begin{enumerate}
\item{\textbf{Heavy tails.}} The $\cN(0,1)$ is replaced with a t-distribution
  with degrees of freedom $\nu \in \{3,5,10,20,40,100\}$, further divided by
  $\sqrt{\nu/(\nu-2)}$ so that the variance is 1, the same as the original
  distribution.
\item {\textbf{Skew-normal.}} The $\cN(0,1)$ is replaced with a skew-Normal
  distribution \citep{azzalini-2013}. The probability density function of a
  skew-Normal random variable $X \sim SN(\alpha, \omega, \xi)$ with shape, scale
  and location parameters $\alpha, \omega$ and $\xi$ is:
$$\frac {2}{\omega {\sqrt {2\pi }}}e^{-{\frac {(x-\xi )^{2}}{2\omega ^{2}}}}\int
_{-\infty }^{\alpha \left({\frac {x-\xi }{\omega }}\right)}{\frac {1}{\sqrt
    {2\pi }}}e^{-{\frac {t^{2}}{2}}}\ dt. $$ For a given choice of $\alpha$, in
order to scale and center $Z$ to have mean 0 and variance 1, we further use
$\omega(\alpha)=\sqrt{\frac{1}{1-2/\pi} \frac{\alpha^2}{1+\alpha^2}}$ as the
scale parameter, and also shift the distribution by
$\omega(\alpha) \cdot (\alpha / \sqrt{1+\alpha^2}) \cdot \sqrt{2/\pi}$. The
values of $\alpha$ we consider are $\{0, 0.5, \cdots, 2\}$.
\end{enumerate}

\begin{figure}
  \centering 
\includegraphics[width=\linewidth]{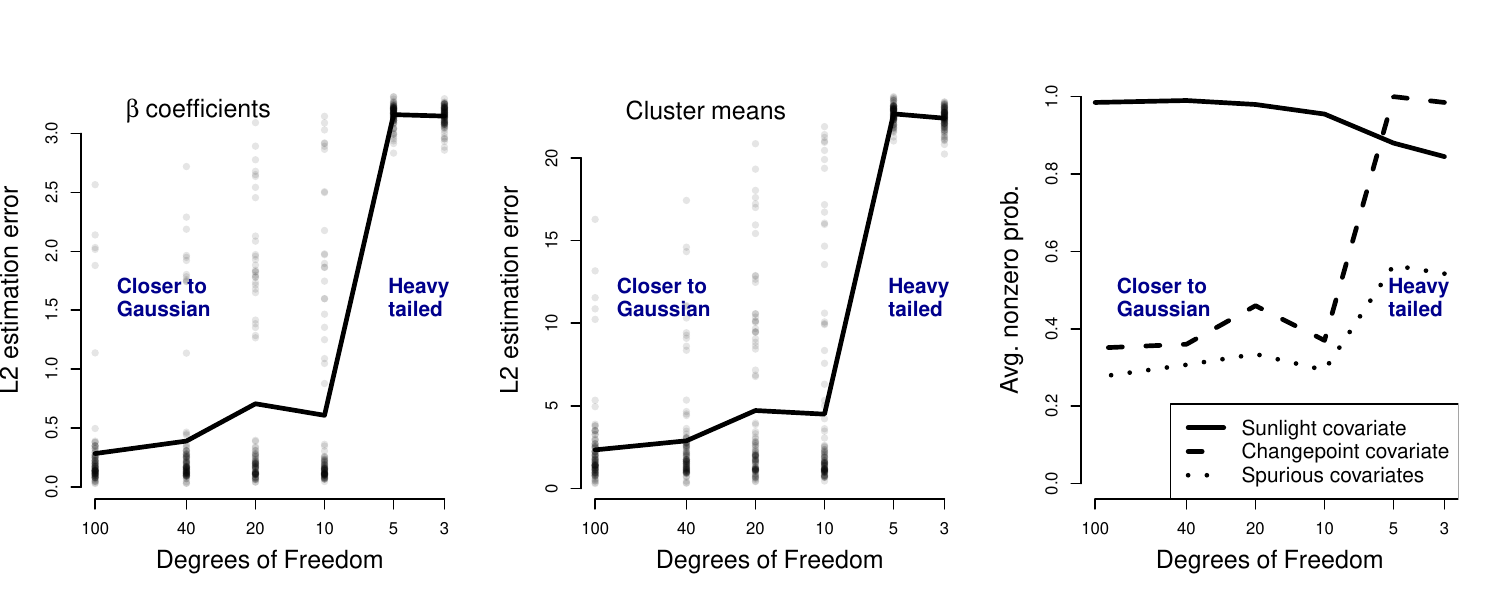} 
  \includegraphics[width=\linewidth]{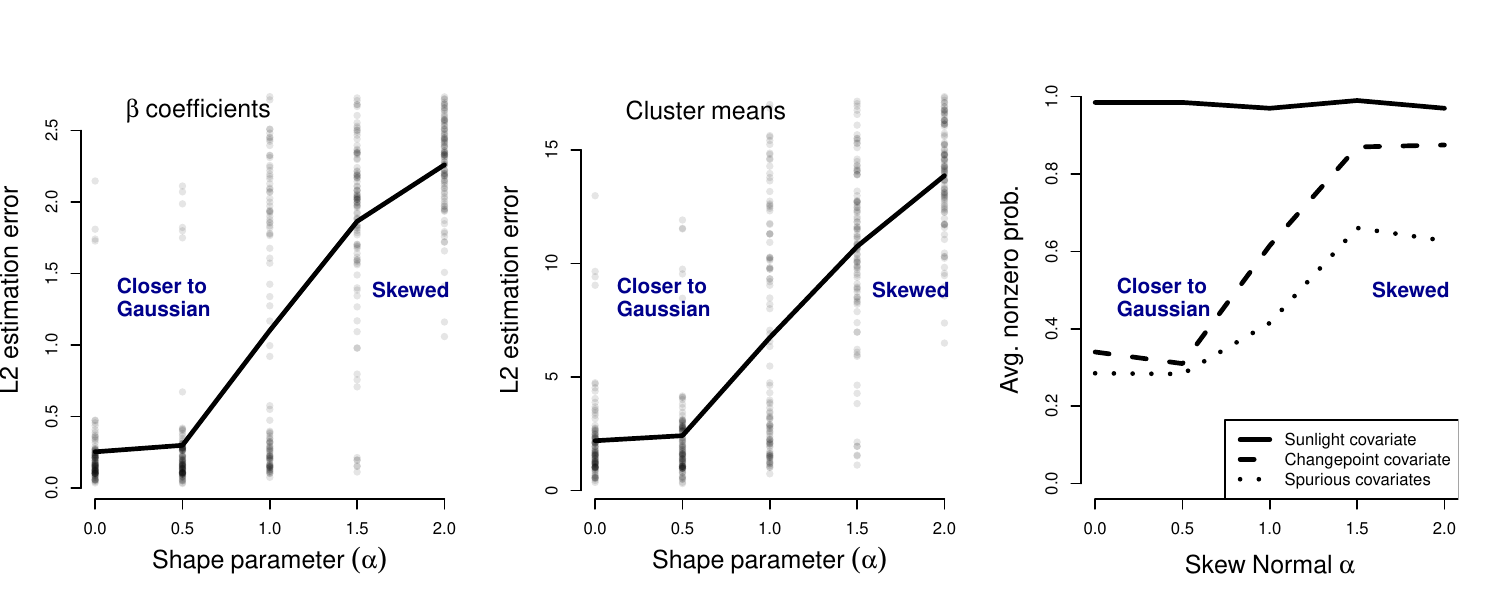} 
  \caption{The top row shows the simulation results of the heavy-tail
    simulations, and the bottom row shows the results of the skewed-data
    simulations, described in Supplement F. In each row, three model performance
    metrics are shown -- the left and middle panel show the entry-wise L2
    estimation error of the $\beta$ coefficients (left panel) and the cluster
    means (middle panel) with the average shown in the thick black line, and the
    right panel shows the variable selection performance. For each simulation
    setup (degrees of freedom or skewness parameter), $100$ different datasets
    were generated and models were estimated on each dataset.}
  \label{fig:heavytail-skewed-model-performance}
\end{figure}

The simulations from the two setups show that the estimated $\beta$ model
coefficients and cluster means predictably become less accurate as the tails
become heavier and the skewness increases (see left and middle panels of
Figure~\ref{fig:heavytail-skewed-model-performance}).

On the other hand, we see some encouraging robustness in variable selection. For
each setup -- shown in the rightmost panels of
Figure~\ref{fig:heavytail-skewed-model-performance} -- we measured the
probability of certain covariates being estimated to be nonzero, averaged across
simulations. Recall from the description in
Section~3.1.1
that the sunlight covariate is solely responsible for mean movement in both
clusters. The nonzero probability of the sunlight covariate remains high and
close to $1$ even for extremely heavy-tailed data.  The changepoint covariates
and the eight spurious covariates -- which play no role in generating the
cluster mean movement -- have nonzero probabilities which are initially low but
increase as the tails get heavier. With extremely heavy tails -- degrees of
freedom equal to $3$ or $5$ -- the nonzero probability of the changepoint
covariate grows to close to $1$, and the nonzero probability of the sunlight
covariate decreases slightly, showing that the estimated model tends to
erroneously predict cluster mean movement to be piece-wise constant instead of
being driven by sunlight.

\bibliographystyle{unsrtnat}
\bibliography{flowmix}